\documentclass[12pt]{article}
 \pdfoutput=1

\usepackage[utf8]{inputenc}
\usepackage{amssymb}
\usepackage{amsmath}
\usepackage{amsthm}
\usepackage{float}
\usepackage{setspace}
\usepackage{subcaption}
\usepackage{siunitx}
\usepackage{bbm}
\usepackage[margin=1in]{geometry}
\usepackage{hyperref}
\usepackage{multicol}
\usepackage{caption}
\usepackage{float}
\usepackage{afterpage}
\usepackage{longtable}
\usepackage{array}
\usepackage{booktabs}
\usepackage{mathrsfs}  
\usepackage{placeins}

\newcolumntype{P}{>{\raggedright\arraybackslash}p{3cm}}
\newcolumntype{T}{>{\raggedright\arraybackslash}p{1.5cm}}
\newcolumntype{R}{>{\raggedright\arraybackslash}p{5cm}}

\usepackage[
    backend=biber,
    maxbibnames=10,
    date=year,
    style=authoryear,
]{biblatex}
\AtEveryBibitem{
    \clearfield{urlyear}
    \clearfield{urlmonth}
}

\DeclareSourcemap{
  \maps[datatype=bibtex]{
    \map[overwrite]{
      \step[fieldsource=doi, final]
      \step[fieldset=url, null]
      \step[fieldset=eprint, null]
    }  
  }
}
\newcommand{\norm}[1]{\left\lVert#1\right\rVert}
\usepackage{xcolor}

\hypersetup{
    colorlinks=true,
    linkcolor=magenta,
    filecolor=magenta,  
    citecolor=magenta,
    urlcolor=magenta,
    pdftitle={Triple Difference Designs with Heterogeneous Treatment Effects}
    }

\theoremstyle{plain}
\newtheorem{proposition}{Proposition}
\theoremstyle{definition}
\newtheorem{proofn}{Proof}
\newtheorem{assumption}{Assumption}

\usepackage{accents}

\usepackage{graphicx}
\author{Laura Caron \footnote{rcaron@haverford.edu} \\ Haverford College }

\title{Triple Difference Designs with Heterogeneous Treatment Effects}
\date{September 2026}
\begin{document}
\maketitle

\begin{abstract}
\singlespacing
\noindent Triple difference designs have become increasingly popular in empirical economics. The advantage of a triple difference design is that, within a treatment group, it allows another subgroup of the population -- potentially less impacted by the treatment -- to serve as a comparison for the subgroup of interest. 
While literature on difference-in-differences has discussed heterogeneity in treatment effects between treated and control groups or over time, relatively little attention has been given to triple difference designs and the implications of heterogeneity in treatment effects in this setting. In this paper, I show that the parameter identified under common triple difference assumptions does not allow for causal interpretation of differences between subgroups when subgroups may differ in their underlying (unobserved) treatment effects. I propose a new parameter of interest, the controlled difference in average treatment effects on the treated, which allows for causal comparisons between subgroups. 
I then propose identification assumptions and doubly-robust estimators for this parameter. I use a simulation study to highlight the desirable finite-sample properties of these estimators, as well as to show the difference between the two parameters. An empirical application shows the importance of considering treatment effect heterogeneity in practical applications.

\end{abstract}

\newpage

\setcounter{section}{0}

\section{Introduction}
Triple difference designs (also known as triple difference-in-difference or difference-in-difference-in-difference designs) are, increasingly, a popular research design for estimating causal effects. Triple difference designs rely on comparisons across three dimensions, for example, across treatment assignment, time, and another characteristic of interest. The simplest design takes on a $2 \times 2 \times 2$ form, with binary variation in each of the three dimensions. Triple difference designs allow researchers to identify causal effects in cases when comparison across only one dimension (for example, a pre-post comparison) or two dimensions (for example, a difference-in-difference design comparing trends over time between treatment and control group) are confounded. In particular, triple difference designs are often useful when the parallel trends assumption required for a difference-in-difference design is not satisfied. 

The use of triple difference designs has been increasing in recent years \parencite{oldenTripleDifferenceEstimator2022}, with more than 1 in 20 papers published in top economics journals in 2025 using triple differences \parencite{goldsmith-pinkhamEmpiricalMethodsTrends2026, goldsmith-pinkhamTrackingCredibilityRevolution2026}. Still, their properties are still little-studied. Prior work has not fully addressed how heterogeneity in treatment effects can shape the parameters identified in triple difference designs. A recent review of the literature on difference-in-difference methods has pointed out that further study of triple difference methods and guidance for researchers is needed \parencite{rothWhatsTrendingDifferenceindifferences2023}. This paper attempts to fill this gap by offering a formal discussion of triple difference designs and estimators under treatment effect heterogeneity. 

First, I discuss the identification and interpretation of a triple difference parameter that is often of interest to researchers but identified under assumptions differing from those in prior econometric literature. I consider a setup with the three sources of variation being time, treatment assignment, and subgroup. For example, this setup would apply to the analysis of a policy implemented at the state level, which is thought to affect one group of individuals, such as married people, more strongly than others. Other work on triple difference designs, such as \textcite{oldenTripleDifferenceEstimator2022,ortiz-villavicencioBetterUnderstandingTriple2025}, has shown that a triple difference design can identify the average treatment effect on the treated (ATT) for the subgroup of interest (eg, married people) by assuming that the comparison subgroup (eg, unmarried people) is unaffected by the treatment. However, in practice, researchers may not be willing to make this assumption and instead are interested in estimating a different parameter: the difference in ATTs between the subgroup of interest and the comparison subgroup, which I call the DATT. In this sense, in this paper, treatment effect heterogeneity is especially important for triple difference designs because the ways treatment effects differ between subgroups may be directly of interest to researchers. I discuss the assumptions needed to identify this parameter.

The distinction between the ATT and DATT is important in many triple difference applications. In some designs, there is no assumption that any subgroup is unaffected by the treatment and, instead, the DATT capturing the relative effect of the treatment between the subgroups is the primary parameter of interest. For example, \textcite{derenoncourtMinimumWagesRacial2020} use a triple difference strategy to study whether an increase in the minimum wage had a larger impact on workers in the US South relative to other regions. In other cases, the unaffected subgroup assumption is made but may need further evidence to support it. For example, \textcite{gruberIncidenceMandatedMaternity1994} compares the impacts of legislation requiring insurance coverage for childbirth costs on married men and women of childbearing age, compared to single men and those older than childbearing age. The comparison subgroup would not be unaffected if there are spillovers, such as employers substituting between groups of workers. A similar point is made in \textcite{baumEffectStateMaternity2003}. When there is doubt about whether one subgroup is truly unaffected by the treatment, the DATT can still be identified.  

Next, I consider the interpretation of the DATT when there are potentially other dimensions of heterogeneity in treatment effects. I show that, when observations' covariates are both correlated with their treatment effects (their sensitivities to the treatment) and their subgroup status, the DATT may not represent the causal effect of belonging to the subgroup of interest relative to the comparison subgroup. For example, this case arises when members of one subgroup would have been more sensitive to the treatment than members of the comparison subgroup \textit{even if they had belonged to the comparison subgroup}. 

To instead identify the causal effect of subgroup status, I propose a new parameter of interest, the controlled DATT or CDATT. I discuss its identification and estimation in the staggered treatment design, addressing the concerns raised in \textcite{strezhnevDecomposingTripleDifferencesRegression2023} and in line with recent difference-in-differences literature (eg, \textcite{rothWhatsTrendingDifferenceindifferences2023}). I introduce and discuss the assumptions necessary to identify the CDATT and compare them with those identifying the DATT. 

In many applications, the CDATT is as much of interest as the DATT. For example, \textcite{gruberIncidenceMandatedMaternity1994} studies whether, following mandates for insurance coverage of childbirth costs, workers' wages fell according to their valuation of the coverage. That is, the author is interested in differences between the subgroups \textit{caused by} likelihood to use the benefits (proxied by demographics, eg, being of childbearing age) rather than, for example, differences caused by differences in occupations between the groups. Similarly, \textcite{matsaFemaleStyleCorporate2013b} use a triple difference strategy to study the impacts of gender quotas for corporate boards in Norway, comparing impacts on publicly listed companies relative to unlisted companies who were exempt. An interesting question is whether the differential impacts of the quota on listed companies are caused by being listed (and, therefore, affected by the regulation) or other differences in firm characteristics, such as firm size, between the subgroups.

I propose estimators for the CDATT and derive their asymptotic properties. When heterogeneity in underlying treatment effects can be modeled using observable characteristics, the CDATT can be identified using an inverse propensity score weighting (IPW) estimator, regression adjustment (RA), or a doubly-robust estimator combining both of these.

Next, I present a Monte Carlo simulation study calibrated to the data and approach in \textcite{gruberIncidenceMandatedMaternity1994} to show the finite-sample properties of these estimators in a case where the DATT and CDATT differ. When treatment effects are correlated with covariates that are also correlated with subgroup status, estimates of the DATT diverge sharply from estimates of the CDATT. 

Finally, I apply these estimators to re-analyze the data in \textcite{gruberIncidenceMandatedMaternity1994}. This analysis attempts to assess the impacts of a policy requiring insurance to cover the costs of childbirth, thereby increasing the cost to employers of hiring workers likely to use these benefits. The author uses a triple difference design to understand whether these costs can be differentially passed on to targeted workers on the basis of likelihood to use benefits (proxied by demographic groups). Although analysis of the DATT would seem to suggest that such cost-shifting is possible, analysis of the CDATT offers weaker evidence of this. This would suggest that any differential cost-shifting may occur on the basis of other personal or job characteristics, rather than demographic group. Researchers estimating DATT parameters should be cautious to avoid interpreting these as CDATT parameters.

\textbf{Contributions.} This work contributes directly to the literature on triple difference designs (eg, \textcite{gruberIncidenceMandatedMaternity1994, oldenTripleDifferenceEstimator2022,ortiz-villavicencioBetterUnderstandingTriple2025}). I focus on the setting where triple difference setups are used to study how treatment effects differ between a subgroup of interest and a comparison subgroup. I discuss two parameters of interest that can be identified in triple difference designs that differ from the $ATT$ discussed by \textcite{oldenTripleDifferenceEstimator2022,ortiz-villavicencioBetterUnderstandingTriple2025}. In both of those papers, the authors assume that a comparison subgroup is available which is entirely unaffected by the treatment and focus on identification of the ATT for the subgroup of interest. My work discusses applications of triple difference designs to cases where this assumption may not be needed or justified and, instead, researchers are interested in understanding the heterogeneity in treatment effects between subgroups. To do this, I focus on the difference in ATTs between the subgroups as the parameter of interest or the only parameter that can be identified, and I discuss the identification and estimation of this type of parameter. 

I also discuss the interpretation of this kind of parameter, which focuses on how treatment effects differ between subgroups, when there is also heterogeneity in potential treatment effects within these subgroups. For example, a researcher studying the impact of a policy on workers in different sectors may find that workers in one sector are highly sensitive to the policy change while workers in another are not. The researcher may wish to understand whether this heterogeneity is related causally to targeting of the policy to a specific sector or instead due to underlying differences in sensitivity to changes (eg, different levels of education or experience among these workers). I highlight that causal statements about the difference between subgroups cannot be made without additional identification assumptions. To address this, I propose an alternative parameter of interest which allows for controlled or causal comparisons between subgroups. This identification strategy can allow researchers to answer different causal questions than those addressed in \textcite{oldenTripleDifferenceEstimator2022,ortiz-villavicencioBetterUnderstandingTriple2025}. 

In this way, this work also relates to a strand of literature that has sought to develop better methods to assess heterogeneity in treatment effects according to individual characteristics in difference-in-difference and other causal frameworks (eg, \textcite{debFlexibleHeterogeneousTreatment2024, chernozhukovFisherSchultzLectureGeneric2023, imaiStatisticalInferenceHeterogeneous2025}). In particular, this work relates to others which have attempted to differentiate treatment effect heterogeneity determined by a mediator of interest as opposed to other observable characteristics (which can be controlled for) (eg, \textcite{pearlDirectIndirectEffects2001,vanderweeleUnificationMediationInteraction2014, acharyaExplainingCausalFindings2016, xuFactorialDifferenceinDifferences2026, callawayDifferenceinDifferencesContinuousTreatment}). In particular, \textcite{xuFactorialDifferenceinDifferences2026} attempt to understand how, under the parallel trends assumption and when all units are exposed to a treatment, causal comparisons can be made between subgroups of the population whose treatment effects are moderated by an additional factor. As discussed below, my approach differs by leveraging both heterogeneity between subgroups and differences between treated and control groups to unpack causal effects---thus allowing applications when the parallel trends assumption is not satisfied.

Finally, this work also contributes to a literature on identification and estimation of difference-in-difference parameters when controls are important for identification and when there is heterogeneity in treatment effects over time (dynamic treatment effects) \parencite{rothEfficientEstimationStaggered2023,santannaDoublyRobustDifferenceindifferences2020, callawayDifferenceinDifferencesMultipleTime2021, abadieSemiparametricDifferenceinDifferencesEstimators2005}. I draw from the growing literature on difference-in-differences and related designs under treatment effect heterogeneity (eg, \textcite{rothWhatsTrendingDifferenceindifferences2023, sunEstimatingDynamicTreatment2021, dechaisemartinTwoWayFixedEffects2020, goodman-baconDifferenceindifferencesVariationTreatment2021, borusyakRevisitingEventStudyDesigns2024,callawayDifferenceinDifferencesMultipleTime2021}). This work adds to this literature by addressing issues that arise in triple difference designs when treatment effects may vary over time, as pointed out by \textcite{strezhnevDecomposingTripleDifferencesRegression2023}. Although previous work in difference-in-differences has mentioned simple ways of extending difference-in-difference results to a triple difference framework (eg, \textcite{rothEfficientEstimationStaggered2023,goodman-baconDifferenceindifferencesVariationTreatment2021}), previous work has not addressed the unique forms of treatment effect heterogeneity that can arise and be leveraged in a triple difference design. I build on this work to describe doubly-robust estimators for the parameters of interest I define for triple difference designs. 

\section{Notation and framework}

I introduce a setup based on a potential outcomes framework \parencite{imbensCausalInferenceStatistics2015}. As in \textcite{callawayDifferenceinDifferencesMultipleTime2021}, define $G_i \in \mathcal{G}$ as the first period in which a unit is treated (which they call the ``cohort"), with $G_i = \infty$ representing units that are never treated. Contemporaneous treatment status for individual $i$ at time $t = 0, \dots, \mathcal{T}$ is given by $W_{it}$, with $W_{it} = 1[G_{i} <= t]$. Let $Y_{it}(g)$ represent the potential outcomes for individual $i$ at time $t$ treated at $g$, and $Y_{it}(\infty)$ represent untreated (or never-treated) potential outcomes. Finally, let individuals belong to subgroup $s_i \in \mathcal{S}$. For example, imagine that individuals may be married or not, with $s_i \in \{\text{married}, \text{not married}\}$. This setup also extends naturally to a case with multiple subgroups, such as $s_i \in \{\text{married}$, $\text{never married}$, $\text{previously married}\}$. 

\begin{assumption}{Irreversible treatment.}
\label{irreversible}
For all $i$ and all $t = 0, \dots, \mathcal{T}$, $W_{it} \geq W_{it-1}$
\end{assumption}
Under this assumption, as in \textcite{callawayDifferenceinDifferencesMultipleTime2021}, once a treatment ``turns on" at a given time $\tau$, the unit remains treated for all $t \geq \tau$. For example, consider the problem of estimating the impacts of a state-level policy. The assumption is satisfied if, once this policy takes effect in a given state, it remains state law for the future. 

The following text will consider individuals to be randomly sampled in a panel data setting. That is, $\{Y_{it}, G_i, W_{it}, X_{i}, S_{i}\}_{t \in \{0, \dots, \mathcal{T} \} }$ are independently and identically distributed (iid).\footnote{In this discussion, I focus on time-invariant covariates $X_i$. However, this approach could be adapted to consider some kinds of time-varying covariates, for example, under the ``covariate exogeneity'' assumption by \textcite{caetanoDifferenceDifferencesTimeVarying2024} or the ``no bad controls'' assumption by \textcite{debFlexibleHeterogeneousTreatment2024}.} As such, the $i$ subscript will be suppressed. The repeated cross-section case is discussed in Appendix \ref{appendix_rc}.

\section{Difference in average treatment effects on the treated}
In this section, I introduce and discuss the identification of the difference in average treatment effects on the treated ($DATT$) parameter. In the canonical triple difference design, such as that studied by \textcite{oldenTripleDifferenceEstimator2022}, there are two time periods, a binary treatment, and two subgroups, one of which is unaffected by the treatment. I extend this design by examining cases where both the subgroup of interest and the comparison subgroup may be affected by the treatment, as well as by allowing for staggered treatment designs.\footnote{In this setting, I use ``comparison'' subgroup to refer to the subgroup that is compared to the subgroup of interest. However, this is not a control group in the sense that units in the comparison subgroup may also be affected by the treatment.} 
I show that identification of the $DATT$ requires an assumption limiting anticipation of the treatment and a parallel-trends-type assumption.

A two-period difference-in-difference (DiD) design identifies the average treatment effect on the treated ($ATT$), which is 
\begin{align*}
ATT = \mathbb{E}[Y_1(1) - Y_{1}(0) | W_1 =1]    
\end{align*}
where $W_1 = 1$ represents units which become treated in the second period. 

In the staggered treatment design case, we allow for many possible $t$, $\mathcal{T} \geq 2$. As highlighted by \textcite{callawayDifferenceinDifferencesMultipleTime2021}, in the staggered treatment event-study DiD case, the parameter of interest is a weighted aggregate of many ATT-type parameters.  They suggest identifying, for $t \geq g$,
$$ATT(g,t) = \mathbb{E}[Y_t(g) - Y_t(\infty) | G = g] $$
where $Y_t(g)$ is the potential outcome at time $t$ had the individual been first treated at $g$ and $Y_t(\infty)$ is the untreated (never-treated) potential outcome. They call the $ATT(g,t)$ the ``group-time average treatment effect". 

Extending this concept to triple difference, the analogous parameters of interest, the group-time difference in average treatment effects on the treated between subgroups, can be defined:
\begin{align*}
   DATT_{s-s'}(g,t) = & \mathbb{E}[Y_{t}(g) - Y_{t}(\infty) | G=g, S =s ] - \mathbb{E}[Y_{t}(g) - Y_{t}(\infty) | G=g, S=s' ] 
\end{align*}

As in \textcite{callawayDifferenceinDifferencesMultipleTime2021}, if desired, these estimated impacts for each cohort and year can then be aggregated with certain weights to summarize across groups or across time. The canonical $2 \times 2 \times 2$ case can be thought of as a case where there is only one $DATT_{s-s'}(g,t)$ parameter.

The following assumptions are used to identify the DATT. 

\begin{assumption}[No treatment anticipation]
\label{no_anticipation_stag}
For all $t < g$, given covariates $X$,
\begin{align*}
 \mathbb{E}[Y_t(g) | G=g, X] = \mathbb{E}[Y_t(\infty) | G=g, X] 
 &
\end{align*} 
\end{assumption}
This assumption is the same as in \textcite{callawayDifferenceinDifferencesMultipleTime2021} and ensures that untreated potential outcomes are observed for all units in the pre-treatment periods. 

\begin{assumption}[Parallel gaps based on not-yet-treated group]
\label{cond_parallel_gaps_stag_ny}
For subgroups $s, s'$, covariates $X$, and for $t \geq g$:
 \begin{align*}
 &\mathbb{E} \Big[Y_{t}(\infty) - Y_{g-1}(\infty) | G=g, S=s, X\Big] - \mathbb{E} \Big[Y_{t}(\infty) - Y_{g-1}(\infty) | G=g, S=s', X \Big] \\
 = & \mathbb{E} \Big[Y_{t}(\infty) - Y_{g-1}(\infty) | W_t = 0, G\neq g, S=s, X \Big] - \mathbb{E} \Big[Y_{t}(\infty) - Y_{g-1}(\infty) | W_t = 0, G \neq g, S=s', X \Big] 
\end{align*} 
\end{assumption}

\begin{assumption}[Parallel gaps based on never-treated group]
\label{cond_parallel_gaps_stag_nev}
For subgroups $s, s'$, covariates $X$, and for $t \geq g$:
 \begin{align*}
 &\mathbb{E} \Big[Y_{t}(\infty) - Y_{g-1}(\infty) | G=g, S=s, X\Big] - \mathbb{E} \Big[Y_{t}(\infty) - Y_{g-1}(\infty) | G=g, S=s', X \Big] \\
 = & \mathbb{E} \Big[Y_{t}(\infty) - Y_{g-1}(\infty) | G= \infty, S=s, X \Big] - \mathbb{E} \Big[Y_{t}(\infty) - Y_{g-1}(\infty) | G =\infty, S=s', X \Big] 
\end{align*} 
\end{assumption}

Assumptions \ref{cond_parallel_gaps_stag_ny} and \ref{cond_parallel_gaps_stag_nev} compare closely with the difference-in-difference parallel trends assumption, and correspond to a parallel trends assumption on the gap in outcomes between subgroups. These assumptions require that the trends in the gap in outcomes between subgroup $s$ and $s'$ be parallel between treated and not-yet-treated or never-treated units.  In the application to identifying the impacts of a policy on married relative to unmarried individuals, this amounts to assuming that, had the policy not been implemented, the difference in outcomes between married and unmarried individuals would have followed the same trend in states that were treated at time $g$ and states that were not-yet-treated at that time (or never treated). This means assuming that there are no other variables that \textit{both} are correlated with how early or late a state adopted its policy \textit{and} the trend in the gap in outcomes between married and unmarried individuals. 

These assumptions relate closely to the identification assumptions outlined by \textcite{callawayDifferenceinDifferencesMultipleTime2021} for difference-in-differences with staggered treatment. However, Assumptions \ref{cond_parallel_gaps_stag_ny} and \ref{cond_parallel_gaps_stag_nev} are written in terms of ``long-differences'', that is, comparing trends from a base period of $g-1$, in contrast to the assumptions used by \textcite{callawayDifferenceinDifferencesMultipleTime2021} and \textcite{ortiz-villavicencioBetterUnderstandingTriple2025}, which are in ``short-differences'', comparing each $t$ to $t-1$. Assumptions \ref{cond_parallel_gaps_stag_ny} and \ref{cond_parallel_gaps_stag_nev} can equivalently be satisfied by repeated applications of the ``short-differences'' assumptions. 

To ensure that the quantities above are well-defined, one more assumption is needed.
\begin{assumption}{(Overlap, DATT).}
    \label{overlap_datt}
Let $C_{ny}$ equal 1 if $W_t = 0, G\neq g$ and 0 otherwise, and let $C_{nev}$ equal 1 if $G=\infty$ and 0 otherwise. For $c \in \{ny, nev\}$ and for each $t \geq 2$ and $g$, there exists $\varepsilon > 0$ such that 
\begin{align*}
    \mathbb{E}[G_gS_s] > \varepsilon, \quad P(G=g |S=s', X) > \varepsilon \text{ a.s.} , \\
    \quad P(C_c = 1 |S=s, X ) > \varepsilon \text{ a.s.}, \quad P(C_c = 1|  S=s' , X) >  \varepsilon \text{ a.s.} 
\end{align*}

\end{assumption}
Assumption \ref{overlap_datt} ensures a positive probability of belonging to the treated group and subgroup of interest $s$. It also ensures, given covariates $X$ and subgroup status $S$, a nonzero probability of being either treated or not-yet/never treated. This assumption extends a similar overlap assumption in \textcite{callawayDifferenceinDifferencesMultipleTime2021} to the triple difference case. 

\begin{proposition}[Identification of $DATT_{s-s'}(g,t)$] \label{id_att_stag}

If Assumptions \ref{irreversible}, \ref{no_anticipation_stag}, \ref{cond_parallel_gaps_stag_ny}, and \ref{overlap_datt} hold for subgroup $s$ and $s'$, the $DATT_{s-s'}(g,t)$ is identified for $t \geq g$. If $X$ is degenerate,  
\begin{align*}
DATT_{s-s'}(g,t) = & \mathbb{E}[Y_{t} - Y_{g-1} |  G=g, S=s] - \mathbb{E}[Y_{t} - Y_{g-1} | G=g, S=s']  \\
      & - \Big(\mathbb{E}[Y_{t} - Y_{g-1} | W_t=0, G \neq g, S=s] - \mathbb{E}[Y_{t} - Y_{g-1} | W_t=0, G \neq g, S=s']  \Big)
\end{align*}

\noindent If Assumptions \ref{irreversible}, \ref{no_anticipation_stag}, \ref{cond_parallel_gaps_stag_nev}, and \ref{overlap_datt} hold for subgroup $s$ and $s'$, the $DATT_{s-s'}(g,t)$ is identified for $t \geq g$. If $X$ is degenerate, 
\begin{align*}
DATT_{s-s'}(g,t) = & \mathbb{E}[Y_{t} - Y_{g-1} | G=g, S=s] - \mathbb{E}[Y_{t} - Y_{g-1} | G=g, S=s']  \\
      & - \Big(\mathbb{E}[Y_{t} - Y_{g-1} | G=\infty,  S=s] - \mathbb{E}[Y_{t} - Y_{g-1} | G=\infty, S=s']  \Big)
\end{align*}
\end{proposition}

This result extends the identification results in \textcite{oldenTripleDifferenceEstimator2022} in three ways. First, it extends to the staggered treatment case using the methods by \textcite{callawayDifferenceinDifferencesMultipleTime2021}. Second, it allows for the possibility that members of subgroup $s'$ with $G=g$ may be affected by the treatment, while \textcite{oldenTripleDifferenceEstimator2022} assume that this subgroup is not affected by the treatment by assuming that we observe untreated potential outcomes for those with $S=s'$ and $G=g$. The assumption of an unaffected subgroup and the recovery of the ATT in this framework is discussed in further detail in Appendix \ref{appendix_unaffected_subgroup}. Third, this result highlights that, in order to make comparisons between multiple subgroups, multiple parallel gaps assumptions are needed. The proof appears in Appendix \ref{appendix_proofs} (Proof \ref{proof:id_att_stag_ny} and \ref{proof:id_att_stag_nev}).

The $DATT_{s-s'}(g,t)$ is a specific parameter that can be identified and potentially of interest in triple difference designs when the comparison subgroup is also potentially affected by the treatment. As in work on heterogeneous treatment effects in randomized or difference-in-difference designs, this parameter captures the difference in response to the treatment between the subgroups $s$ and $s'$. However, unlike the case studied by \textcite{debFlexibleHeterogeneousTreatment2024} in difference-in-differences or more generally by \textcite{chernozhukovFisherSchultzLectureGeneric2023}, the difference in treatment effects between subgroups can be identified even when the treatment is not conditionally random or the parallel trends assumption is not satisfied. For example, when $g$ is a state-level policy variation, the $DATT_{s-s'}(g,t)$ can be identified even when there are state-specific time trends that are correlated with $g$, as long as these trends do not affect the comparison between subgroups.

This definition of the $DATT_{s-s'}(g,t)$ relates closely to the ``effect moderation'' parameter defined by \textcite{xuFactorialDifferenceinDifferences2026} in factorial difference-in-difference designs. Both the $DATT_{s-s'}(g,t)$ and the effect moderation parameter permit comparisons between a subgroup of interest and a comparison subgroup who may differ in their response to a treatment or event. However, they are identified under different assumptions. In \textcite{xuFactorialDifferenceinDifferences2026}, identification relies on a parallel trends assumption and occurs in a setting where all units are exposed to the treatment. In contrast, the triple difference design studied in this paper uses the parallel gaps assumption to also make comparisons with an untreated control group which is not exposed to the treatment. As such, the $DATT_{s-s'}(g,t)$ can be studied in different cases than the effect moderation parameter. \\

\noindent \textbf{Remark}. In both cases above, the researcher may make comparisons between more than two subgroups, for $s \in \mathcal{S}$. This does not require mutually exclusive subgroups; however, researchers should note that identifying variation will come from the difference in subgroups, so that comparisons between groups that are too similar may lack power. 
A full treatment of triple difference with a continuous subgroup variable is outside the scope of this paper. Difference-in-differences with a continuous treatment is discussed in \textcite{callawayDifferenceinDifferencesContinuousTreatment}. \\

\noindent \textbf{Remark}. Many researchers, including \textcite{oldenTripleDifferenceEstimator2022,ortiz-villavicencioBetterUnderstandingTriple2025}, are interested in using triple difference designs to recover an $ATT$, rather than the $DATT_{s-s'}$. Identification of the $ATT$ requires an additional assumption and is discussed in Appendix \ref{appendix_unaffected_subgroup}.

\section{Controlled \texorpdfstring{$DATT_{s-s'}$}{DATT} }\label{section:cdatt}
In this section, I show that the interpretation of the $DATT$ parameters discussed in the previous section may be affected by other forms of treatment effect heterogeneity, even if the above parameters are identified. I then introduce and discuss a new parameter of interest, the controlled difference in average treatment effects on the treated. 

\subsection{Interpretation of the controlled \texorpdfstring{$DATT_{s-s'}$}{DATT}}

To clarify the interpretation of the $DATT_{s-s'}$, I rewrite potential outcomes in terms of both potential treatment status \textit{and} potential subgroup status. Define, for any subgroups $s$ and $s'$, the potential outcomes $Y_t(G,S)$, which represent the potential outcome under (possibly counterfactual) treatment cohort $G$ and (possibly counterfactual) subgroup $S$. 

Using these quantities, I decompose the $DATT_{s-s'}$ parameter:
\begin{align*}
   D&ATT_{s-s'}(g,t)  =  
   \\
   = & \mathbb{E}[Y_{t}(g,s) - Y_{t}(\infty,s) | G=g, S=s ] -  \mathbb{E}[Y_{t}(g,s') - Y_{t}(\infty,s') | G=g, S=s'] \\
   = & \underbrace{\mathbb{E}[Y_{t}(g,s) - Y_{t}(\infty,s) | G=g, S=s ] -  \mathbb{E}[Y_{t}(g,s') - Y_{t}(\infty,s') | G=g, S=s]}_\text{controlled $DATT_{s-s'}(g,t)$} \\
& + \underbrace{\mathbb{E}[Y_{t}(g,s') - Y_{t}(\infty,s') | G=g, S=s ] -  \mathbb{E}[Y_{t}(g,s') - Y_{t}(\infty,s') | G=g, S=s']}_\text{due to treatment effect heterogeneity} 
\end{align*}

The first term, the \textit{controlled} $DATT_{s-s'}$ or $CDATT_{s-s'}$, captures the difference in treatment effects arising from a change in subgroup status, among those who (in reality) belonged to subgroup $s$. 
The second term, which is due to treatment effect heterogeneity correlated with subgroup status, captures how the treatment effects would differ if all individuals had belonged to the same subgroup, with any differences arising from differences in the underlying treatment effects for the individuals that selected into each subgroup. In words, this arises when subgroup status is correlated with treatment effects.\footnote{This decomposition parallels the one in \textcite{callawayDifferenceinDifferencesContinuousTreatment}, who discuss challenges with comparing treatment effects in difference-in-difference designs with continuous treatments.}

For example, consider a triple difference design analyzing the impacts of a policy change on the employment of married individuals ($s$) and unmarried ($s'$) individuals. Treatment effect heterogeneity will affect the $DATT_{s-s'}$ to the extent that those who actually were married in this sample have different underlying treatment effects than those who actually were not married. For example, suppose that married people tend to be in different occupations than unmarried people, and suppose that individuals in those occupations are more sensitive to a policy change. Then, the average treatment effect on the treated of the married group, had they been unmarried (but still in those occupations), differs from that of the unmarried group. Meanwhile, the true difference in treatment effects controlling for differences in characteristics besides marital status would be estimated by the $CDATT_{s-s'}$.

Given this interpretation, I argue that the $CDATT_{s-s'}$ can be interpreted as a difference in treatment effects \textit{caused by} subgroup status. 

However, in assigning a causal interpretation to differences in subgroup characteristics---which, in practice, are often demographic groups---researchers must contend with established notions of causality. For example, \textcite{hollandStatisticsCausalInference1986} famously argued that causal effects can only be understood when a cause can be imagined to be manipulated in an experiment. In the example above, one can imagine an experiment assigning (or allowing) some individuals to be married and some not, allowing marriage to be a valid cause in Holland's sense. However, this notion of causality also poses challenges when authors are interested in studying differences based on demographic groups that are difficult to imagine manipulating in an experiment, like race, gender, or sex. In response to this challenge, authors including \textcite{senRaceBundleSticks2016} have argued for understanding race as a composite of many more plausibly manipulable differences, such as name, perceived race, or dialect, that can be studied in isolation. A similar argument can be made for other demographic characteristics. 

At the same time, other authors in statistics have maintained seemingly immutable characteristics like race and gender as ``causes''. For example, \textcite{pearlDirectIndirectEffects2001} discusses sex/gender as a cause in hiring decisions and references legal frameworks defining discrimination as disparate treatment on the basis of a given characteristic, if all other characteristics were held equal. A possible alternative interpretation of these frameworks that would align with \textcite{hollandStatisticsCausalInference1986} is to conceptualize disparate impacts as caused by the different (disparate) treatment each group experiences (rather than by their group characteristics).

Given this, researchers interested in creating causally-interpretable triple difference designs should carefully consider the role of demographic subgroup statuses and whether these can be thought of as representing a particular, more manipulable attribute (such as perceived demographic group or likelihood to receive a certain treatment by individuals or policy). For example, as discussed below, in many observational studies, authors may study subgroups based demographic characteristics because these determine eligibility for a particular program, such as maternity leave. In these cases, the question of manipulability may be addressed by imagining manipulating eligibility for the policy rather than demographics. Then, the causal effect of belonging to the given demographic group can be thought of as the causal effect of eligibility (relative, for example, to spillovers on the ineligible subgroup).

In cases where subgroup identity cannot be thought of as representing a manipulable attribute, caution should be taken in interpreting the $CDATT_{s-s'}$ as fully causal. A more general interpretation of the $CDATT_{s-s'}$ is to consider it to be the differential effect of the treatment on subgroup $s$, holding other characteristics constant. Although I emphasize the causal interpretation of the $CDATT_{s-s'}$ throughout this paper, the questions discussed can also be thought of as questions about controlled differences, holding other relevant characteristics fixed.

Four applications below highlight cases in which researchers might be interested in studying the $CDATT_{s-s'}$ and how it might diverge from the $DATT_{s-s'}$. 

\textcite{gruberIncidenceMandatedMaternity1994} uses a triple difference design to evaluate the difference in impacts of a policy requiring insurance companies to provide coverage for childbirth costs. The author compares the effect of being in a group likely to use these benefits (married individuals and single women of childbearing age) relative to groups that are not likely to use these benefits (single men and older adults). The $CDATT_{s-s'}$ would allow the researcher to understand whether differences in impacts between these subgroups are driven by differences in likelihood to use the benefits (proxied by demographics), holding all other characteristics constant, or by other characteristics correlated with subgroup, such as occupation. 

\textcite{baumEffectStateMaternity2003} uses a similar design to investigate the impacts of maternity leave benefits on the wages and employment of women with children and women of childbearing age relative to single men. The $CDATT_{s-s'}$ would allow the author to interpret differences in the impacts of maternity leave benefits for these subgroups as being caused by subgroup status (ie, caused by women being eligible for the benefits) rather than by other characteristics that differ between the subgroups.\footnote{One corresponding imagined experiment would involve a policy experiment with different eligibility for parental leave.} Understanding whether the differences between subgroups are caused by eligibility for parental leave may have important implications for policy. 

\textcite{matsaFemaleStyleCorporate2013b} study the impacts of quotas for women's representation on corporate boards in Norway. To do this, they use a triple difference design which compares the effects of the policy on listed (public) companies and unlisted companies, which were exempt from the quota. Showing that the quota's impact on the listed companies was caused by the regulation requires a causal statement about the comparison between subgroups. That is, the researcher might be interested to understand whether listed companies were impacted because they were listed (and, therefore, bound by the regulation), or because of another characteristic that differs between listed and unlisted companies, such as firm size. 

In another application of a triple difference design, \textcite{derenoncourtMinimumWagesRacial2020} investigate whether the impacts of the minimum wages imposed by the Fair Labor Standards Act were larger for workers in the US South compared to other regions. The authors' motivation for studying this is to unpack the impact of this policy on racial inequality, and so an interesting investigation would be whether differential impacts are related to being in the South (eg, due to regional histories) or to differences in other characteristics correlated with region (eg, worker or firm characteristics).\footnote{Although the attribute of whether a region is in the US South cannot be thought of as manipulable in the sense discussed by \textcite{hollandStatisticsCausalInference1986}, one can imagine an experiment moving workers to particular regions in order to understand the causal effect of being located in a particular region.}

The treatment effect heterogeneity arising between subgroups may be particularly concerning in cases where the policy studied can \textit{cause} selection into the subgroups analyzed based on treatment effects. For example, if a policy intervention causes those who are more sensitive to its impacts to get married or have children, then those subgroups will disproportionately contain individuals who are more sensitive to the treatment. A similar concern in the difference-in-difference setting is discussed by \textcite{blundellAlternativeApproachesEvaluation2009}, who consider the challenge of estimating the returns to education when individuals select into education based on the returns they expect to receive.

Finally, a common assumption behind triple difference designs, discussed more formally in Appendix \ref{appendix_unaffected_subgroup}, involves treating one subgroup as a ``control subgroup", which is assumed to be unaffected by a given policy change. Even under this assumption, a causal interpretation of the differences between subgroups requires an additional assumption on the potential outcomes of the subgroup of interest.

\subsection{Identification of \texorpdfstring{$CDATT$}{CDATT}}

As introduced above, neither Assumption \ref{cond_parallel_gaps_stag_ny} nor Assumption \ref{cond_parallel_gaps_stag_nev} is enough to identify the $CDATT$ when treatment effects are heterogeneous. Its identification requires an assumption on treatment effect heterogeneity. In this section, I describe two identification assumptions that can be used, along with the assumptions above, to identify the $CDATT$. 

\begin{assumption}[No subgroup selection, staggered design]
\label{no_subgroup_selection_stag}
For each subgroup $s,s'$,
   \begin{align*}
\mathbb{E} \Big[Y_{t}(g,s') - Y_{t}(\infty,s') | G=g, S=s\Big]  & = \mathbb{E} \Big[Y_{t}(g,s') - Y_{t}(\infty,s') | G=g, S=s'\Big] \\
& = \mathbb{E} \Big[Y_{t}(g) - Y_{t}(\infty) | G=g, S=s'\Big]    
\end{align*}
\end{assumption}

\begin{assumption}[Observable subgroup selection, staggered design]
\label{cond_subgroup_selection_stag}
For each subgroup $s,s'$, given a set of control variables $X$,
   \begin{align*}
\mathbb{E} \Big[Y_{t}(g,s') - Y_{t}(\infty,s') | G=g, S=s, X\Big]  & = \mathbb{E} \Big[Y_{t}(g,s') - Y_{t}(\infty,s') | G=g, S=s', X\Big] \\
& = \mathbb{E} \Big[Y_{t}(g) - Y_{t}(\infty) | G=g, S=s', X\Big]    
\end{align*}
\end{assumption}
Assumptions \ref{no_subgroup_selection_stag} and \ref{cond_subgroup_selection_stag} limit the treatment effect heterogeneity between the subgroups. Assumption \ref{no_subgroup_selection_stag} implies that underlying average treatment effects on the treated would have been the same for members of subgroup $s$ and $s'$, had they all belonged to subgroup $s'$. In a sense, this assumption says that, had the identities of those in each subgroup been swapped (that is, had those in subgroup $s$ actually belonged to $s'$), the average treatment effect on the treated for this subgroup would be unchanged. This assumption is violated whenever the subgroups differ in their sensitivity to the treatment for reasons other than their subgroup status.

Assumption \ref{cond_subgroup_selection_stag} weakens this assumption by requiring that there is no difference in average treatment effects on the treated between the subgroups for any reason besides subgroup status, after conditioning on control variables $X$. Assumption \ref{cond_subgroup_selection_stag} is a more general form of Assumption \ref{no_subgroup_selection_stag} if $X$ is allowed to be degenerate. 

For example, in an application comparing the outcomes of married women with those of single men, suppose that married women tend to be more sensitive to a labor market shock because they tend to work in more sensitive occupations. In this case, the heterogeneity between the groups can be described in terms of heterogeneity in job characteristics, satisfying Assumption \ref{cond_subgroup_selection_stag} by taking $X$ to be occupation. After conditioning on occupation, there is no difference in sensitivity between the groups. \\

\noindent \textbf{Remark.} A special case of Assumption \ref{cond_subgroup_selection_stag} is one in which the researcher assumes an average treatment effect on the treated of zero for both subgroups, had they been in subgroup $s'$. For subgroup $s$, this is a statement about counterfactual treatment effects, since they were not observed in subgroup $s'$. In this way, this is an additional assumption beyond the unaffected subgroup assumption discussed in Appendix \ref{appendix_unaffected_subgroup}, which only makes an assumption on the treatment effects of those in subgroup $s'$. In the example where the subgroups are married and unmarried individuals, Assumption \ref{control_subgroup} requires assuming that the average treatment effect on the treated for unmarried individuals is zero. Adding Assumption \ref{no_subgroup_selection_stag} requires the assumption that, had the subgroup assignments been swapped so that those were actually married were not married, their average treatment effect on the treated would also have been zero. While Assumption \ref{control_subgroup} relates only to the potential outcomes for the comparison subgroup, Assumptions \ref{no_subgroup_selection_stag} and \ref{cond_subgroup_selection_stag} also relate to the potential outcomes for the subgroup of interest. \\

As with the $DATT_{s-s'}$, identification of the $CDATT_{s-s'}$ also requires an overlap assumption.
\begin{assumption}{(Overlap, CDATT).}
    \label{overlap_cdatt}
Let $C_{ny}$ equal 1 if $W_t = 0, G\neq g$ and 0 otherwise, and let $C_{nev}$ equal 1 if $G=\infty$ and 0 otherwise. For $c \in \{ny, nev\}$ and for each $t \geq 2$ and $g$, there exists $\varepsilon > 0$ such that 
\begin{align*}
    \mathbb{E}[G_gS_s] > \varepsilon, \quad P(G=g ,S=s'| X) > \varepsilon \text{ a.s.} , \\
    \quad P(C_c = 1, S=s | X ) > \varepsilon \text{ a.s.}, \quad P(C_c = 1, S=s' | X) >  \varepsilon \text{ a.s.} 
\end{align*}

\end{assumption}
Assumption \ref{overlap_cdatt} differs from Assumption \ref{overlap_datt} because it requires positive probability of belonging to each subgroup-treatment combination, after conditioning on covariates $X$. In contrast, Assumption \ref{overlap_datt} only requires positive probability of being treated or not-yet/never treated after conditioning on subgroup and $X$. These differ, for example, when covariates $X$ lack overlap between the subgroups.

\begin{proposition}[Identification of $CDATT_{s-s'}(g,t)$ with no treatment effect heterogeneity] \label{id_satt_stag}
If Assumptions \ref{irreversible}, \ref{no_anticipation_stag},  \ref{cond_parallel_gaps_stag_ny}, \ref{no_subgroup_selection_stag}, and \ref{overlap_cdatt} or Assumptions \ref{irreversible}, \ref{no_anticipation_stag},  \ref{cond_parallel_gaps_stag_nev}, \ref{no_subgroup_selection_stag}, and \ref{overlap_cdatt} hold for subgroup $s$ and $s'$, the $CDATT_{s-s'}(g,t)$ is identified, and 
\begin{align*}
CDATT_{s-s'}(g,t) = & DATT_{s-s'}(g,t)
 \end{align*}
\end{proposition}

This proposition implies that, when there is no treatment effect heterogeneity, the $DATT_{s-s'}$ and the $CDATT_{s-s'}$ are equivalent. In effect, Assumption \ref{no_subgroup_selection_stag} affects the interpretation of the parameters of interest, but not their estimation. The proof appears in Appendix \ref{appendix_proofs} (Proofs \ref{proof:id_cdatt_stag_ny} and \ref{proof:id_cdatt_stag_nev}). In this case, controls $X$ related to subgroup status $S$ may still be needed to satisfy the parallel gaps assumption, but, under Assumption \ref{no_subgroup_selection_stag}, they do not play a role in differentiating between the $DATT_{s-s'}$ and $CDATT_{s-s'}$.

To discuss the identification of the $CDATT_{s-s'}$ when treatment effect heterogeneity between the subgroups is captured by observable characteristics, I introduce additional notation. Let $S_s = 1$ if $S =s$ and 0 otherwise. Let $G_g = 1$ if $G=g$ and 0 otherwise. Finally, let $C_{ny} = 1$ if $W_t=0, G\neq g$ and 0 otherwise, and let $C_{nev} = 1$ if $G = \infty$ and 0 otherwise.

Now, define propensity scores 
\begin{align*}
\pi_{g,s}(X) & = P(G=g, S=s |X)  \\
\pi_{g,s'}(X) & = P(G=g, S=s' |X)  \\
\pi_{ny,s}(X) & = P(W_t = 0, G\neq g, S=s |X ) \\
\pi_{nev,s}(X) & = P(G=\infty, S=s |X) \\
\pi_{ny,s'}(X) & = P(W_t = 0, G\neq g, S=s' |X) \\
\pi_{nev,s'}(X) & = P(G=\infty, S=s' |X)
\end{align*}
and outcome functions 
\begin{align*}
\mu^{s'}_{g,t}(X) & = \mathbb{E}[Y_{t}-Y_{g-1} | G=g, S=s', X] \\
\mu^{s}_{ny, g, t}(X) & =\mathbb{E}[Y_{t}-Y_{g-1} | G\neq g, W_t = 0, S=s, X]  \\
\mu^{s}_{nev,g,t}(X) & =\mathbb{E}[Y_{t}-Y_{g-1} | G=\infty, S=s, X] \\
\mu^{s'}_{ny, g, t}(X) & =\mathbb{E}[Y_{t}-Y_{g-1} | G\neq g, W_t = 0, S=s', X]  \\
\mu^{s'}_{nev,g,t}(X) & =\mathbb{E}[Y_{t}-Y_{g-1} | G=\infty, S=s', X] 
\end{align*}

Then, define the following parameters, for $c \in \{ny, nev\}$ : 
{ 
\begin{align*}
 CDATT^{IPW, c}_{s-s'}(g,t) = & \mathbb{E}\Big[ \Big( w_1(G_g, S_s) - w_2(G_g, S_{s'}, X)\Big)(Y_{t} - Y_{g-1})\Big]  \\
 &- \mathbb{E}\Big[ \Big(w_3^c(C_c, S_s, X) - w_4^c(C_c, S_{s'}, X)\Big) (Y_{t} - Y_{g-1})\Big]  \\
CDATT^{RA, c}_{s-s'}(g,t) = & \mathbb{E}\Big[ \frac{G_g S_s}{\mathbb{E}[G_g S_s]} \Big(Y_{t} - Y_{g-1} - \mu_{g,t}^{s'}(X) - \mu_{c,g,t}^{s}(X)+ \mu_{c,g,t}^{s'}(X) \Big)\Big] \\
CDATT^{DR, c}_{s-s'}(g,t) = & \mathbb{E}\Big[ \frac{G_g S_s}{\mathbb{E}[G_g S_s]} \Big(Y_{t} - Y_{g-1} - \mu_{g,t}^{s'}(X) - \mu_{c,g,t}^{s}(X)+ \mu_{c,g,t}^{s'}(X) \Big)\Big] \\
&- \mathbb{E}\Big[w_2(G_g, S_s, X) (Y_t - Y_{g-1} - \mu_{g,t}^{s'}(X))\Big] \\
&- \mathbb{E}\Big[w_3^c(C_c, S_s, X)  (Y_t - Y_{g-1}- \mu_{c,g,t}^{s}(X))\Big] \\
&+\mathbb{E}\Big[w_4^c(C_c, S_{s'}, X)(Y_t-Y_{g-1}- \mu_{c,g,t}^{s'}(X))\Big]
\end{align*}
where 
\begin{align*}
w_1(G_g, S_s) & = \frac{G_gS_s}{\mathbb{E}[G_gS_s]} \\
w_2(G_g, S_{s'}, X) & = \frac{G_gS_{s'} \pi_{g,s}(X)}{\pi_{g,s'}(X)} \Big/ \mathbb{E} \Big[\frac{G_gS_{s'} \pi_{g,s}(X)}{\pi_{g,s'}(X)} \Big] \\
w_3^c(C_c, S_s, X) & = \frac{C_c S_{s} \pi_{g,s}(X)}{\pi_{c,s}(X)} \Big/ \mathbb{E} \Big[\frac{C_c S_{s} \pi_{g,s}(X)}{\pi_{c,s}(X)}  \Big]\\
w_4^c(C_c, S_{s'}, X) & = \frac{C_c S_{s'} \pi_{g,s}(X)}{\pi_{c,s'}(X)} \Big/ \mathbb{E} \Big[\frac{C_c S_{s'} \pi_{g,s}(X)}{\pi_{c,s'}(X)}  \Big]
\end{align*}

}

This setup is general in the sense that $X$  may be degenerate.

\begin{proposition}[Identification of $CDATT_{s-s'}(g,t)$ with observable treatment effect heterogeneity] \label{id_satt_stag_cond}

If Assumptions \ref{irreversible}, \ref{no_anticipation_stag}, \ref{cond_parallel_gaps_stag_ny} (applied to the $s'$ potential outcomes), \ref{cond_subgroup_selection_stag}, and \ref{overlap_cdatt}, and hold for subgroup $s$ and $s'$, then, the $CDATT_{s-s'}(g,t)$ is identified for $t \geq 2$, and 
\begin{align*}
CDATT_{s-s'}(g,t) = CDATT^{IPW, ny}_{s-s'}(g,t) = CDATT^{RA, ny}_{s-s'}(g,t) = CDATT^{DR, ny}_{s-s'}(g,t)
\end{align*}

If Assumptions \ref{irreversible}, \ref{no_anticipation_stag}, \ref{cond_parallel_gaps_stag_nev} (applied to the $s'$ potential outcomes), \ref{cond_subgroup_selection_stag}, and \ref{overlap_cdatt} hold for subgroup $s$ and $s'$, then, the $CDATT_{s-s'}(g,t)$ is identified for $t \geq 2$, and 
\begin{align*}
CDATT_{s-s'}(g,t) = CDATT^{IPW, nev}_{s-s'}(g,t) = CDATT^{RA, nev}_{s-s'}(g,t) = CDATT^{DR, nev}_{s-s'}(g,t)
\end{align*}
\end{proposition}

The proof appears in Appendix \ref{appendix_proofs} (Proofs \ref{proof_ipw}, \ref{proof_ra}, and \ref{proof_dr}).\\

The identification of the $CDATT_{s-s'}(g,t)$ emphasizes the ways it compares to and contrasts with other related parameters in the literature. The $CDATT_{s-s'}(g,t)$ parallels the ``causal moderation'' parameter defined by \textcite{xuFactorialDifferenceinDifferences2026} for factorial difference-in-difference designs. However, it is worth noting that they are identified under different assumptions. First, the main parallel trends assumption for factorial difference-in-difference includes only two differences (that is, over time and between subgroups), as all units are considered to be treated. In contrast, Assumptions \ref{cond_parallel_gaps_stag_ny} and Assumption \ref{cond_parallel_gaps_stag_nev} also take the difference between treated and untreated units. Second, the factorial parallel trends assumption required by \textcite{xuFactorialDifferenceinDifferences2026} to identify the causal moderation parameter is imposed on average trends (before-after differences) on four combinations of treatment and subgroup (both subgroups and both treated and untreated potential outcomes). On the other hand, Assumptions \ref{no_subgroup_selection_stag} and \ref{cond_subgroup_selection_stag} allow average treated and untreated potential outcomes ($Y_t(g,s')$ and $Y_t(\infty,s')$) to differ between the subgroups $s$ and $s'$, as long as the difference (gap) between these is equivalent. In addition, Assumptions \ref{no_subgroup_selection_stag} and \ref{cond_subgroup_selection_stag} restrict potential outcomes only under (counterfactual) subgroup status $s'$ and at time $t$ (rather than for both subgroups and in trends). 

The parameters I discuss also have parallels in the literature on controlled direct effects (see, eg, \textcite{robinsIdentifiabilityExchangeabilityDirect1992, pearlDirectIndirectEffects2001, acharyaExplainingCausalFindings2016}). This literature also uses potential outcomes to separate the direct effects of a treatment on an outcome from indirect effects via a mediating factor. If subgroup status is re-conceptualized as a mediating factor, the $DATT$ can be thought of as parallel to a total effect, while the $CDATT$ mirrors a controlled direct effect (that is, by capturing an effect, holding some characteristics constant). The decomposition of the $DATT$ may be compared to that in \textcite{vanderweeleUnificationMediationInteraction2014}, who argues that the total effect can be decomposed into a controlled direct effect, an interaction between treatment and mediator, and two other indirect effects that arise from effects of the treatment on the mediator. Similar to Assumption \ref{cond_subgroup_selection_stag}, identification of controlled direct effects relies on an assumption of (conditional or sequential) unconfoundedness between potential outcomes and the mediator \parencite{vanderweeleUnificationMediationInteraction2014, pearlDirectIndirectEffects2001, acharyaExplainingCausalFindings2016}. However, in contrast to the parallel gaps assumption identifying the $CDATT$, identification of these parameters also requires unconfoundedness between potential outcomes and treatment assignment. \\

\noindent \textbf{Remark}. If neither Assumption \ref{no_subgroup_selection_stag} nor Assumption \ref{cond_subgroup_selection_stag} holds, it may not be possible to recover the $CDATT$. However, an alternative approach is to bound the parameter by making more limited assumptions on treatment effect heterogeneity. For example, one possible assumption, in the spirit of \textcite{manskiMonotoneInstrumentalVariables2000a}, is that individuals in $s'$ have selected into the subgroup in which they will experience the larger treatment effect. Under this assumption, researchers can interpret the $DATT_{s-s'}$ as a lower bound for $CDATT_{s-s'}$.

\section{Properties, estimation, and inference}

In this section, I discuss the estimation of both the $DATT$ and $CDATT$ and the properties of these estimators. 

\subsection{Estimation of \texorpdfstring{$DATT_{s-s'}(g,t)$}{DATT(g,t)}}

As we have seen above, the $DATT_{s-s'}(g,t)$ can be thought of as the difference between two difference-in-difference $ATT$ parameters, estimation and  inference for which is provided in \textcite{callawayDifferenceinDifferencesMultipleTime2021} and \textcite{santannaDoublyRobustDifferenceindifferences2020}. When conditioning on controls, as \textcite{callawayDifferenceinDifferencesMultipleTime2021} point out, several methods may be appropriate, including regression adjustment, inverse probability weighting, and doubly robust approaches.  

For example, consider parametric models for the propensity scores $\pi_{g}^b = P(G=g | S=b)$ and $\pi_{c}^b = P(C_c = 1 | S=b)$ that take the form $e_{a}^{b}(X; \theta_{g}^{b})$ for $a \in \{g,c\}$ and $b \in \{s, s'\}$. Consider also parametric models for the outcome functions $\mu_{c,t}^{b} = \mathbb{E}[Y_t - Y_{g-1} | C_c = 1, S=b]$ of the form $m_{c,t}^{b}(X; \beta_{c,t}^{b})$ for $b \in\{s, s'\}$. These parametric models are estimated by $e_{a}^{b}(X; \widehat\theta_{g}^{b})$ and $m_{c,t}^{b}(X; \widehat\beta_{c,t}^{b})$, respectively. 

Then, the analogous doubly-robust estimator takes the form:

\begin{align*}
\widehat {DATT}^{DR, c}_{s-s'}(g,t) = & \mathbb{E}_n\Big[\Big(\widehat w_1^{DATT}(G_g,S_s)- \widehat w_2^{DATT,c}(G_g, S_s, X;\widehat\theta_{g}^{s})\Big)\Big(Y_t - Y_{g-1} - m_{c,t}^s(X;\widehat\beta_c^s) \Big) \Big] \\
& - \mathbb{E}_n\Big[\Big(\widehat w_3^{DATT}(G_g, S_s) - \widehat w_4^{DATT,c}(G_g, S_s, X;\widehat\theta_{g}^{s'}) \Big)\Big(Y_t- Y_{g-1} - m_{c,t}^{s'}(X;\widehat\beta_c^{s'}) \Big) \Big]  
\end{align*}
where 
\begin{align*}
\widehat w_1^{DATT}(G_g, S_s) & = \frac{G_gS_s}{\mathbb{E}_n[G_gS_s]} \\
\widehat w_2^{DATT,c}(G_g, S_s, X;\widehat\theta_{g}^{s}) & = \frac{C_c S_{s} e_{g}(X; \widehat\theta_{g}^{s})}{e_{c}(X;\widehat\theta_{g}^{s})} \Big/ \mathbb{E}_n \Big[\frac{C_c S_{s} e_{g}(X; \widehat\theta_{g}^{s})}{e_{c}(X;\widehat\theta_{g}^{s})}  \Big] \\
\widehat w_3^{DATT}(G_g, S_s) & =  \frac{G_gS_{s'}}{\mathbb{E}_n[G_gS_{s'}]}\\
\widehat w_4^{DATT,c}(G_g, S_s, X;\widehat\theta_{g}^{s'}) & = \frac{C_c S_{s'} e_{g}(X; \widehat\theta_{g}^{s'})}{e_{c}(X;\widehat\theta_{g}^{s'})} \Big/ \mathbb{E}_n \Big[\frac{C_c S_{s'} e_{g}(X; \widehat\theta_{g}^{s'})}{e_{c}(X;\widehat\theta_{g}^{s'})}  \Big]
\end{align*}

\noindent \textbf{Remark}.
As pointed out by \textcite{goodman-baconDifferenceindifferencesVariationTreatment2021}, it is also possible to obtain an estimate of the triple difference $DATT$ by estimating a difference-in-differences model on treatment cohort-level gaps between subgroup $s$ and $s'$. For example, if $G$ is assigned at the state level, one may collapse the data to the state level and estimate a weighted difference-in-difference model on the gap in average outcomes between subgroup $s$ and $s'$ in each state. However, aggregating the data from the individual to the state level may reduce the precision of the estimates and preclude the inclusion of individual-level controls. \\

\noindent \textbf{Remark}.
\textcite{ortiz-villavicencioBetterUnderstandingTriple2025} point out that, when estimating the $ATT$ using a triple difference design, taking the difference between two DiD $ATT$ may not recover the desired parameter. This is because the two DiDs average $X$ over different populations of interest (in the setup above, the first expectation averages over subgroup $s$, while the second averages over subgroup $s'$). In contrast, in this setting, the estimator estimates the $DATT$, conditional on being in each subgroup; as such, averaging $X$ over these two subgroups is not a concern for estimation of the $DATT$.

\subsection{Estimation and inference for \texorpdfstring{$CDATT^{DR}_{s-s'}(g,t)$}{CDATT(g,t)}}
Consider parametric models for the propensity scores $\pi_{g,s}, \pi_{g,s'}, \pi_{c,s}$ and $\pi_{c,s'}$ that take the form $e_{a,b}(X; \theta_{g})$ for $a \in \{g,c\}$ and $b \in \{s,s'\}$. Consider also parametric models for the outcome functions $\mu_{g,t}^{s'}$ and $\mu_{c,g,t}^{s'}$ of the form $m_{a,t}^{s'}(X; \beta_{a}^{s'})$ for $a \in\{g,(c,g)\}$. These parametric models are estimated by $e_{a,b}(X; \widehat\theta_{g})$ and $m_{a,t}^{s'}(X; \widehat\beta_{a}^{s'})$, respectively. 

Let $\mathbb{E}_n [X] = \frac{1}{n} \sum_{i=1}^n X_i$. Then, the following estimates $CDATT^{DR,c}_{s-s'}(g,t)$, for $c \in \{ny, nev\}$: 
\begin{align*}
\widehat{CDATT}^{DR, c}_{s-s'}(g,t) = & \mathbb{E}_n\Big[ \widehat w_1(G_g,S_s) \Big(Y_{t} - Y_{g-1} - m_{g,t}^{s'}(X; \widehat\beta_{g,s'}) - m_{c,g,t}^{s}(X; \widehat\beta_{c,g,s})+ m_{c,g,t}^{s'}(X; \widehat\beta_{c,g,s'}) \Big)\Big] \\
&- \mathbb{E}_n\Big[\widehat w_2(G_g, S_s, X; \widehat\theta_{g}) (Y_t - Y_{g-1} - m_{g,t}^{s'}(X; \widehat\beta_{g,s'}))\Big] \\
&- \mathbb{E}_n\Big[\widehat w_3^c(C_c, S_s, X; \widehat\theta_{g})  (Y_t - Y_{g-1}- m_{c,g,t}^{s}(X; \widehat\beta_{c,g,s}))\Big] \\
&+\mathbb{E}_n\Big[\widehat w_4^c(C_c, S_{s'}, X; \widehat\theta_{g})(Y_t-Y_{g-1}- m_{c,g,t}^{s'}(X; \widehat\beta_{c,g,s'}))\Big]
\end{align*}
where 
\begin{align*}
\widehat w_1(G_g, S_s) & = \frac{G_gS_s}{\mathbb{E}_n[G_gS_s]} \\
\widehat w_2(G_g, S_s, X;\widehat\theta_{g}) & = \frac{G_gS_{s'} e_{g,s}(X; \widehat\theta_{g})}{e_{g,s'}(X;\widehat\theta_{g})} \Big/ \mathbb{E}_n \Big[\frac{G_gS_{s'} e_{g,s}(X; \widehat\theta_{g})}{e_{g,s'}(X;\widehat\theta_{g})} \Big] \\
\widehat w_3^c(C_c, S_s, X;\widehat\theta_{g}) & = \frac{C_c S_{s} e_{g,s}(X; \widehat\theta_{g})}{e_{c,s}(X;\widehat\theta_{g})} \Big/ \mathbb{E}_n \Big[\frac{C_c S_{s} e_{g,s}(X; \widehat\theta_{g})}{e_{c,s}(X;\widehat\theta_{g})}  \Big]\\
\widehat w_4^c(C_c, S_{s'}, X;\widehat\theta_{g}) & = \frac{C_c S_{s'} e_{g,s}(X; \widehat\theta_{g})}{e_{c,s'}(X;\widehat\theta_{g})} \Big/ \mathbb{E}_n \Big[\frac{C_c S_{s'} e_{g,s}(X; \widehat\theta_{g})}{e_{c,s'}(X;\widehat\theta_{g})}  \Big]
\end{align*}

Estimation of the $\widehat{CDATT}_{s-s'}^{DR,c}$ requires specifying both the outcome models and the propensity score models. Even when both the treatment and subgroup status are binary, the propensity score models require estimating the probability of belonging to 4 groups (ie, $P(G=1, S=s |X)$, $P(G=1, S=s'|X)$, $P(G=\infty, S=s|X)$, $P(G=\infty, S=s'|X)$). For this, an approach like a multinomial logit or probit estimation would often be suitable. Because of the dependence of the propensity scores, $\widehat\theta_g$ represents the vector of parameters potentially used to estimate them jointly.

This form of the estimator makes clear that the $CDATT$ and the $DATT$ differ in how they use subgroup status. In the same way that the $CDATT$ seeks to ensure that outcomes are compared always conditioning on belonging to subgroup $s$, note that the $CDATT$ requires estimating propensity scores that predict both subgroup status and treatment status, while the propensity scores used to estimate the $DATT$ only condition on subgroup status rather than predict it. In addition, the $DATT$ includes one fewer regression adjustment term, because the average outcome for the treated units in subgroup $s'$ are taken as given, rather than adjusted to match subgroup $s$. 

Next, I discuss the asymptotic properties of $\widehat{CDATT}^{DR,c}_{s-s'}(g,t)$. Derivation of the asymptotic properties depends on a regularity assumption, which is further laid out in Appendix \ref{proof:asymptotic_properties} (Assumption \ref{regularity}). This assumption is standard in the literature, eg, \textcite{santannaDoublyRobustDifferenceindifferences2020} and \textcite{callawayDifferenceinDifferencesMultipleTime2021}. It puts some smoothness restrictions on the form of the working models, which are satisfied by common models such as linear models and multinomial logit. 

For ease of notation when considering multiple time periods, let $\widehat{CDATT}^{DR,c}_{t \geq g}$ denote the vector of all $\widehat{CDATT}^{DR,c}_{s-s'}(g,t)$ with $t \geq g$. Define $CDATT_{t\geq g}$ and $\eta_{t \geq g}^{DR, c}$ analogously.  
\begin{proposition}{Asymptotic distribution of $\widehat{CDATT}^{DR,c}_{s-s'}$}
\label{asym_dist}
    Under Assumptions \ref{irreversible}, \ref{no_anticipation_stag}, \ref{cond_parallel_gaps_stag_ny}, \ref{cond_subgroup_selection_stag}, and \ref{regularity},
    $$\sqrt{n}(\widehat{CDATT}^{DR, ny}_{t \geq g} - CDATT_{t \geq g}) \to \mathcal{N}(0, \Sigma^{ny}) $$
    with
    $$\Sigma^{ny} = \mathbb{E}\Big[\eta_{t \geq g}^{DR, ny}(\eta_{t \geq g}^{DR, ny})' \Big]  $$
where $\eta_{t \geq g}^{DR, ny}$ is as described in Appendix \ref{proof:asymptotic_properties}.

Under Assumptions \ref{irreversible}, \ref{no_anticipation_stag}, \ref{cond_parallel_gaps_stag_nev}, \ref{cond_subgroup_selection_stag}, and \ref{regularity},
$$\sqrt{n}(\widehat{CDATT}^{DR, nev}_{ t \geq g} - CDATT_{t \geq g}) \to \mathcal{N}(0, \Sigma^{nev}) $$
with
$$\Sigma^{nev} = \mathbb{E}\Big[\eta_{t \geq g}^{DR, nev}(\eta_{t \geq g}^{DR, nev})' \Big]  $$
where $\eta_{t \geq g}^{DR, nev}$ is as described in Appendix \ref{proof:asymptotic_properties}.

\end{proposition}

The proof appears in Appendix \ref{proof:asymptotic_properties}. 

Proposition \ref{asym_dist} shows that, under the given assumptions, the proposed estimator is asymptotically normally distributed. 
The asymptotic variance of the estimator can be estimated using a straightforward plug-in estimator, that is, by replacing population values with their sample analogues. 

In related work on two-period difference-in-differences, \textcite{santannaDoublyRobustDifferenceindifferences2020} derive the semiparametric efficiency bound and show that a doubly-robust-type estimator is semiparametrically efficient, and \textcite{chenEfficientDifferenceinDifferencesEvent2025} expand this to include staggered treatment designs. Future work should explore how these efficiency results operate for estimation of the $CDATT$ in the staggered triple difference setting.

\section{Application}
In this section, I use the data and approach by \textcite{gruberIncidenceMandatedMaternity1994} to illustrate the impacts of estimating the DATT and CDATT by the methods I propose. First, I design a realistic Monte Carlo simulation study calibrated to the data and application in that paper to show the properties of my proposed estimators in several cases where the DATT and CDATT diverge. Next, I re-analyze the data to show that, in this analysis of the impacts of mandated insurance coverage of childbirth costs on labor market outcomes, the DATT and the CDATT can offer different conclusions. 

\subsection{Background and data}
\textcite{gruberIncidenceMandatedMaternity1994} studies the impacts of legislation requiring the cost of childbirth (``maternity benefits") to be covered by employers' health insurance policies. The author is interested in studying whether employers' costs of providing these benefits are shifted to the workers likely to need them. Specifically, the paper studies whether employers are able to pass on group-specific costs on the basis of demographics, or whether other frictions, such as anti-discrimination legislation, prevent this and make some workers more costly for employers to hire. 

The paper exploits two main sources of policy variation: state-level and federal-level mandates which required insurance companies to cover childbirth costs on a basis equal to their coverage of other medical conditions. In 1978, the federal Pregnancy Discrimination Act (PDA) made this requirement national. 

The author uses a triple difference design and is interested in comparing multiple subgroups. The subgroups of interest are those more likely to use the coverage for childbirth costs: married women age 20-40, single women age 20-40, and married men age 20-40 (whose may have wives covered by their insurance policy). The comparison individuals are single men age 20-40 and people over age 40, who are unlikely to use these benefits. The parallel gaps assumption requires that, in the absence of the mandates, the difference between the comparison subgroup and each of the subgroups of interest would have evolved similarly across states. That is, for example, any macroeconomic or state-level shocks correlated with the policy would have affected the subgroups similarly. 

The data for this analysis are drawn from the May Current Population Survey 1974-1978 \parencite{usbureauofthecensusCurrentPopulationSurvey1992a,usbureauofthecensusCurrentPopulationSurvey1992b,usbureauofthecensusCurrentPopulationSurvey1992c,usbureauofthecensusCurrentPopulationSurvey1992}. In my re-analysis, I focus on the state-level mandates. Between July 1, 1976 and January 1, 1977, three states enacted such a mandate to provide coverage for childbirth. Thus, for this analysis, the treated states are Illinois, New Jersey, and New York. The untreated comparison states are Ohio, Indiana, Connecticut, Massachusetts, and North Carolina. The study window covers two years before the mandate (1974 and 1975) and two years after (1977 and 1978).\footnote{For simplicity in this analysis, and consistent with the original paper, I pool these into a pre- and post-period.}

The author's conclusions suggest that employers can, and do, pass on group-specific costs to workers. The paper first documents that, before these mandates came into effect, many people did not have full coverage for the costs of childbirth and that adding this policy to insurance packages was likely to be very costly. The author then attempts to evaluate causal impacts of the mandates on the hourly wages, hours worked, and employment of the subgroups. The author concludes that, for targeted workers likely to use the childbirth coverage, wages decreased significantly.  

\subsection{Design of calibrated simulation study}
\label{section:simulation_design}
In this section, I evaluate the finite-sample performance of the proposed estimators for the CDATT using a Monte Carlo simulation study. I design a setup calibrated to the empirical application in \textcite{gruberIncidenceMandatedMaternity1994} to highlight a clear case where the DATT and CDATT diverge and in which failing to differentiate the two when interpreting results could lead to misleading conclusions. I show the performance of my estimator under different forms of treatment effect heterogeneity. 

Let $Y$ represent the log of hourly wage, $W$ represent whether the individual lives in a state with a mandate or no mandate, $S$ represent whether the individual is a member of the targeted subgroup (married and single women age 20-40, as well as married men age 20-40) or a member of the untargeted subgroup (for this simulation, unmarried men age 20-40),\footnote{Although the original design also includes individuals over age 40 in the untargeted group, these are excluded in this simulation in order to improve the overlap between the groups with respect to age.} and $X$ represent the individual's covariates (education, a quadratic in age, white/non-white, union/non-union, and white-collar/not white collar job). Since age is highly correlated with marital status, there may be limited overlap between subgroups with respect to age. To address this, when estimating the CDATT, I trim observations with any propensity score less than 0.025. A detailed discussion of overlap and robustness to trimming in this simulation appears in Appendix \ref{appendix_overlap}.

The simulation sample is constructed as follows: for individual $i$, a vector $X_i$ is drawn randomly from the data, without replacement, so that the original distribution of covariates is maintained. To match the formal results outlined above on panel data, a panel is created by taking all pre-period observations and assuming that covariates do not change over time.

Then, subgroup status is assigned according to the following propensity score models: 
\begin{align}
   P(G=1, S=1) & = \frac{\exp(\alpha^{ws} X_i)}{1 + \exp(\alpha^{ws} X_i) + \exp(\alpha^{ws'} X_i) + \exp(\alpha^{cs} X_i)} \\
   P(G=1, S=0) & = \frac{\exp(\alpha^{ws'} X_i)}{1 + \exp(\alpha^{ws} X_i) + \exp(\alpha^{ws'} X_i) + \exp(\alpha^{cs} X_i)}  \\
   P(G=\infty, S=1) & = \frac{\exp(\alpha^{cs} X_i)}{1 + \exp(\alpha^{ws} X_i) + \exp(\alpha^{ws'} X_i) + \exp(\alpha^{cs} X_i)}  \\
   P(G=\infty, S=0) & = 1 - P(G=1, S=1) - P(G=1, S=0) - P(G=\infty, S=1)
\end{align}
The coefficients $\alpha^{ws}, \alpha^{ws'},$ and $\alpha^{cs}$ are chosen to create a realistic data generating process (DGP) by using the coefficients for the same regression in the original, unaltered dataset. The coefficients appear in Appendix Table \ref{tab:sim_study_param_ps}. This strategy results in subgroup shares that mirror those in the original data. However, the results are robust to balancing the subgroup shares (so that, for example, 25\% of observations are in each subgroup), as shown in  Appendix \ref{appendix_simulation_balanced}.

Then, outcomes are constructed according to an outcome regression: 
\begin{align}
    Y_0 & = \beta X_i + u_i^{pre} \\
    Y_1 & = \beta X_i + W_i R_i + u_i^{post} 
\end{align}
with $Y_0$ assigned to observations in the pre-period and $Y_1$ assigned to observations in the post-period. Again, the coefficients $\beta$ are chosen to give a realistic DGP by using the coefficients for the same regression in the unaltered, original dataset. The error terms $u_i^{pre}, u_i^{post} \sim \mathcal{N}(0, \sigma_u)$, where $\sigma_u$ is the standard deviation of the residuals of this regression in the original dataset. The coefficients and standard deviation of the residual appear in Appendix Table \ref{tab:sim_study_mod_y}.

The crucial feature of this outcome regression is $R_i$, which represents a random variable distributed according to $\mathcal{N}(\beta X_i, \gamma \sigma_{\beta X_i})$, where $\sigma_{\beta X_i}$ is the standard deviation of $\beta X_i$ across observations (see Appendix Table \ref{tab:sim_study_mod_y}). In this setup, the $W_i R_i$ term introduces a non-zero average treatment effect on the treated (because outcomes depend causally on $W_i$), the magnitude of which is heterogeneous across individuals (because $R_i$ is random) and, importantly, across subgroups (because $X_i$ is correlated with $R_i$ and $S_i$ given the propensity score models). The parameter $\gamma$ controls the strength of these correlations. For comparison, I also show a scenario where this term is omitted. 

This setup clearly showcases a case where the DATT and CDATT differ. The outcome model is constructed so that the true CDATT is zero, because there is no causal effect of subgroup status. However, the true DATT is not zero, because there is a causal effect of treatment $W$ which differs between the subgroups. More formally, for each individual, 
\begin{align*}
    Y_{i1}(0, s) & = Y_{i1}(0,s') = \beta X_i + u_i^{post} \\
    Y_{i1}(1, s) & = Y_{i1}(1,s') = \beta X_i + W_i R_i + u_i^{post} \\
\end{align*}
Thus, individual treatment effects are given by 
\begin{align*}
S_iY_{i1}(1, s)  + (1-S_i) Y_{i1}(1,s') - (S_i Y_{i1}(0, s) + (1-S_i)Y_{i1}(0,s')) = W_i R_i 
\end{align*}
which is correlated with $S_i$ via $X_i$. 

Given this setup, the outcome regressions are correctly specified when they are linear in $X_i$ and the propensity score models are correctly specified by a multinomial logistic regression. For comparison, examples will be shown when these models are misspecified. When models are misspecified, the only covariate used to estimate the parameters of interest will be a nonlinear function of education, where 
\begin{align*}
\Tilde X_i = \ln(X_i+1) + \text{sign}(X_i)|X_i|^\nu \text{ where }\nu \sim U (2, 5)
\end{align*}
That is, the models specified by the researcher will be misspecified because they omit the other elements of $X_i$ and because of the nonlinear transformation of education. 

\subsection{Results of calibrated simulation study}
\label{section:simulation_results}
In this section, I highlight the properties of my proposed estimators in the calibrated simulation described above. The simulations highlight a clear case where the DATT and CDATT can diverge. They show how the performance of the estimators for the DATT and CDATT compare depending on the variance in treatment effects.

\begin{figure}[h]
    \caption{Treatment effect heterogeneity under different data generating processes}
    \label{fig:het_by_gamma}
    \centering
    \begin{subfigure}{.3\textwidth}
        \includegraphics[width=\textwidth]{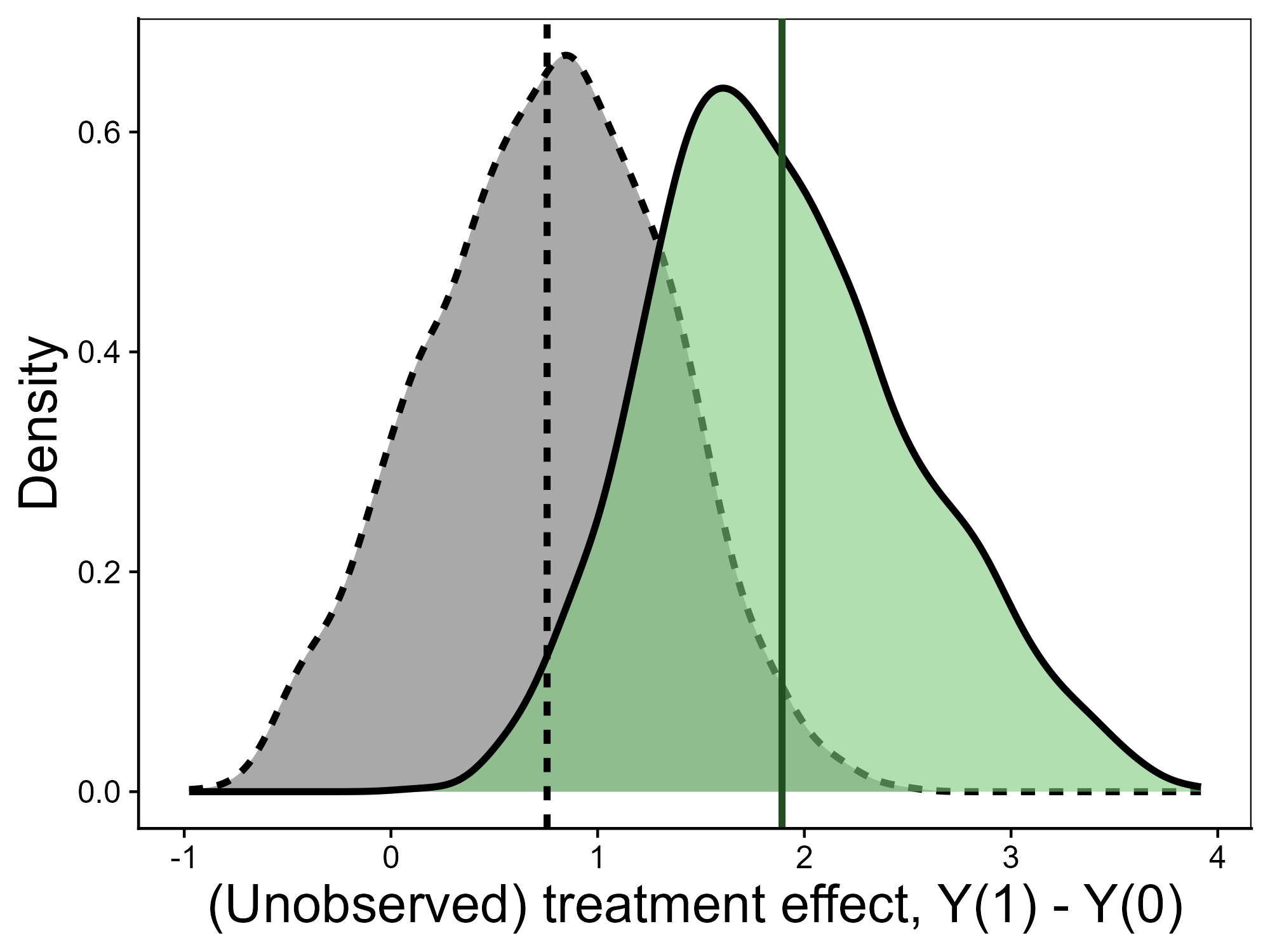}
        \caption{$\gamma = 0$}
    \end{subfigure}
    \begin{subfigure}{.3\textwidth}
        \includegraphics[width=\textwidth]{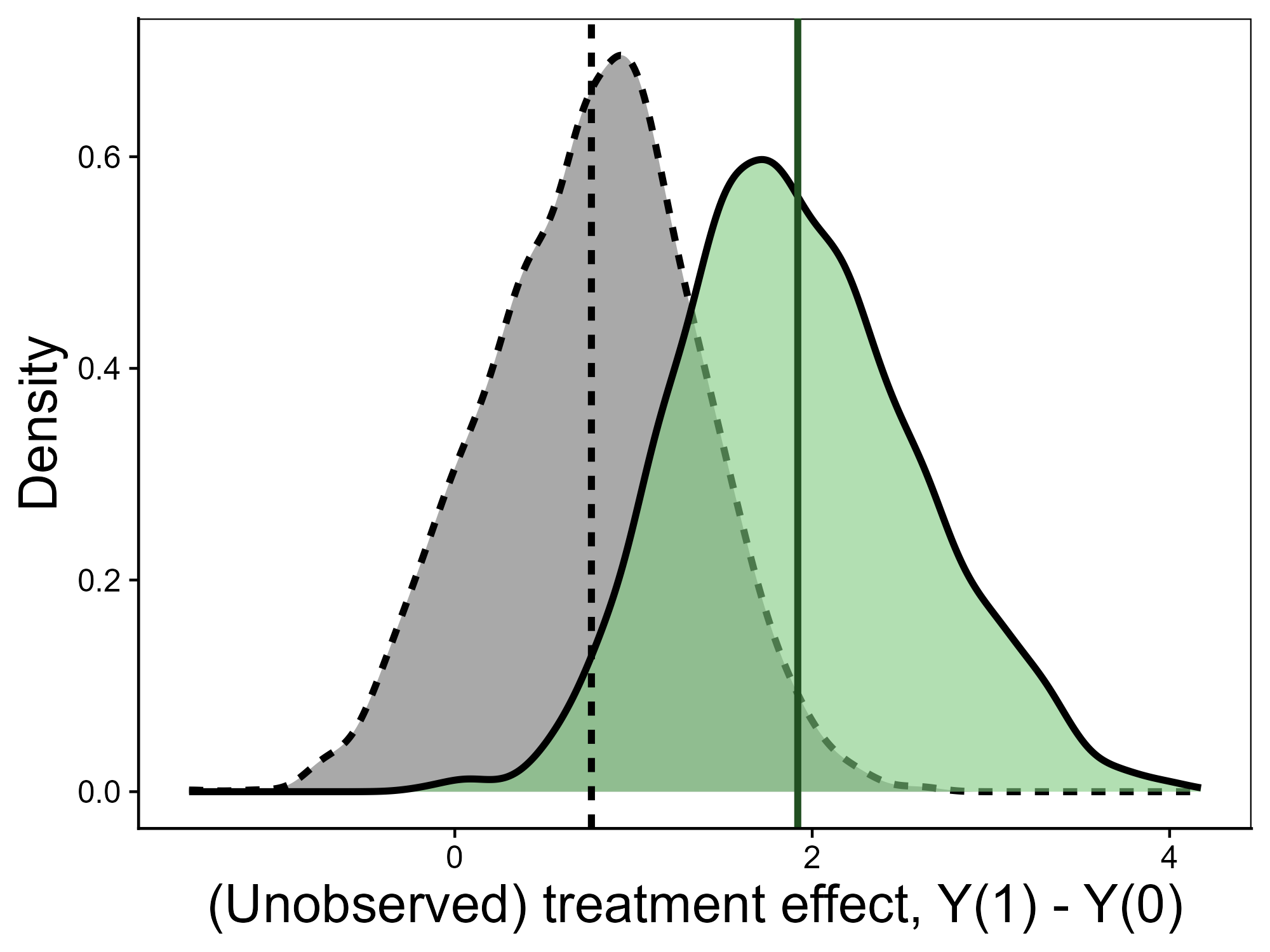}
        \caption{$\gamma = 0.2$}
    \end{subfigure}

    \begin{subfigure}{.3\textwidth}
        \includegraphics[width=\textwidth]{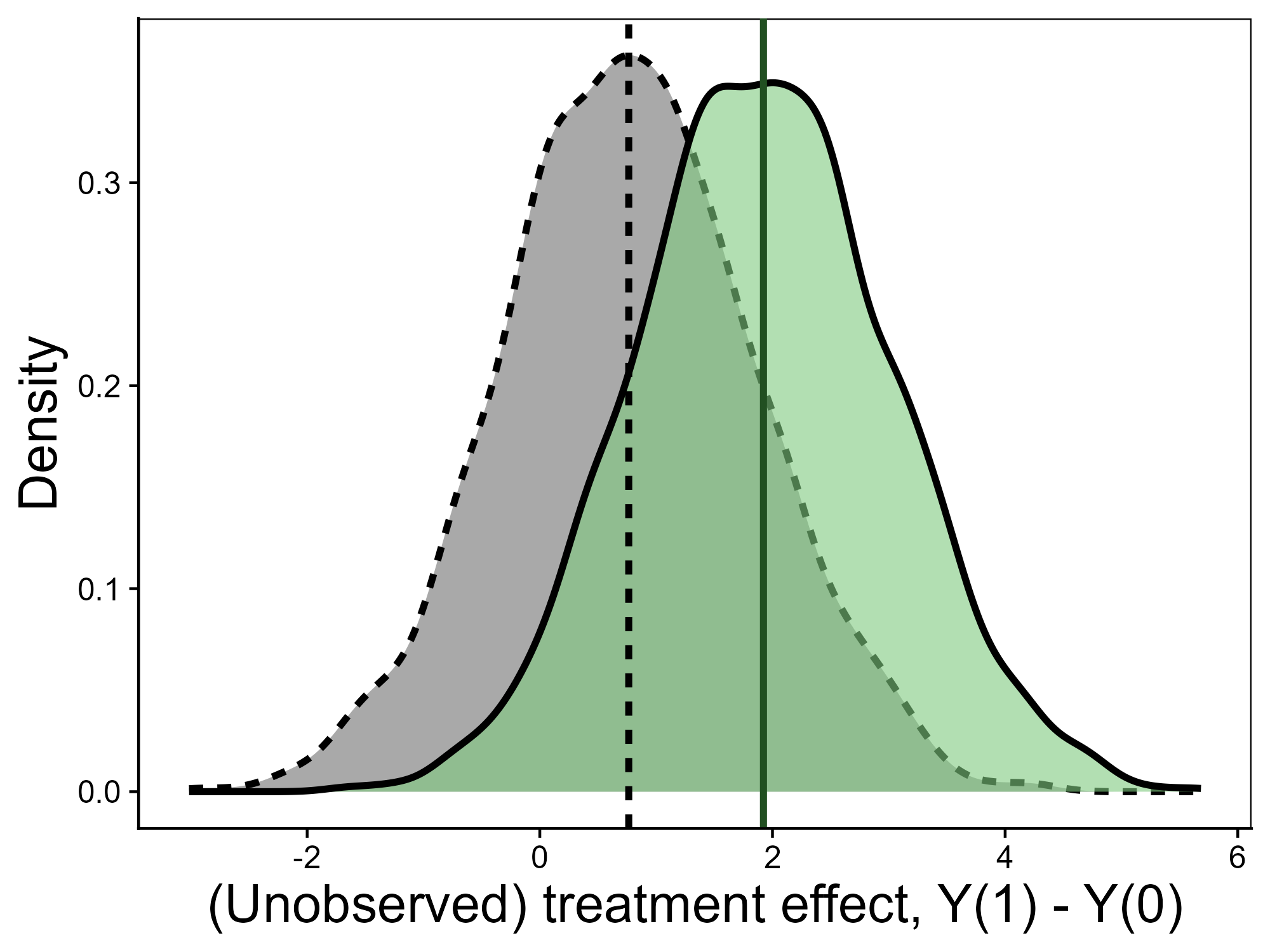}
        \caption{$\gamma = 1$}
    \end{subfigure}
    \begin{subfigure}{.3\textwidth}
        \includegraphics[width=\textwidth]{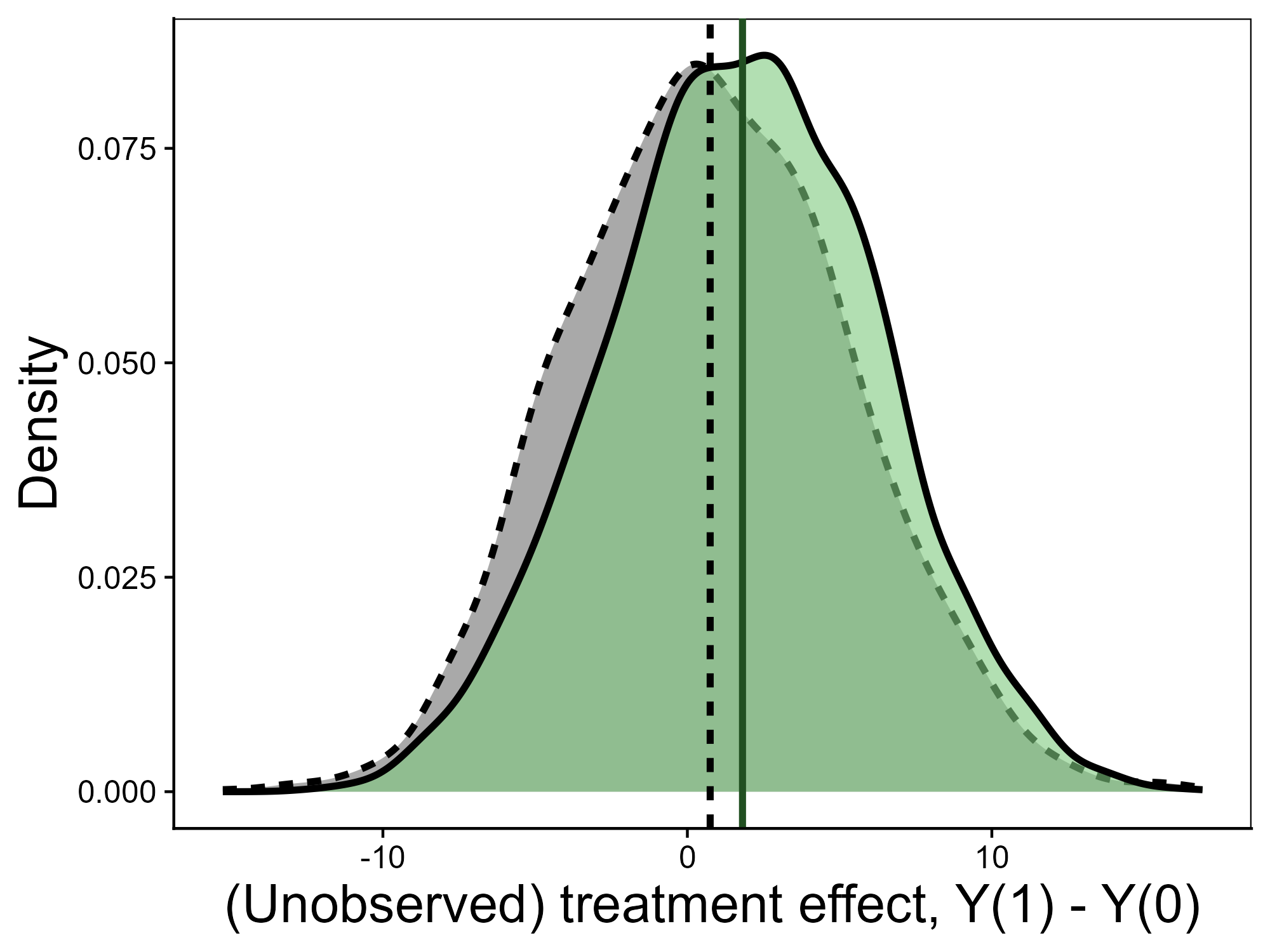}
        \caption{$\gamma = 5$}
    \end{subfigure}

    \begin{subfigure}{\textwidth}
        \centering 
        \includegraphics[width=.3\textwidth]{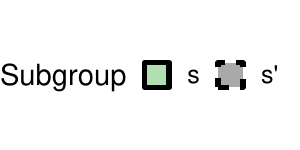}
    \end{subfigure}
    {\\ \raggedright \textbf{Notes:} The figure shows the distribution of true treatment effects in one iteration of the simulation design described above. Vertical lines indicate the average treatment effect on the treated for each subgroup. \par }
\end{figure}

Figure \ref{fig:het_by_gamma} introduces the scenarios studied, each with a different distribution of treatment effects, controlled by the parameter $\gamma$. As described above, $\gamma$ controls the variance of the distribution of treatment effects between individuals. The figures highlight that, as $\gamma$ increases and the variance of treatment effects increases, the the difference between subgroups becomes more difficult to distinguish. 

\begin{figure}[h]
    \caption{Estimator performance under various treatment effect heterogeneity scenarios}
    \centering
    \includegraphics[width=0.9\linewidth]{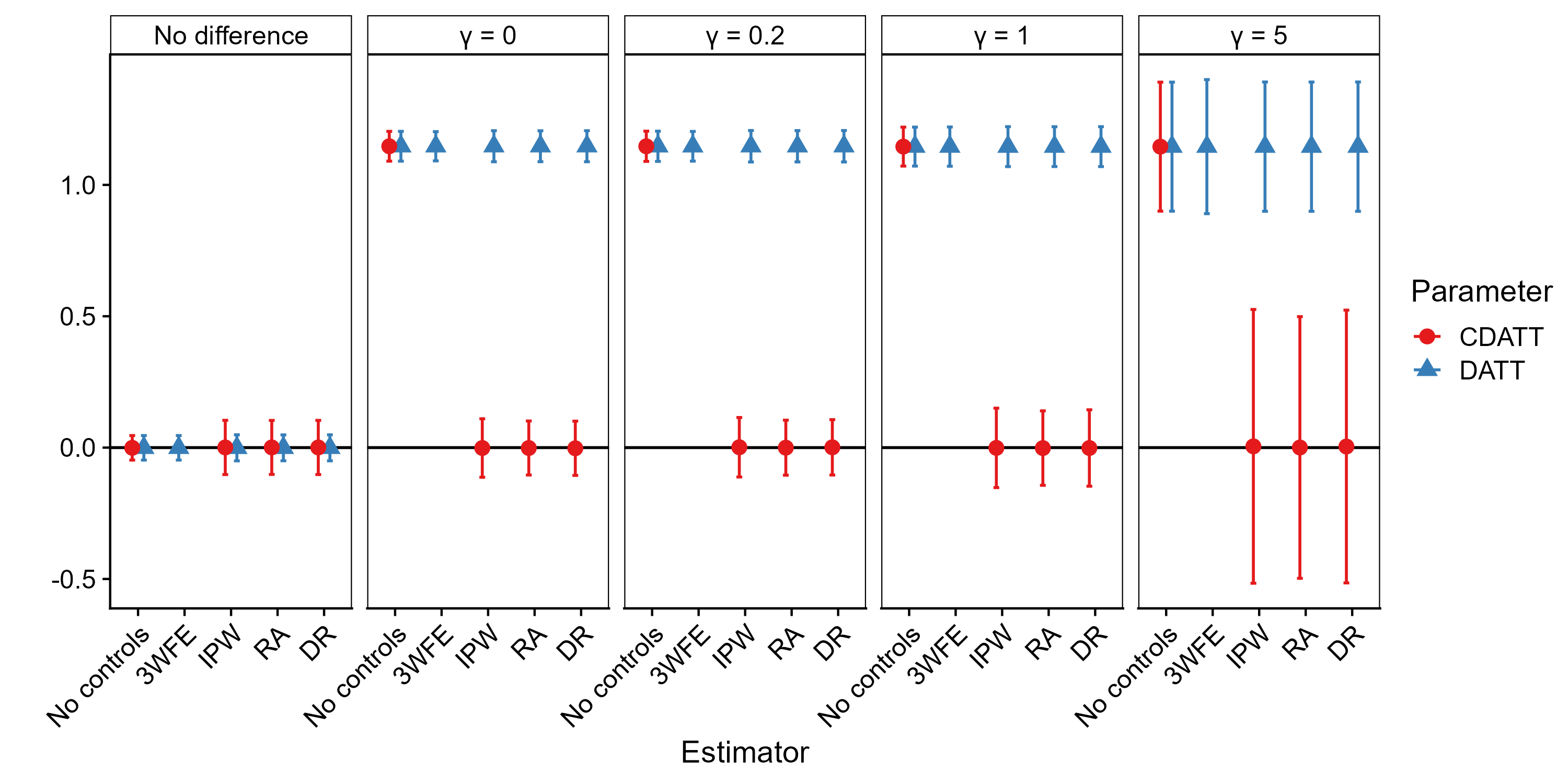}
    \label{fig:calibrated_by_het_with_trimming}
    {\\ \raggedright \textbf{Notes:} This figure presents the results of 2000 Monte Carlo simulations using the data-generating processes described above. The points represent the average value of the estimate across 2000 simulations. 95\% confidence intervals are shown. ``No difference'' refers to a scenario with no average treatment effect, and no difference between subgroups. \par }

\end{figure}

Table \ref{tab:sim_study_att_by_het_with_trimming} and Figure \ref{fig:calibrated_by_het_with_trimming} highlight two key features of the performance of the estimators for the DATT and the CDATT under the various scenarios for treatment effect heterogeneity.

First, the results underscore the divergence between the DATT and the CDATT when the underlying treatment effects are correlated with subgroup status. When there is no difference in the treatment effects between the groups, as in the first panel, the estimates of the DATT and CDATT are virtually identical. However, when the treatment effects vary with subgroup status, the DATT differs strongly from the CDATT. For example, an analysis of the DATT could lead a researcher to conclude that the treatment caused individuals in the targeted subgroup to experience an effect of 1.1 relative to their untargeted counterparts. However, an analysis of the CDATT would highlight that this differential impact on the targeted subgroup was not causally due to their targeted-subgroup status, but rather explained by other factors, like age and education, that differ between the subgroups. The simulations make clear that, regardless of which estimator is used for the DATT, conditioning on covariates does not help the researcher recover an estimate of the CDATT. 

Second, the results show that, as the variance in treatment effects increases, the estimators lose power and precision. When $\gamma = 5$, the variance of the underlying treatment effects is large, and both estimators have substantially larger standard errors and confidence intervals relative to the other scenarios. The loss in precision is somewhat greater for the estimators of the CDATT due to the increased difficulty of distinguishing between the two subgroups. In this setting, the CDATT is most precisely estimated using the regression adjustment approach, which leverages the (correct) assumption of that outcome models are linear in $X$, while the IPW and DR estimators have slightly larger standard errors but allow for more flexible relationships between $X$ and $Y$ (as in \textcite{santannaDoublyRobustDifferenceindifferences2020}). 

\begin{table}[H]
    \caption{Simulation study results based on 2000 Monte Carlo trials}
    \label{tab:sim_study_att_by_het_with_trimming}
    \centering
    \resizebox{0.6\textwidth}{!}{
    \begin{tabular}{lccccccc}
    \multicolumn{5}{l}{\textbf{No average treatment effect on the treated}} \\ \hline
    % latex table generated in R 4.4.3 by xtable 1.8-4 package
% Fri Sep  4 02:35:09 2026
Estimator & Avg. bias & Med. bias & RMSE & SE & Cover & CI Len \\ 
  \hline
No controls & -0.001 & -0.001 & 0.024 & 0.024 & 0.952 & 0.093 \\ 
  $DATT^{3WFE}$ & -0.001 & -0.001 & 0.024 & 0.024 & 0.953 & 0.093 \\ 
  $DATT^{RA}$ & -0.001 & -0.002 & 0.026 & 0.025 & 0.951 & 0.098 \\ 
  $DATT^{IPW}$ & -0.001 & -0.001 & 0.026 & 0.025 & 0.950 & 0.099 \\ 
  $DATT^{DR}$ & -0.001 & -0.001 & 0.026 & 0.025 & 0.949 & 0.099 \\ 
  $CDATT^{RA}$ & 0.001 & 0.000 & 0.053 & 0.052 & 0.945 & 0.205 \\ 
  $CDATT^{IPW}$ & 0.001 & 0.000 & 0.053 & 0.053 & 0.949 & 0.207 \\ 
  $CDATT^{DR}$ & 0.001 & 0.000 & 0.053 & 0.053 & 0.948 & 0.206 \\ 
   \hline
 \\
    \multicolumn{5}{l}{\textbf{$\mathbf{\gamma = 0}$ }} \\ \hline
    % latex table generated in R 4.4.3 by xtable 1.8-4 package
% Fri Sep  4 02:35:09 2026
Estimator & Avg. bias & Med. bias & RMSE & SE & Cover & CI Len \\ 
  \hline
No controls & 1.147 & 1.147 & 1.147 & 0.029 & 0.000 & 0.114 \\ 
  $DATT^{3WFE}$ & 1.147 & 1.147 & 1.147 & 0.028 & 0.000 & 0.111 \\ 
  $DATT^{RA}$ & 1.147 & 1.148 & 1.148 & 0.030 & 0.000 & 0.118 \\ 
  $DATT^{IPW}$ & 1.147 & 1.148 & 1.148 & 0.030 & 0.000 & 0.118 \\ 
  $DATT^{DR}$ & 1.147 & 1.148 & 1.148 & 0.030 & 0.000 & 0.118 \\ 
  $CDATT^{RA}$ & -0.001 & -0.001 & 0.054 & 0.052 & 0.953 & 0.206 \\ 
  $CDATT^{IPW}$ & -0.002 & -0.002 & 0.056 & 0.057 & 0.957 & 0.223 \\ 
  $CDATT^{DR}$ & -0.003 & -0.003 & 0.052 & 0.053 & 0.959 & 0.207 \\ 
   \hline
 \\
    \multicolumn{5}{l}{\textbf{$\mathbf{\gamma = 0.2}$}} \\ \hline
    % latex table generated in R 4.4.3 by xtable 1.8-4 package
% Fri Sep  4 02:35:09 2026
Estimator & Avg. bias & Med. bias & RMSE & SE & Cover & CI Len \\ 
  \hline
No controls & 1.146 & 1.146 & 1.146 & 0.038 & 0.000 & 0.148 \\ 
  $DATT^{3WFE}$ & 1.146 & 1.146 & 1.146 & 0.038 & 0.000 & 0.149 \\ 
  $DATT^{RA}$ & 1.146 & 1.146 & 1.146 & 0.039 & 0.000 & 0.152 \\ 
  $DATT^{IPW}$ & 1.146 & 1.146 & 1.146 & 0.039 & 0.000 & 0.152 \\ 
  $DATT^{DR}$ & 1.146 & 1.146 & 1.146 & 0.039 & 0.000 & 0.152 \\ 
  $CDATT^{RA}$ & -0.002 & -0.000 & 0.071 & 0.072 & 0.952 & 0.283 \\ 
  $CDATT^{IPW}$ & -0.001 & 0.000 & 0.078 & 0.077 & 0.951 & 0.302 \\ 
  $CDATT^{DR}$ & -0.001 & 0.000 & 0.074 & 0.074 & 0.948 & 0.291 \\ 
   \hline
 \\
    \multicolumn{5}{l}{\textbf{$\mathbf{\gamma = 1}$}} \\ \hline
    % latex table generated in R 4.4.3 by xtable 1.8-4 package
% Fri Sep  4 02:35:09 2026
Estimator & Avg. bias & Med. bias & RMSE & SE & Cover & CI Len \\ 
  \hline
No controls & 1.147 & 1.146 & 1.147 & 0.029 & 0.000 & 0.115 \\ 
  $DATT^{3WFE}$ & 1.147 & 1.146 & 1.147 & 0.029 & 0.000 & 0.113 \\ 
  $DATT^{RA}$ & 1.147 & 1.147 & 1.148 & 0.030 & 0.000 & 0.119 \\ 
  $DATT^{IPW}$ & 1.147 & 1.147 & 1.148 & 0.031 & 0.000 & 0.120 \\ 
  $DATT^{DR}$ & 1.147 & 1.147 & 1.148 & 0.031 & 0.000 & 0.120 \\ 
  $CDATT^{RA}$ & -0.000 & -0.001 & 0.054 & 0.053 & 0.944 & 0.209 \\ 
  $CDATT^{IPW}$ & 0.001 & 0.000 & 0.058 & 0.058 & 0.949 & 0.226 \\ 
  $CDATT^{DR}$ & 0.001 & 0.001 & 0.055 & 0.054 & 0.943 & 0.211 \\ 
   \hline
 \\
    \multicolumn{5}{l}{\textbf{$\mathbf{\gamma = 5}$}} \\ \hline
    % latex table generated in R 4.4.3 by xtable 1.8-4 package
% Fri Sep  4 02:35:09 2026
Estimator & Avg. bias & Med. bias & RMSE & SE & Cover & CI Len \\ 
  \hline
No controls & 1.146 & 1.145 & 1.152 & 0.125 & 0.000 & 0.491 \\ 
  $DATT^{3WFE}$ & 1.146 & 1.145 & 1.152 & 0.130 & 0.000 & 0.510 \\ 
  $DATT^{RA}$ & 1.146 & 1.146 & 1.152 & 0.126 & 0.000 & 0.492 \\ 
  $DATT^{IPW}$ & 1.146 & 1.146 & 1.152 & 0.126 & 0.000 & 0.492 \\ 
  $DATT^{DR}$ & 1.146 & 1.145 & 1.152 & 0.126 & 0.000 & 0.492 \\ 
  $CDATT^{RA}$ & 0.000 & 0.002 & 0.255 & 0.254 & 0.947 & 0.996 \\ 
  $CDATT^{IPW}$ & 0.005 & 0.010 & 0.265 & 0.266 & 0.948 & 1.042 \\ 
  $CDATT^{DR}$ & 0.004 & 0.011 & 0.266 & 0.265 & 0.948 & 1.038 \\ 
   \hline
 \\
    \end{tabular}
}
{\\ \raggedright \footnotesize \textbf{Notes:} This table presents the results of 2000 Monte Carlo simulations using the data-generating processes described above. Avg. bias refers to the average value of the estimate across 2000 simulations. Median bias refers to the median estimate. RMSE refers to the root mean squared error. SE refers to the average standard error across the trials. Cover refers to the 95\% confidence interval coverage rate. CI Len refers to the average length of the 95\% confidence interval. \par }
\end{table}
\newpage

\begin{figure}[h]
    \caption{Simulation study results based on 2000 Monte Carlo trials}
    \centering
    \includegraphics[width=0.75\linewidth]{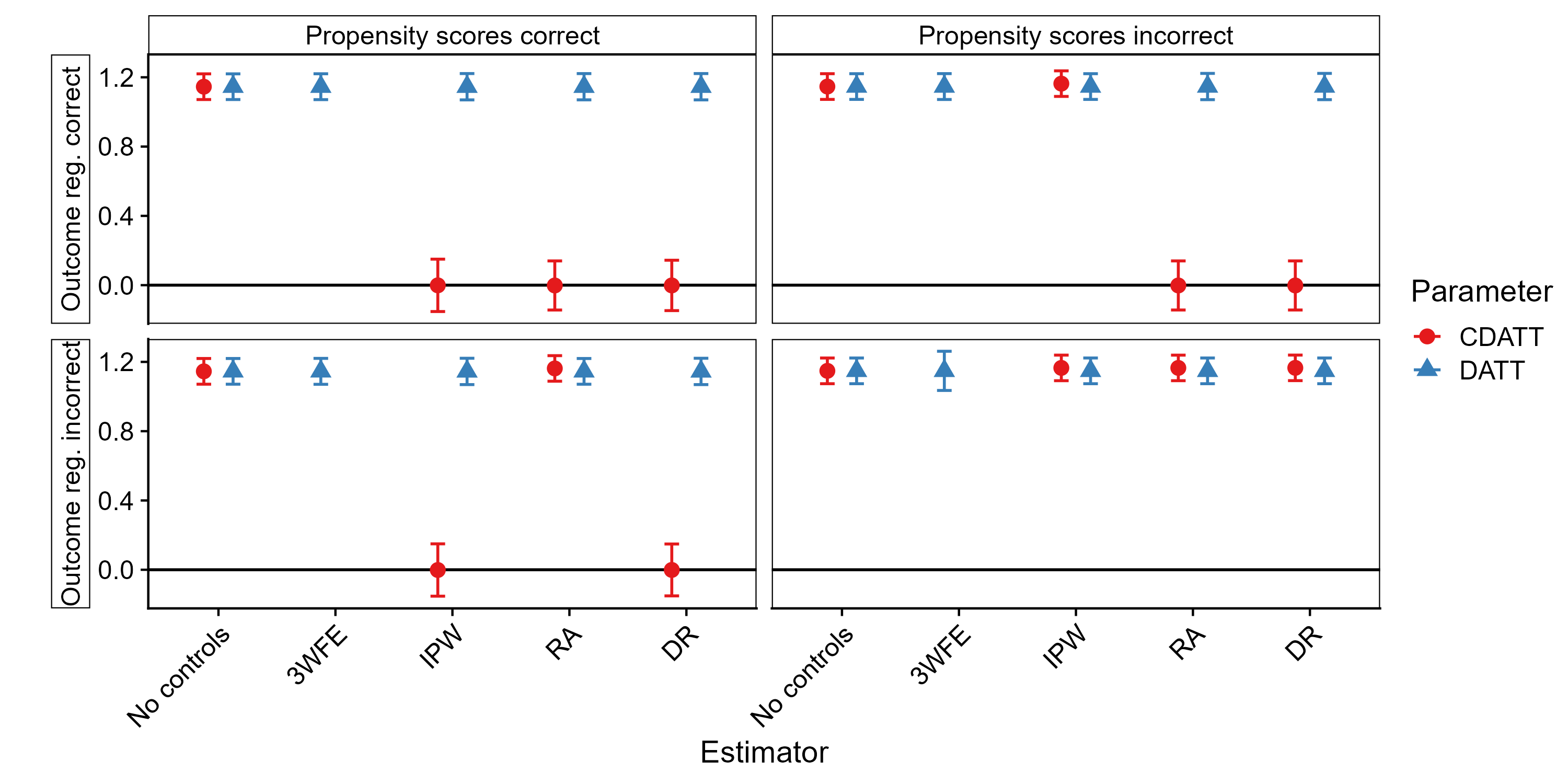}
    \label{fig:calibrated_psor_with_trimming}
    {\\ \raggedright \textbf{Notes:} This figure presents the results of 2000 Monte Carlo simulations using the data-generating processes described above with $\gamma=1$. The points represent the average value of the estimate across 2000 simulations for each estimator studied. 95\% confidence intervals are shown. \par }

\end{figure}

Next, Table \ref{tab:sim_study_att_psor_with_trimming} and Figure \ref{fig:calibrated_psor_with_trimming} compare the performance of the different estimators of the CDATT under various misspecification scenarios. The results highlight the double-robustness of the doubly-robust estimator.

The results show that the doubly-robust estimator of the CDATT performs well when at least one (but not necessarily both) set of working models is correctly specified. When both sets of working models are correctly specified, as in Case 1 in Table \ref{tab:sim_study_att_psor_with_trimming}, the regression adjustment, IPW, and doubly-robust estimators all perform well, with minimal bias. All three estimators attain nearly correct coverage of the 95\% confidence interval. When only the outcome regressions are specified correctly but the propensity scores are misspecified, as in Case 2, the regression adjustment and doubly-robust estimators have minimal bias, while the IPW estimator is biased. On the other hand, when the propensity score models are correct but the outcome regressions are misspecified, as in Case 3 and as expected, the IPW and doubly-robust methods are nearly unbiased, while the regression adjustment estimator has substantial bias. When all working models are misspecified, all available estimators have substantial bias.

\begin{table}[H]
    \caption{Simulation study results under misspecification}
    \label{tab:sim_study_att_psor_with_trimming}
    \centering
    \resizebox{0.9\textwidth}{!}{
    \begin{tabular}{lccccccc}
    \multicolumn{5}{l}{\textbf{Case 1: PS and OR correct}} \\ \hline
    % latex table generated in R 4.4.3 by xtable 1.8-4 package
% Fri Sep  4 02:35:13 2026
Estimator & Avg. bias & Med. bias & RMSE & Asym. SE & Cover & CI Len \\ 
  \hline
No controls & 1.146 & 1.146 & 1.146 & 0.038 & 0.000 & 0.148 \\ 
  $CDATT^{OR}$ & -0.002 & -0.000 & 0.071 & 0.072 & 0.952 & 0.283 \\ 
  $CDATT^{IPW}$ & -0.001 & 0.000 & 0.078 & 0.077 & 0.951 & 0.302 \\ 
  $CDATT^{DR}$ & -0.001 & 0.000 & 0.074 & 0.074 & 0.948 & 0.291 \\ 
   \hline
 \\
    \multicolumn{5}{l}{\textbf{Case 2: OR correct, PS incorrect}} \\ \hline
    % latex table generated in R 4.4.3 by xtable 1.8-4 package
% Fri Sep  4 02:35:13 2026
Estimator & Avg. bias & Med. bias & RMSE & Asym. SE & Cover & CI Len \\ 
  \hline
No controls & 1.147 & 1.147 & 1.147 & 0.038 & 0.000 & 0.148 \\ 
  $CDATT^{OR}$ & -0.002 & -0.002 & 0.074 & 0.072 & 0.946 & 0.283 \\ 
  $CDATT^{IPW}$ & 1.163 & 1.164 & 1.164 & 0.038 & 0.000 & 0.148 \\ 
  $CDATT^{DR}$ & -0.002 & -0.003 & 0.074 & 0.072 & 0.947 & 0.283 \\ 
   \hline
 \\
    \multicolumn{5}{l}{\textbf{Case 3: PS correct, OR incorrect}} \\ \hline
    % latex table generated in R 4.4.3 by xtable 1.8-4 package
% Fri Sep  4 02:35:13 2026
Estimator & Avg. bias & Med. bias & RMSE & Asym. SE & Cover & CI Len \\ 
  \hline
No controls & 1.145 & 1.145 & 1.146 & 0.038 & 0.000 & 0.148 \\ 
  $CDATT^{OR}$ & 1.162 & 1.162 & 1.163 & 0.038 & 0.000 & 0.148 \\ 
  $CDATT^{IPW}$ & -0.001 & 0.000 & 0.079 & 0.077 & 0.938 & 0.302 \\ 
  $CDATT^{DR}$ & -0.001 & 0.002 & 0.078 & 0.076 & 0.942 & 0.300 \\ 
   \hline
 \\
    \multicolumn{5}{l}{\textbf{Case 4: OR and PS incorrect}} \\ \hline
    % latex table generated in R 4.4.3 by xtable 1.8-4 package
% Fri Sep  4 02:35:13 2026
Estimator & Avg. bias & Med. bias & RMSE & Asym. SE & Cover & CI Len \\ 
  \hline
No controls & 1.148 & 1.148 & 1.149 & 0.038 & 0.000 & 0.148 \\ 
  $CDATT^{OR}$ & 1.165 & 1.164 & 1.166 & 0.038 & 0.000 & 0.148 \\ 
  $CDATT^{IPW}$ & 1.165 & 1.164 & 1.166 & 0.038 & 0.000 & 0.148 \\ 
  $CDATT^{DR}$ & 1.166 & 1.164 & 1.166 & 0.038 & 0.000 & 0.148 \\ 
   \hline
 \\
    \end{tabular}
}
{\\ \raggedright \textbf{Notes:} This table presents the results of 2000 Monte Carlo simulations using the data-generating processes described above with $\gamma =1$. Avg. bias refers to the average value of the estimate across 2000 simulations. Median bias refers to the median estimate. RMSE refers to the root mean squared error. SE refers to the average standard error across the trials. Cover refers to the 95\% confidence interval coverage rate. CI Len refers to the average length of the 95\% confidence interval. \par }
\end{table}

Finally, in Appendix \ref{appendix_overlap}, I explore the importance of the overlap assumption in the performance of these estimators. The overlap assumption is required for both DATT and CDATT, but operates in different ways due to the different ways the estimators utilize the propensity scores. As described above, when estimating the DATT, the propensity scores do not estimate the probability of belonging to a particular subgroup. In this calibrated simulation, the greatest threat to overlap occurs between the subgroups rather than between treated and control states, as subgroup status is highly correlated with age. As such, the IPW and DR estimates of the CDATT may be affected by limited overlap in this calibration. As described above, to address the possibility of limited overlap, the main results presented here include trimming of observations with propensity to belong to any given subgroup of less than 0.025. Appendix \ref{appendix_overlap} demonstrates that, without trimming of extreme propensity scores, coverage for these estimators falls below 0.95 as the first-order asymptotic approximation is no longer suitable. In Appendix \ref{appendix_overlap}, I also test various trimming thresholds and show that the estimators perform well under multiple possible levels of trimming.

\subsection{Empirical approach for re-analysis of Gruber (1994)}
As in the original paper, I use a triple difference design to estimate the impacts of coverage for childbirth costs on three subgroups of interest relative to a subgroup unlikely to use these benefits. I extend the analysis by implementing my proposed estimators of the CDATT. 

The specification in the paper is as follows: 
\begin{align*}
    Y_{ijt} = & \alpha + \beta_1 X_{ijt} + \beta_2 \tau_t + \beta_3 \delta_j + \beta_4 s_i + \beta_5(\delta_j \times \tau_t) \\
    & + \beta_6 (\tau_t \times s_i) + \beta_7(\delta_j \times s_i) + \beta_8 (\delta_j \times \tau_t \times s_i) + u_{ijt}
\end{align*}
where $Y_{ijt}$ represents the outcome for individual $i$ in state $j$ in year $t$, $\delta_j$ represents state fixed effects, $\tau_t$ represents year fixed effects, $s_i$ represents the demographic subgroup, and $\times$ represents the interaction between two variables. Control variables are represented by $X_{ijt}$ and, in the original paper, include education, experience and experience$^2$, sex, marital status, an interaction between sex and marital status, a binary variable for white/non-white, union/non-union, and indicators for 15 major industries. In this model, which I will call the three way fixed effects (3WFE) model, the coefficient of interest is given by $\beta_8$.\footnote{Although, as in difference-in-differences, 3WFE may not be a valid approach when studying staggered treatments (see, eg, \textcite{strezhnevDecomposingTripleDifferencesRegression2023}), the treatment is not staggered in this case, so this issue does not arise.}

\begin{figure}[h]
    \caption{Descriptive statistics by subgroup}
    \centering
    \includegraphics[width=0.7\linewidth]{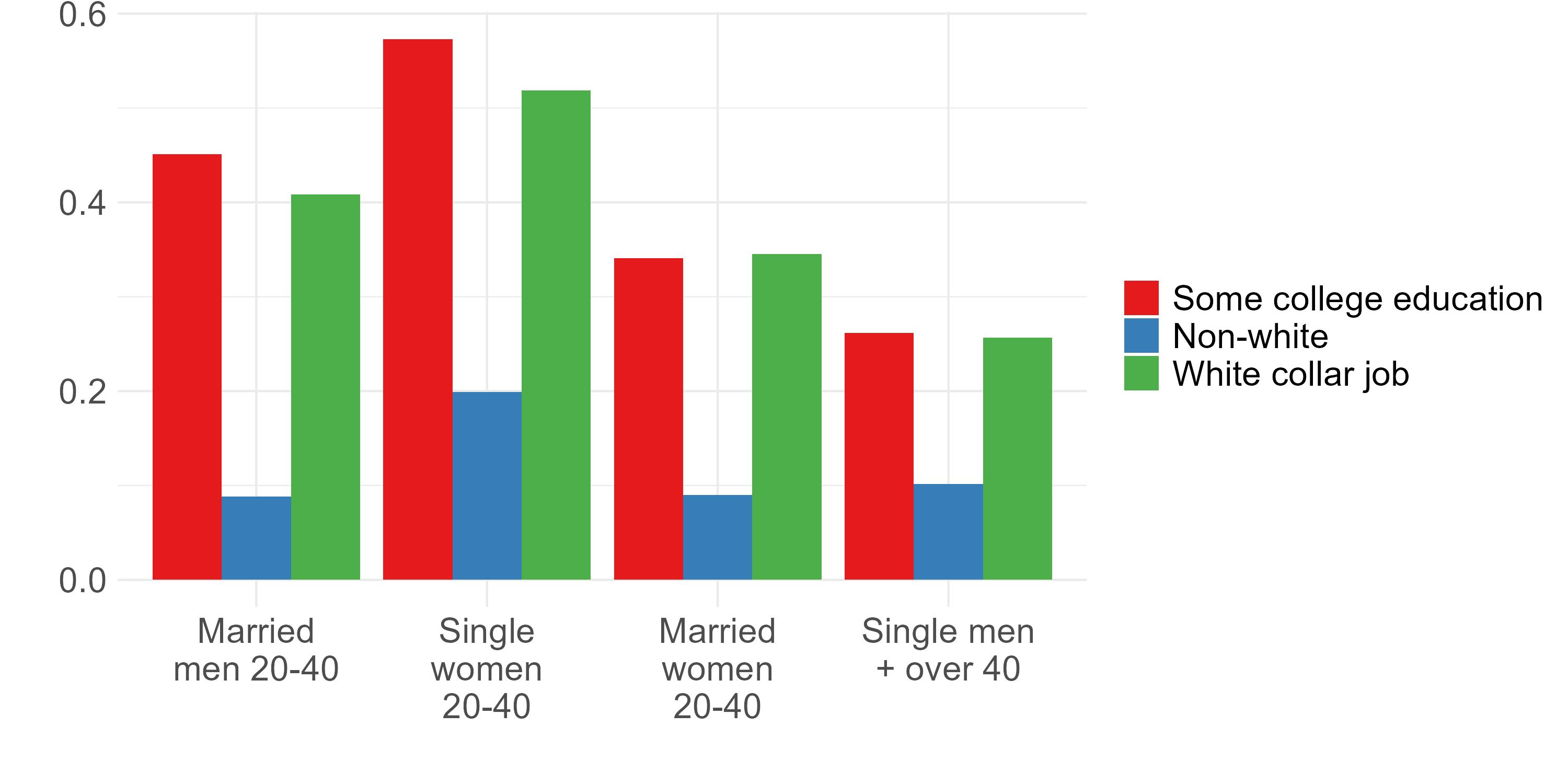}
    \label{fig:gruber_demographics}

{\raggedright \textbf{Notes:} This figure shows descriptive statistics (means) for the subgroups studied in \textcite{gruberIncidenceMandatedMaternity1994}, with data drawn from May CPS from 1974-75 and 1978-79. Some college education is an indicator for whether the individual has at least some college education, non-white is an indicator for whether the individual reports non-white race, and white collar job is an indicator for whether the person's occupation is classified as white collar. \par }
\end{figure}

To motivate the inclusion of covariates and their role in this analysis, I first highlight the divergence between demographic groups in these covariates. Figure \ref{fig:gruber_demographics} shows, for example, that single women age 20-40 tend to have higher education than the other demographic groups, are more likely to be non-white, and are more likely to work in a white-collar job. This motivates the desire to separate between the DATT and CDATT parameters. If certain jobs, such as white-collar jobs, are more or less sensitive to the mandated benefits, then any difference in ATTs between the demographic groups might be explained by their different job characteristics rather than their eligibility for the coverage of childbirth costs. 

As described above, the DATT and the CDATT coincide under the assumption that any untargeted workers would be unaffected by the policy. That is, if single men age 20-40 and people over age 40 are assumed to be entirely unaffected by the coverage of childbirth costs, regardless of other characteristics, then these two parameters will be equivalent. However, this assumption does not seem justified. For example, if the mandates make some workers more desirable (less costly) than others, then there will be spillovers on this untargeted group as employers substitute towards them. Further, if employers cannot pass on group-specific costs, then outcomes for all workers will be affected by the mandates. 

In my analysis, I use the following covariates: education, bins for age groups (under 25, 26-30, 31-35, and over 35), white/non-white, and white-collar/not white-collar.

I address treatment effect heterogeneity in this re-analysis in two ways. First, I replicate the prior work and show how the DATT and CDATT diverge even when using a similar approach and control variables. Second, I expand on the original analysis by estimating dynamic event study effects to study how treatment effects may vary over time.

\subsection{Results of Gruber (1994) re-analysis}
Estimating the DATT, as in the published paper, I find that the impacts of mandating childbirth coverage differ between the targeted subgroups (married women age 20-40, married men age 20-40, and single women age 20-40). However, when estimating the CDATT, I find evidence that these differences between the subgroups should not be interpreted as caused by subgroup status (that is, by likelihood to use the benefits), but instead appear to be due to differences in other covariates between the groups. 

\begin{figure}[h]
    \caption{Pooled results from re-analyzing impacts of childbirth coverage}
    \centering
\begin{subfigure}{.45\textwidth}
    \caption{Impacts on log of hourly wages}
    \centering
    \includegraphics[width=\textwidth]{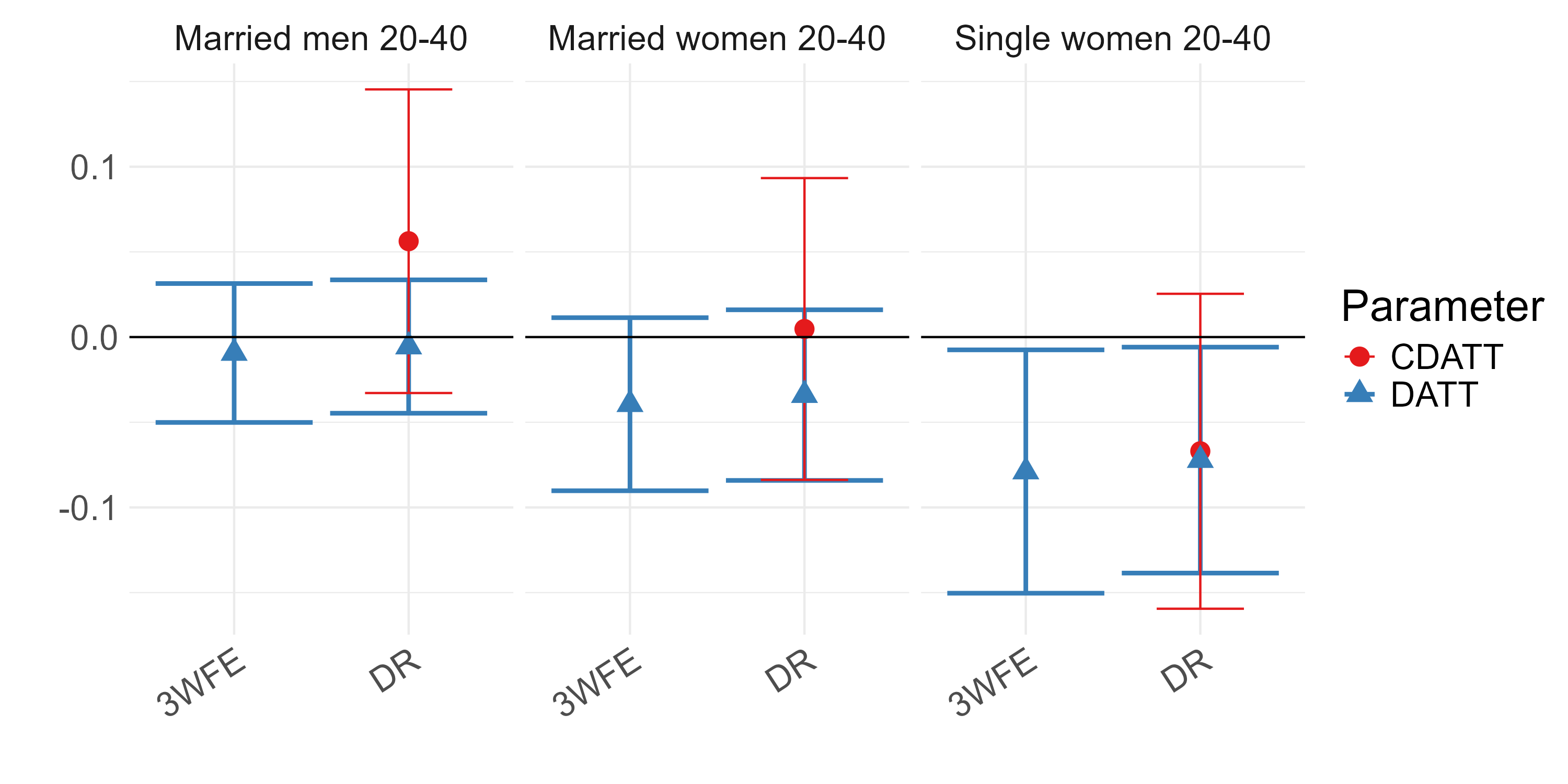}
    \label{fig:gruber_wages_DR}
\end{subfigure}
\begin{subfigure}{.45\textwidth}
    \caption{Impacts on log of hours worked}
    \centering
    \includegraphics[width=\textwidth]{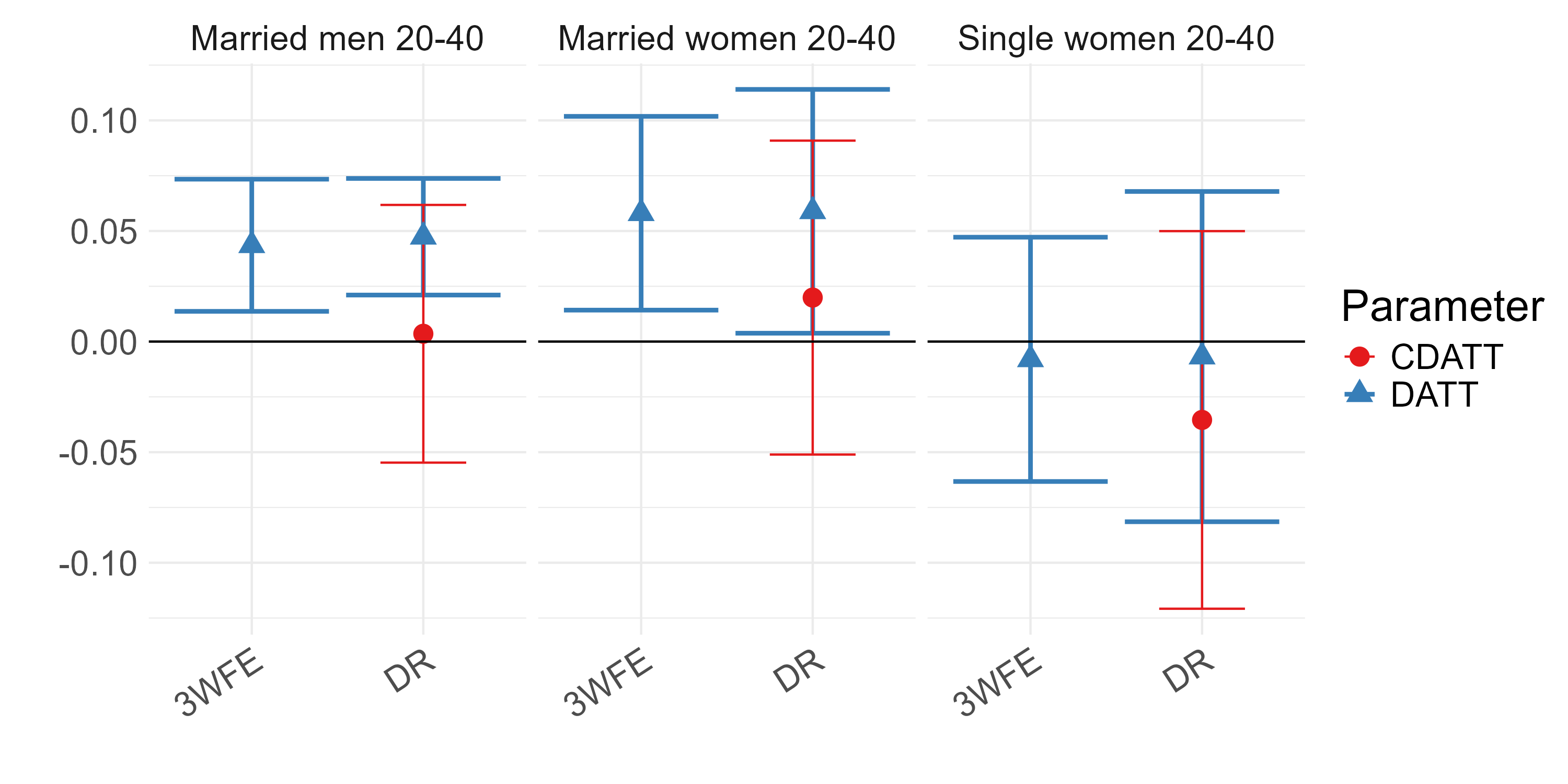}
    \label{fig:gruber_hours_DR}
\end{subfigure}
\begin{subfigure}{.45\textwidth}
    \caption{Impacts on employment}
    \centering
    \includegraphics[width=\textwidth]{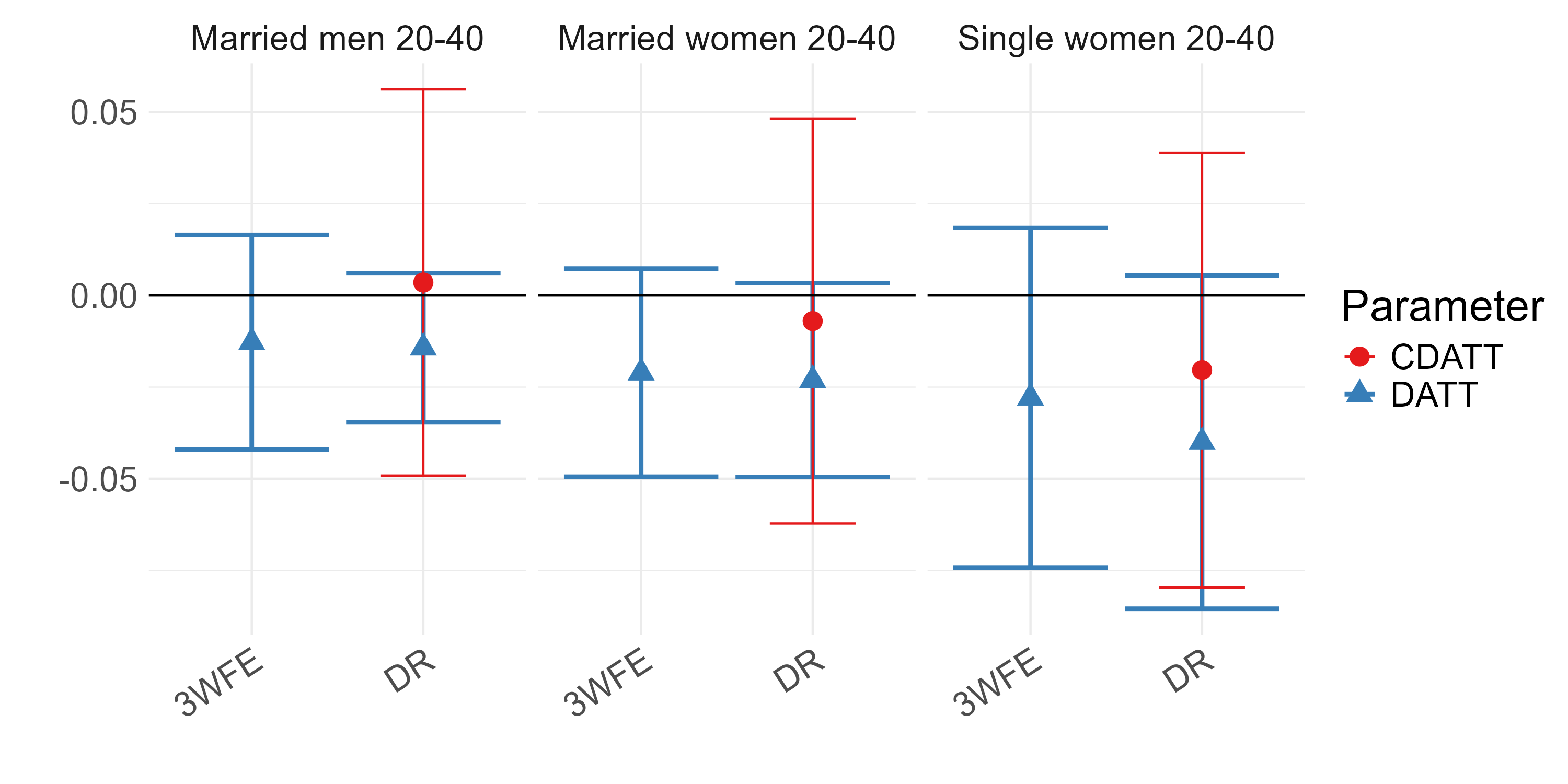}
    \label{fig:gruber_emp_DR}
\end{subfigure}

{\raggedright \textbf{Notes:} This figure plots estimated effects from the re-analysis of \textcite{gruberIncidenceMandatedMaternity1994} using pooled years as in the original paper. 3WFE represents the 3WFE estimator most closely replicating results from the original paper. DR represents the doubly-robust estimators of the DATT and CDATT described in this paper. Effects are plotted for each of the three subgroups of interest (married men 20-40, married women 20-40, and single women 20-40) relative to the comparison subgroup, and for each of three outcomes (the log of hourly wages, the log of hours worked, and employment). 95\% confidence intervals are shown. \par }

\end{figure}

Figure \ref{fig:gruber_wages_DR} highlights that, when estimating the DATT on the log of hourly wages, results are consistent with the original published results, regardless of which estimator of the DATT is used. When not including controls, the CDATT and the DATT coincide. When using a 3WFE design similar to that in the original paper, the results suggest an impact of -0.009, -0.039, and -0.079 log points in wages for married men age 20-40, married women age 20-40, and single women age 20-40, respectively. These estimates are similar to those presented in the original paper, which suggested impacts of -0.009, -0.043, and -0.042 for these groups, respectively (see Table 4 in \textcite{gruberIncidenceMandatedMaternity1994}). When including controls using the doubly-robust estimator, the magnitudes of the estimates of the DATT are largely unchanged and estimates are slightly more precise. These estimates suggest a reduction in wages for all targeted groups, which is significant regardless of estimator for single women age 20-40. The published result found significant declines for both married women age 20-40 and single women age 20-40. 

Although analysis of the DATT suggests that the benefits mandates reduced wages for the targeted subgroups relative to the untargeted subgroup, the CDATT suggests this may not be causally due to their likelihood to use the benefits. Figure \ref{fig:gruber_wages_DR} shows how the estimates of the CDATT diverge from the DATT for both married men and married women. For married men, the doubly-robust estimate is large and positive (although insignificant). For married women, it is slightly positive (insignificant), and for single women, it is negative (insignificant). Although the DATT is consistently negative for all three groups, the CDATT offers a qualitatively different result, with impacts that are less negative and are insignificant. These qualitative findings are robust to the choice of estimator, as shown in Appendix Figure \ref{fig:gruber_full_static}. 

Turning to the impacts on hours worked, the DATT and the CDATT again may have different interpretations. For married men age 20-40, the 3WFE specification reveals a significant increase in hours worked per week, by about 0.044 log points (compared to the published estimate of 0.030 log points). The magnitude of this impact is similar when including controls via the doubly-robust estimator of the DATT (0.047 log points). However, it is nearly zero when estimating the CDATT using the doubly-robust estimator. For married women, again, both estimates of the DATT reveal a positive impact of 0.058 and 0.059 log points (compared to the published impact of 0.049 log points). However, the estimate of the CDATT is smaller, only 0.019 log points, and insignificant. Finally, there is no significant impact in any case for single women, as in the published results. 

The results on employment suggest less divergence between the two estimators, and weaker overall evidence of an impact. As in the published paper, estimates of the DATT are generally negative and insignificant. Estimates of the CDATT are less negative and also insignificant. However, it is not surprising that the DATT and CDATT diverge less in this case. The difference between the two arises when treatment effects are correlated with subgroup status, and, if treatment effects are truly small or null, the magnitude of the difference between the parameters will be smaller. 

Given that treatment effects may also vary over time, I next consider each year separately to estimate event study coefficients (rather than pooling years as in the original paper). I estimate individual $DATT(g,t)$ and $CDATT(g,t)$ parameters using each year of data from the original analysis, and I add additional data from 1976 \parencite{censusCurrentPopulationSurvey1992}.\footnote{This year was omitted in \textcite{gruberIncidenceMandatedMaternity1994} because it is a partial treatment year. I treat it as the year of treatment.} 

These dynamic estimates complement the conclusions of the static analysis by both allowing for the study of treatment effects over time and for a test lending evidence to support the parallel gaps assumption by testing for evidence of parallel gaps in the pre-period. 

\begin{figure}[h]
    \caption{Event-study estimates of impacts of childbirth coverage on the log of hourly wages}
    \label{fig:gruber_wages_es_DR}
    \centering
\begin{subfigure}{.45\textwidth}
    \caption{Married men 20-40}
    \centering
    \includegraphics[width=\textwidth]{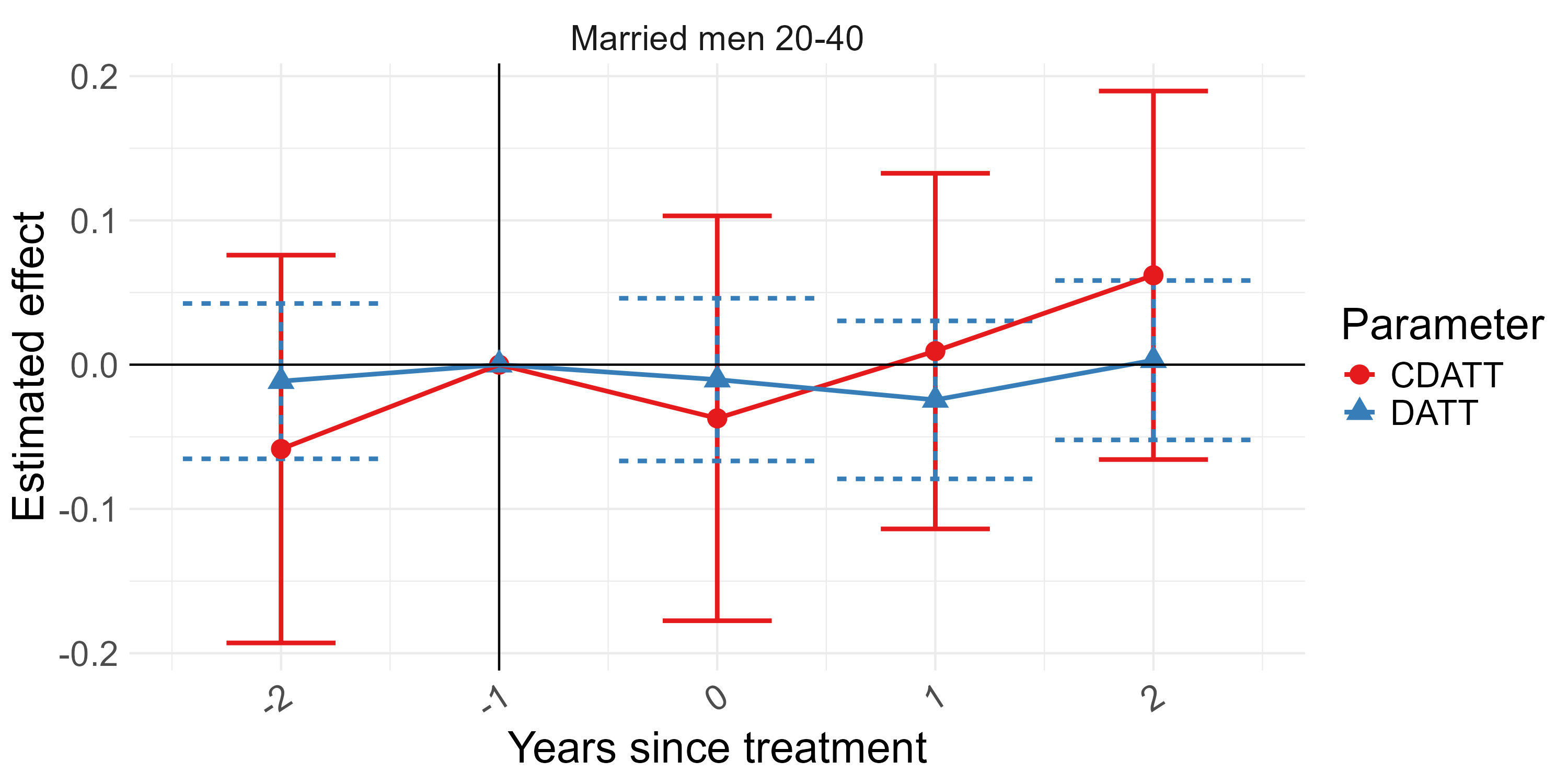}
    \label{fig:gruber_wages_es_DR_3}
\end{subfigure}
\begin{subfigure}{.45\textwidth}
    \caption{Married women 20-40}
    \centering
    \includegraphics[width=\textwidth]{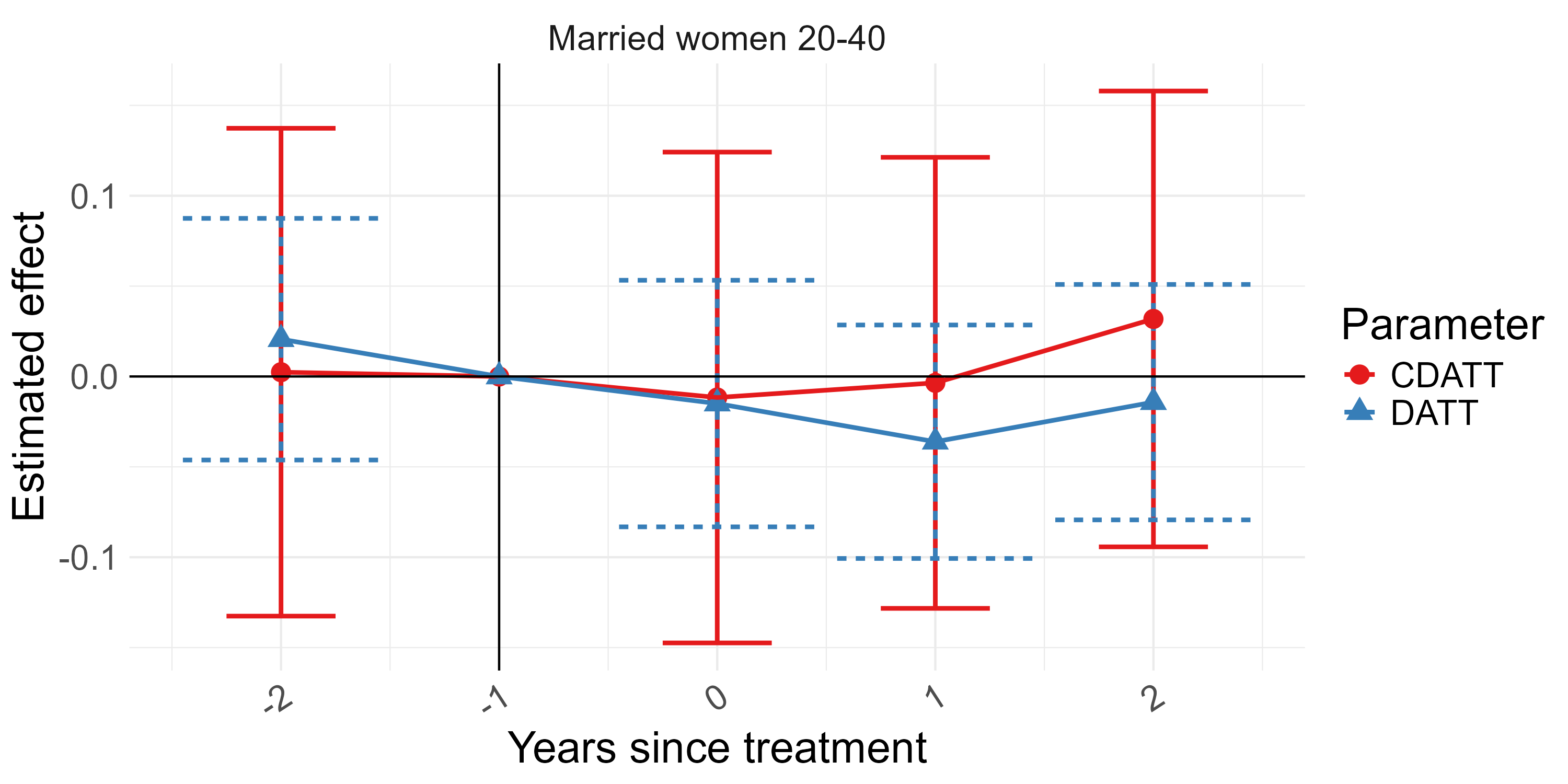}
    \label{fig:gruber_wages_es_DR_1}
\end{subfigure}
\begin{subfigure}{.45\textwidth}
    \caption{Single women 20-40}
    \centering
    \includegraphics[width=\textwidth]{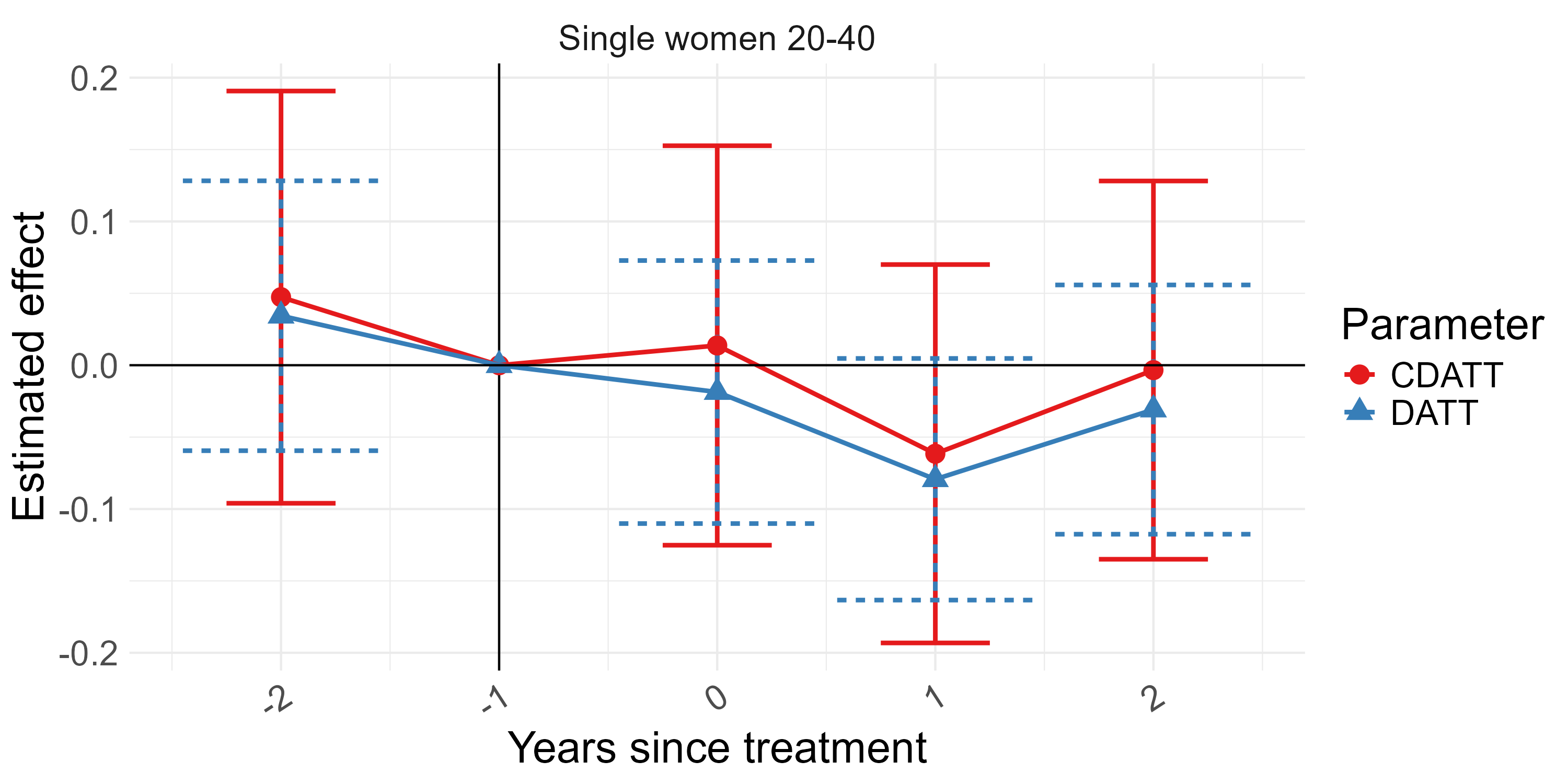}
    \label{fig:gruber_wages_es_DR_2}
\end{subfigure}

{\raggedright \textbf{Notes:} This figure plots estimated effects of childbirth coverage mandates on the log of hourly wages using an event study approach. Points represent event study estimates drawn from the doubly-robust estimators of the DATT and CDATT described in this paper. Effects are plotted for each of the three subgroups of interest (married men 20-40, married women 20-40, and single women 20-40). 95\% confidence intervals are shown. \par }

\end{figure}

On wages, across all three groups and both DATT and CDATT, I find little evidence of a violation of the parallel gaps assumption before the treatment. This can be see in Figure \ref{fig:gruber_wages_es_DR}, which plots doubly-robust estimators of the DATT and CDATT for each subgroup of interest. In all cases, estimated effects at time -2 (that is, two years before the treatment) are close to zero and insignificant.

Meanwhile, the dynamic results on wages highlight that the divergence between the CDATT and the DATT appears in almost every post-period. It is important to note that effects are less precisely estimated when disaggregated by event time. Even so, Figure \ref{fig:gruber_wages_es_DR} shows that the divergence between the two parameters found in the pooled analysis is present across nearly every post-period. At the same time, the plots suggest that any negative effects attenuate over time. For example, studying the DATT, for all three subgroups, the largest negative effects are found in the year following the policy, while these effects are smaller two years following the policy. For comparison, Appendix Figure \ref{fig:gruber_es_3WFE} plots the 3WFE estimates, which (as in the pooled analysis) generally align with the DR estimates of the DATT. 

\begin{figure}[h]
    \caption{Event-study estimates of impacts of childbirth coverage on the log of hours worked}
    \label{fig:gruber_hours_es_DR}
    \centering
\begin{subfigure}{.45\textwidth}
    \caption{Married men 20-40}
    \centering
    \includegraphics[width=\textwidth]{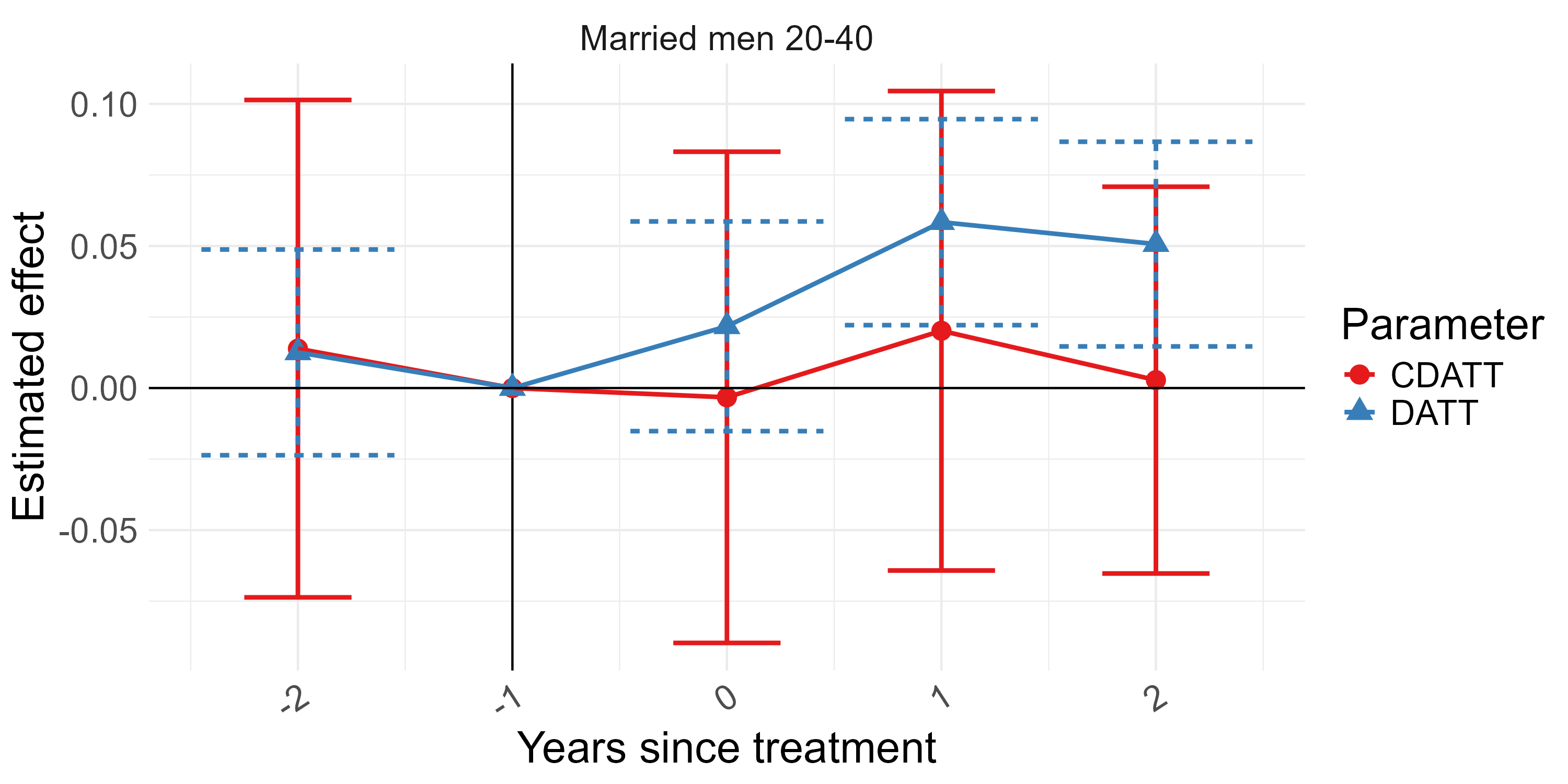}
    \label{fig:gruber_hours_es_DR_3}
\end{subfigure}
\begin{subfigure}{.45\textwidth}
    \caption{Married women 20-40}
    \centering
    \includegraphics[width=\textwidth]{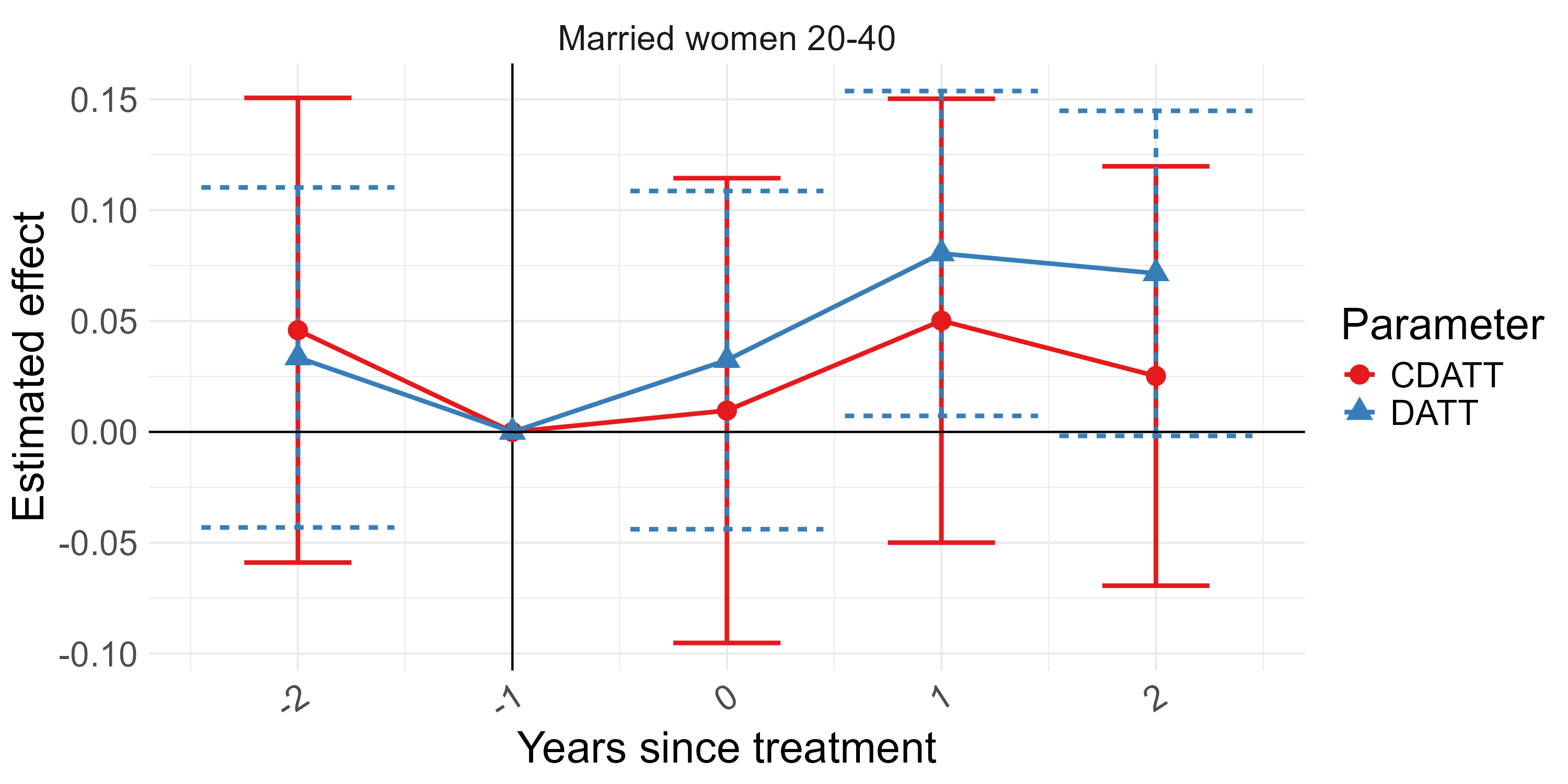}
    \label{fig:gruber_hours_es_DR_1}
\end{subfigure}
\begin{subfigure}{.45\textwidth}
    \caption{Single women 20-40}
    \centering
    \includegraphics[width=\textwidth]{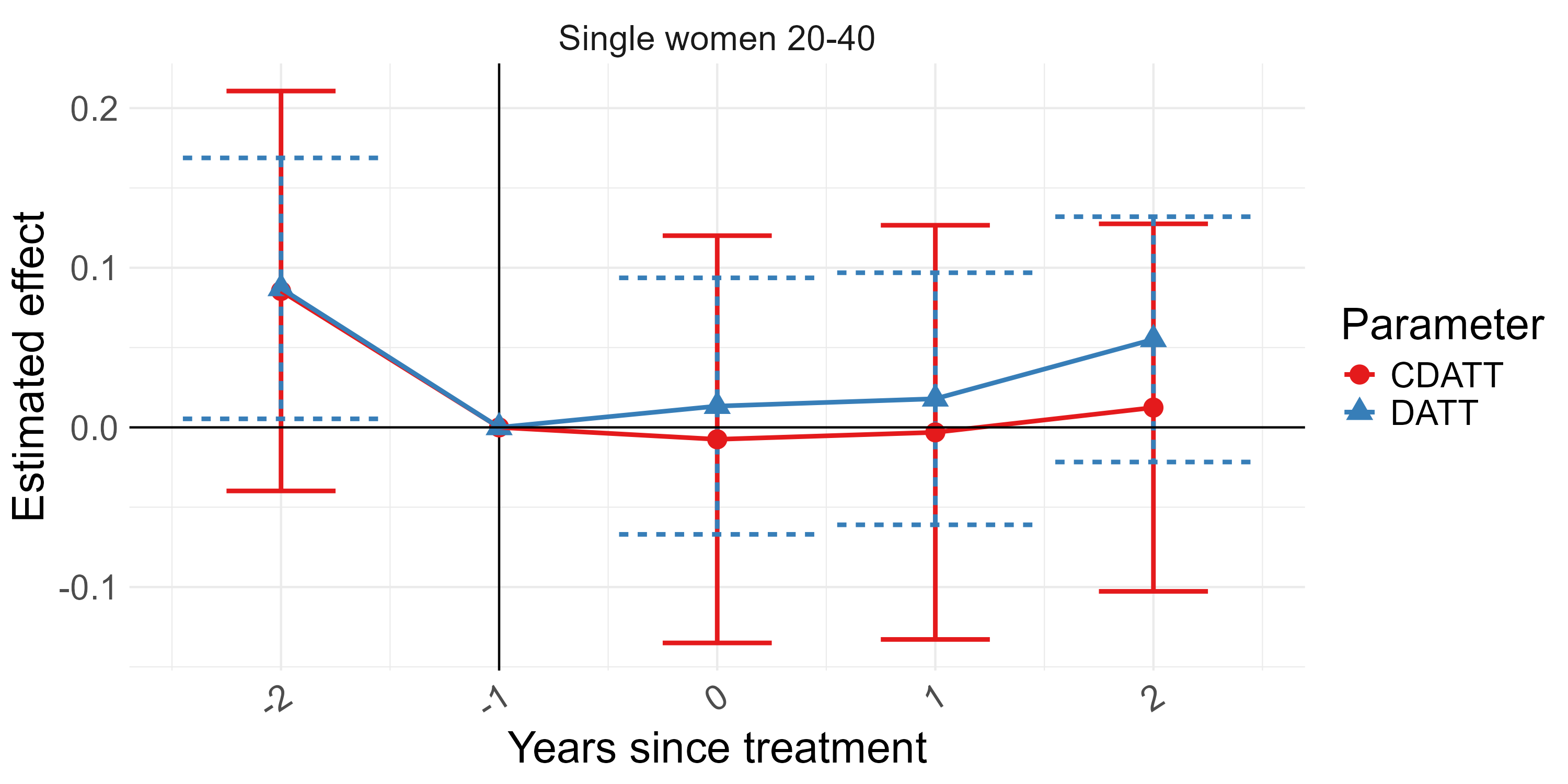}
    \label{fig:gruber_hours_es_DR_2}
\end{subfigure}

{\raggedright \textbf{Notes:} This figure plots estimated effects of childbirth coverage mandates on the log of hours worked using an event study approach. Points represent event study estimates drawn from the doubly-robust estimators of the DATT and CDATT described in this paper. Effects are plotted for each of the three subgroups of interest (married men 20-40, married women 20-40, and single women 20-40). 95\% confidence intervals are shown. \par }
\end{figure}

A similar story arises when studying hours worked. Figure \ref{fig:gruber_hours_es_DR} plots the doubly-robust estimators for the DATT and the CDATT. For married men and married women, there is no evidence of a violation of the parallel gaps assumption in the pre-period, as estimates are small and insignificant. For single women, the estimated pre-period coefficient is large, nearly a 10\% increase in hours, and is significant at the 95\% level for the DATT. This evidence would raise a concern that, especially for single women, there may have been underlying pre-treatment differences driving their outcomes relative to the comparison subgroup, even before the mandate was in effect. As such, we should interpret results about this subgroup with particular caution. 

As in the pooled results, the estimated impacts on hours worked are much smaller when using the CDATT than the DATT. When studying the DATT, both married women and married men of childbearing age experience significant increases in hours worked in years 1 and 2 after the policy change relative to the comparison subgroup. However, the CDATT estimates suggest much smaller causal impacts. This suggests that the increase in hours for these subgroups relative to the comparison subgroup does not persist when holding other characteristics constant, as in the CDATT approach. 

\begin{figure}[h]
    \caption{Event-study estimates of impacts of childbirth coverage on employment}
    \label{fig:gruber_emp_es_DR}
    \centering
\begin{subfigure}{.45\textwidth}
    \caption{Married men 20-40}
    \centering
    \includegraphics[width=\textwidth]{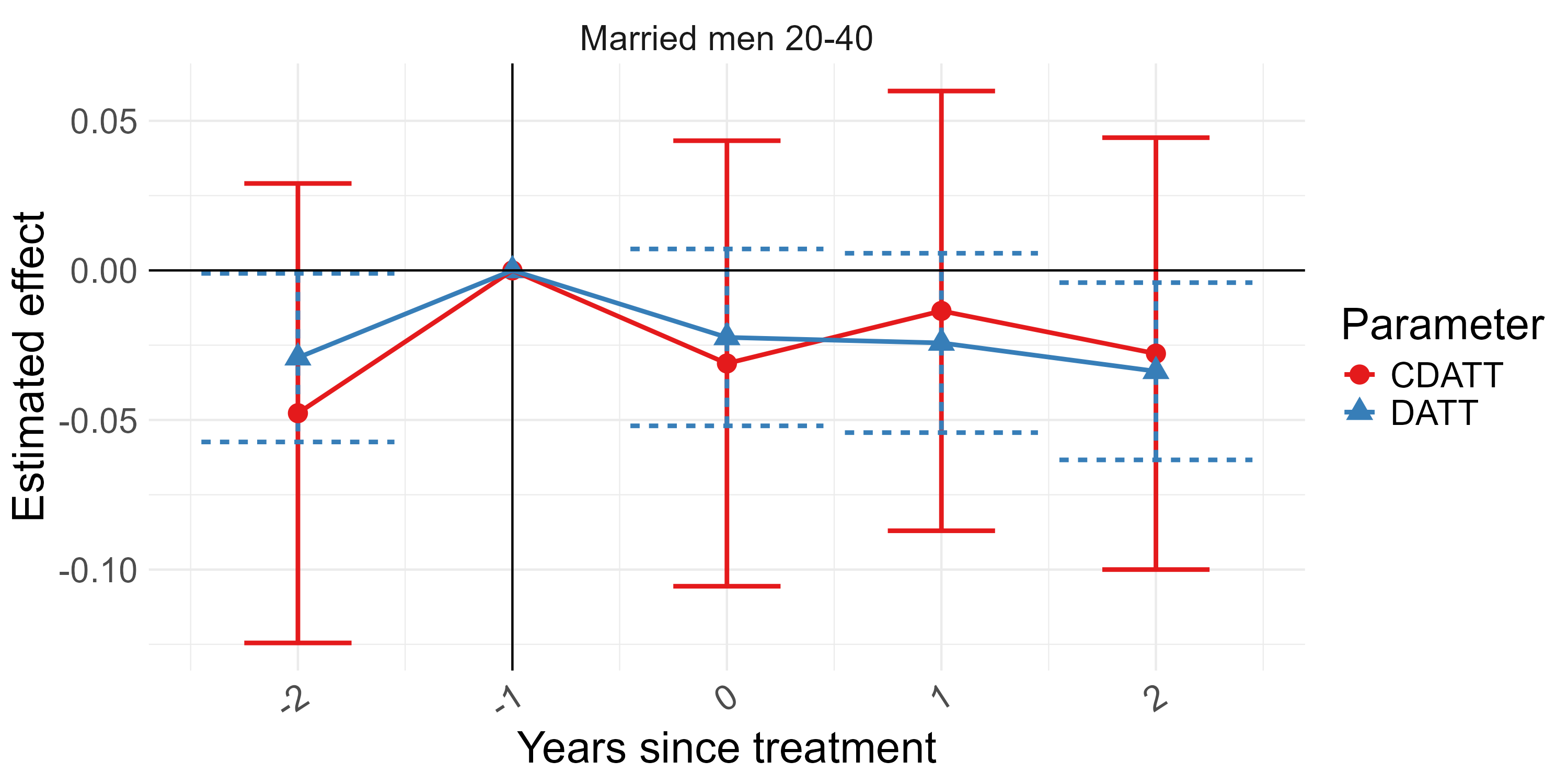}
    \label{fig:gruber_emp_es_DR_3}
\end{subfigure}
\begin{subfigure}{.45\textwidth}
    \caption{Married women 20-40}
    \centering
    \includegraphics[width=\textwidth]{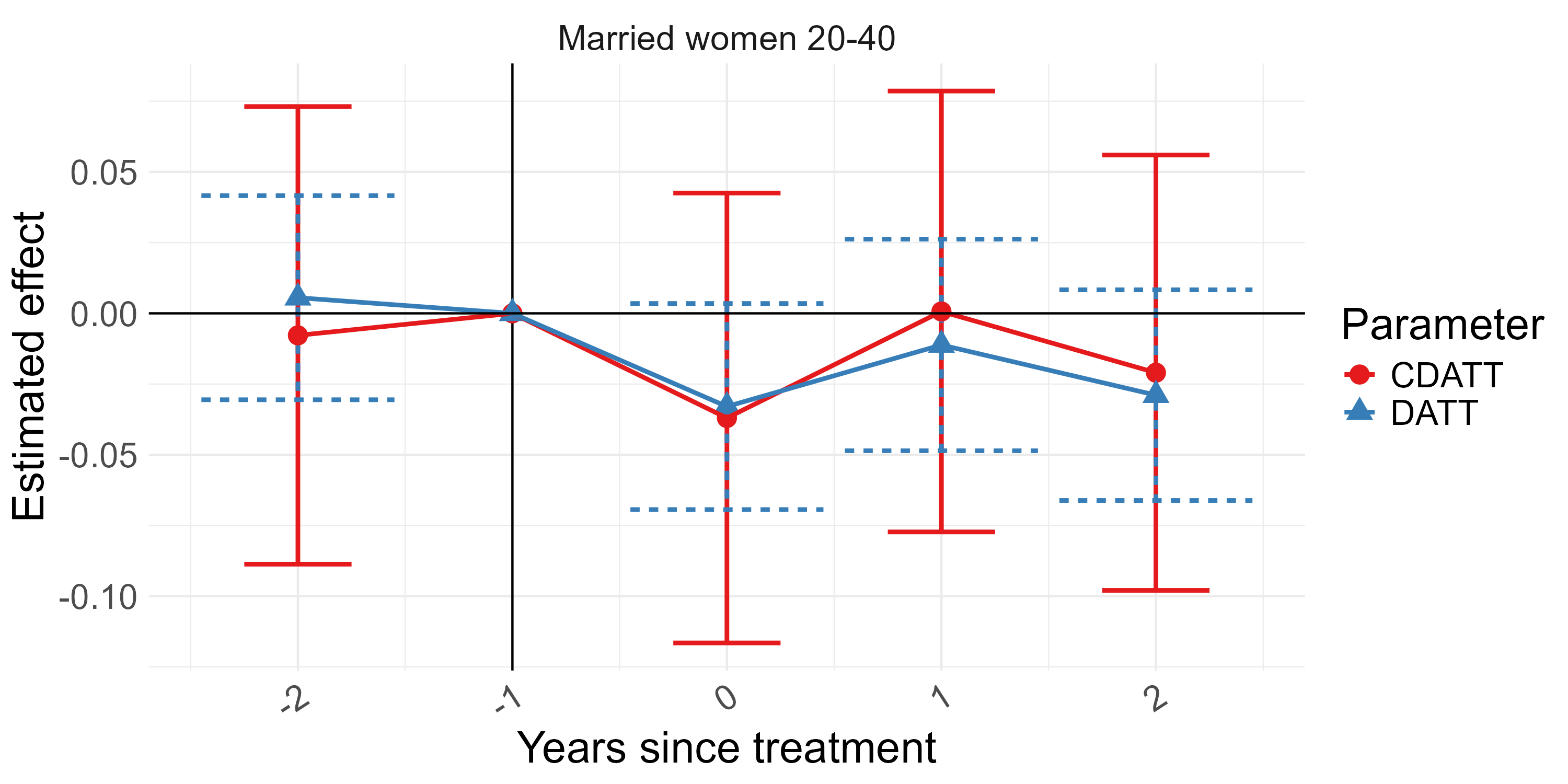}
    \label{fig:gruber_emp_es_DR_1}
\end{subfigure}
\begin{subfigure}{.45\textwidth}
    \caption{Single women 20-40}
    \centering
    \includegraphics[width=\textwidth]{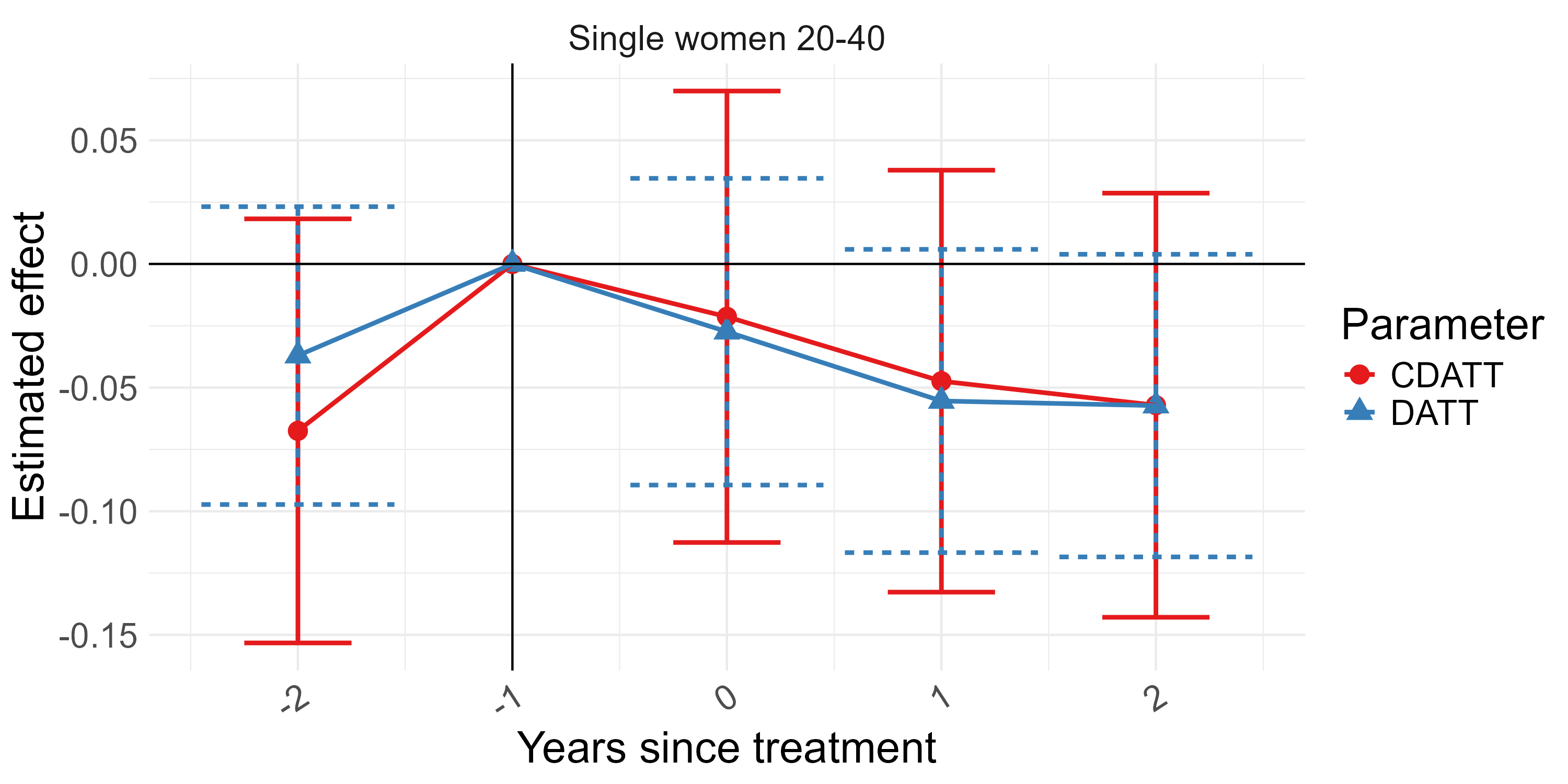}
    \label{fig:gruber_emp_es_DR_2}
\end{subfigure}

{\raggedright \textbf{Notes:} This figure plots estimated effects of childbirth coverage mandates on employment using an event study approach. Points represent event study estimates drawn from the doubly-robust estimators of the DATT and CDATT described in this paper. Effects are plotted for each of the three subgroups of interest (married men 20-40, married women 20-40, and single women 20-40). 95\% confidence intervals are shown. \par }
\end{figure}

On employment, as above, there is little evidence of an overall effect when either using the DATT or the CDATT, and the divergence between the two estimators is smaller. Figure \ref{fig:gruber_emp_es_DR} shows that, for married men of childbearing age and single women of childbearing age, in particular, there is evidence of a potential violation of the parallel gaps assumption for this outcome. Both the DATT and CDATT estimators result in large estimated effects in the pre-period, with magnitudes that are nearly as large as any estimated effects in the post-period. For all subgroups, both DATT and CDATT estimates in the post-period are largely insignificant and negative. That said, in this analysis, we are limited in power to detect even large employment effects.

\FloatBarrier
\section{Conclusion}

In this paper, I have addressed identification and estimation in triple difference designs when treatment effects are heterogeneous. I begin by discussing two parameters of interest, the difference in ATTs between subgroups (DATT) and the controlled difference in ATTs between subgroups (CDATT). When treatment effects are heterogeneous, these two parameters may differ in important ways, and caution is warranted to avoid interpreting a DATT as a CDATT.

I show that the DATT can be identified under an assumption on the trend in the gap between subgroups' outcomes. I then show that identification of the CDATT requires additional assumptions on treatment effect heterogeneity, for example, the assumption that treatment effect heterogeneity is captured by observable characteristics. 

Next, a realistic simulation study calibrated to the application in 
\textcite{gruberIncidenceMandatedMaternity1994} highlights these results. The simulations first highlight the divergence between the DATT and the CDATT, and show that estimators for the DATT can provide misleading results if they are interpreted causally in cases where subgroups differ in their covariates correlated with their treatment effects. They also show that the proposed estimators for the CDATT are unbiased and perform well in finite samples. 

An application of these estimators to the analysis in \textcite{gruberIncidenceMandatedMaternity1994} highlights a case where heterogeneity in sensitivity to a treatment can correlate with subgroup status, affecting the interpretation of the triple difference estimates. The analysis addresses the question of whether employers can pass on group-specific costs on the basis of demographic characteristics (that is, due to their different likelihoods of using the benefits). The paper uses a triple difference design to study the impacts of state-level mandates requiring insurance to cover costs for childbirth. Since these benefits are likely to be used by certain subgroups of the population (ie, women age 20-40 and married men age 20-40), the triple difference design allows to test for differential effects for these subgroups relative to other segments of the population. Although an analysis of the DATT suggests that there is significant differential cost-shifting for these subgroups, estimating the CDATT offers little evidence that this difference is caused by their different likelihoods to use the benefits (as proxied by their demographic groups). Researchers should take treatment effect heterogeneity into account in the interpretation and estimation of treatment effects in triple difference designs.

\section*{Acknowledgments}
The author acknowledges the support of the National Science Foundation Graduate Research Fellowship Program. I thank the editor %, Bo Honoré, 
and two anonymous referees for their many helpful suggestions. I am grateful to Alessandra Casella, Simon Lee, and Ebonya Washington, as well as Bitsy Perlman and Erwin Tiongson, for their excellent feedback and support. I also thank Alyssa Bilinski, Clément de Chaisemartin, Peng Ding, Serena Ng, Jonathan Roth, Bernard Salanié, Pedro Sant'Anna, Jesse Shapiro, and Anton Strezhnev for their encouraging comments and many helpful suggestions for improvement. I acknowledge attendees of the Applied Micro Methods and Econometrics colloquia at Columbia University in March 2023, October 2023, April 2024, and November 2024 for their comments and questions. All remaining errors are my own.

\printbibliography

@ARTICLE{abadieSemiparametricDifferenceinDifferencesEstimators2005,
  AUTHOR = {Abadie, Alberto},
  DATE = {2005-01},
  DOI = {10.1111/0034-6527.00321},
  ISSN = {0034-6527},
  JOURNALTITLE = {The Review of Economic Studies},
  NUMBER = {1},
  PAGES = {1--19},
  TITLE = {Semiparametric {{Difference-in-Differences Estimators}}},
  URLDATE = {2023-03-19},
  VOLUME = {72},
}

@ARTICLE{acharyaExplainingCausalFindings2016,
  AUTHOR = {Acharya, Avidit and Blackwell, Matthew and Sen, Maya},
  DATE = {2016-08},
  DOI = {10.1017/S0003055416000216},
  ISSN = {0003-0554, 1537-5943},
  JOURNALTITLE = {American Political Science Review},
  LANGID = {english},
  NUMBER = {3},
  PAGES = {512--529},
  SHORTTITLE = {Explaining {{Causal Findings Without Bias}}},
  TITLE = {Explaining {{Causal Findings Without Bias}}: {{Detecting}} and {{Assessing Direct Effects}}},
  URLDATE = {2026-08-13},
  VOLUME = {110},
}

@ARTICLE{baumEffectStateMaternity2003,
  AUTHOR = {Baum, Charles L.},
  DATE = {2003-10},
  DOI = {10.1016/S0927-5371(03)00037-X},
  ISSN = {0927-5371},
  JOURNALTITLE = {Labour Economics},
  NUMBER = {5},
  PAGES = {573--596},
  TITLE = {The Effect of State Maternity Leave Legislation and the 1993 {{Family}} and {{Medical Leave Act}} on Employment and Wages},
  URLDATE = {2024-09-30},
  VOLUME = {10},
}

@ARTICLE{blundellAlternativeApproachesEvaluation2009,
  AUTHOR = {Blundell, Richard and Dias, Monica Costa},
  PUBLISHER = {University of Wisconsin Press},
  CHAPTER = {Articles},
  DATE = {2009-07},
  DOI = {10.3368/jhr.44.3.565},
  ISSN = {0022-166X, 1548-8004},
  JOURNALTITLE = {Journal of Human Resources},
  LANGID = {english},
  NUMBER = {3},
  PAGES = {565--640},
  TITLE = {Alternative {{Approaches}} to {{Evaluation}} in {{Empirical Microeconomics}}},
  URLDATE = {2025-04-05},
  VOLUME = {44},
}

@ARTICLE{borusyakRevisitingEventStudyDesigns2024,
  AUTHOR = {Borusyak, Kirill and Jaravel, Xavier and Spiess, Jann},
  DATE = {2024-11},
  DOI = {10.1093/restud/rdae007},
  ISSN = {0034-6527},
  JOURNALTITLE = {The Review of Economic Studies},
  NUMBER = {6},
  PAGES = {3253--3285},
  SHORTTITLE = {Revisiting {{Event-Study Designs}}},
  TITLE = {Revisiting {{Event-Study Designs}}: {{Robust}} and {{Efficient Estimation}}},
  URLDATE = {2026-08-12},
  VOLUME = {91},
}

@MISC{caetanoDifferenceDifferencesTimeVarying2024,
  AUTHOR = {Caetano, Carolina and Callaway, Brantly and Payne, Stroud and Rodrigues, Hugo Sant'Anna},
  PUBLISHER = {arXiv},
  DATE = {2024-06},
  DOI = {10.48550/arXiv.2202.02903},
  EPRINTCLASS = {econ},
  EPRINTTYPE = {arXiv},
  NUMBER = {arXiv:2202.02903},
  TITLE = {Difference in {{Differences}} with {{Time-Varying Covariates}}},
  URLDATE = {2025-01-06},
}

@ARTICLE{callawayDifferenceinDifferencesContinuousTreatment,
  AUTHOR = {Callaway, Brantly and {Goodman-Bacon}, Andrew and Sant'Anna, Pedro H. C.},
  DATE = {forthcoming},
  DOI = {10.1257/aer.20240137},
  ISSN = {0002-8282},
  JOURNALTITLE = {American Economic Review},
  LANGID = {english},
  TITLE = {Difference-in-{{Differences}} with a {{Continuous Treatment}}},
  URLDATE = {2026-09-06},
}

@ARTICLE{callawayDifferenceinDifferencesMultipleTime2021,
  AUTHOR = {Callaway, Brantly and Sant'Anna, Pedro H. C.},
  DATE = {2021-12},
  DOI = {10.1016/j.jeconom.2020.12.001},
  ISSN = {0304-4076},
  JOURNALTITLE = {Journal of Econometrics},
  LANGID = {english},
  NUMBER = {2},
  PAGES = {200--230},
  SERIES = {Themed {{Issue}}: {{Treatment Effect}} 1},
  TITLE = {Difference-in-{{Differences}} with Multiple Time Periods},
  URLDATE = {2023-03-19},
  VOLUME = {225},
}

@MISC{censusCurrentPopulationSurvey1992,
  AUTHOR = {of the Census, United States Bureau},
  PUBLISHER = {Inter-university Consortium for Political and Social Research},
  DATE = {1992-02},
  DOI = {10.3886/ICPSR07939.v1},
  LANGID = {english},
  TITLE = {Current {{Population Survey}}, {{May}} 1976},
  URLDATE = {2026-08-10},
}

@MISC{chenEfficientDifferenceinDifferencesEvent2025,
  AUTHOR = {Chen, Xiaohong and Sant'Anna, Pedro H. C. and Xie, Haitian},
  PUBLISHER = {arXiv},
  DATE = {2025},
  DOI = {10.48550/ARXIV.2506.17729},
  LANGID = {english},
  TITLE = {Efficient {{Difference-in-Differences}} and {{Event Study Estimators}}},
  URLDATE = {2026-09-04},
}

@MISC{chernozhukovFisherSchultzLectureGeneric2023,
  AUTHOR = {Chernozhukov, Victor and Demirer, Mert and Duflo, Esther and {Fernández-Val}, Iván},
  PUBLISHER = {arXiv},
  DATE = {2023-10},
  DOI = {10.48550/arXiv.1712.04802},
  EPRINTCLASS = {stat.ML},
  EPRINTTYPE = {arXiv},
  LANGID = {english},
  NUMBER = {arXiv:1712.04802},
  SHORTTITLE = {Fisher-{{Schultz Lecture}}},
  TITLE = {Fisher-{{Schultz Lecture}}: {{Generic Machine Learning Inference}} on {{Heterogenous Treatment Effects}} in {{Randomized Experiments}}, with an {{Application}} to {{Immunization}} in {{India}}},
  URLDATE = {2026-08-12},
}

@ARTICLE{crumpDealingLimitedOverlap2009a,
  AUTHOR = {Crump, Richard K. and Hotz, V. Joseph and Imbens, Guido W. and Mitnik, Oscar A.},
  DATE = {2009-03},
  DOI = {10.1093/biomet/asn055},
  ISSN = {0006-3444},
  JOURNALTITLE = {Biometrika},
  NUMBER = {1},
  PAGES = {187--199},
  TITLE = {Dealing with Limited Overlap in Estimation of Average Treatment Effects},
  URLDATE = {2026-08-16},
  VOLUME = {96},
}

@ARTICLE{dechaisemartinTwoWayFixedEffects2020,
  AUTHOR = {{de Chaisemartin}, Clément and D'Haultf{œ}uille, Xavier},
  DATE = {2020-09},
  DOI = {10.1257/aer.20181169},
  ISSN = {0002-8282},
  JOURNALTITLE = {American Economic Review},
  LANGID = {english},
  NUMBER = {9},
  PAGES = {2964--2996},
  TITLE = {Two-{{Way Fixed Effects Estimators}} with {{Heterogeneous Treatment Effects}}},
  URLDATE = {2022-04-23},
  VOLUME = {110},
}

@MISC{debFlexibleHeterogeneousTreatment2024,
  AUTHOR = {Deb, Partha and Norton, Edward C. and Wooldridge, Jeffrey M. and Zabel, Jeffrey E.},
  PUBLISHER = {National Bureau of Economic Research},
  DATE = {2024-10},
  DOI = {10.3386/w33026},
  EPRINTTYPE = {National Bureau of Economic Research},
  NUMBER = {33026},
  SERIES = {Working {{Paper Series}}},
  TITLE = {A {{Flexible}}, {{Heterogeneous Treatment Effects Difference-in-Differences Estimator}} for {{Repeated Cross-Sections}}},
  TYPE = {Working {{Paper}}},
  URLDATE = {2026-08-12},
}

@ARTICLE{derenoncourtMinimumWagesRacial2020,
  AUTHOR = {Derenoncourt, Ellora and Montialoux, Claire},
  DATE = {2020-12},
  DOI = {10.1093/qje/qjaa031},
  ISSN = {0033-5533, 1531-4650},
  JOURNALTITLE = {The Quarterly Journal of Economics},
  LANGID = {english},
  NUMBER = {1},
  PAGES = {169--228},
  TITLE = {Minimum {{Wages}} and {{Racial Inequality}}},
  URLDATE = {2024-08-25},
  VOLUME = {136},
}

@ARTICLE{desiereHowEffectiveAre2022,
  AUTHOR = {Desiere, Sam and Cockx, Bart},
  DATE = {2022-05},
  DOI = {10.2478/izajolp-2022-0003},
  JOURNALTITLE = {IZA Journal of Labor Policy},
  LANGID = {english},
  NUMBER = {1},
  SHORTTITLE = {How Effective Are Hiring Subsidies in Reducing Long-Term Unemployment among Prime-Aged Jobseekers?},
  TITLE = {How Effective Are Hiring Subsidies in Reducing Long-Term Unemployment among Prime-Aged Jobseekers? {{Evidence}} from {{Belgium}}},
  URLDATE = {2025-04-04},
  VOLUME = {12},
}

@MISC{dornHowMuchWeak2025,
  AUTHOR = {Dorn, Jacob},
  PUBLISHER = {arXiv},
  DATE = {2025-04},
  DOI = {10.48550/arXiv.2504.13273},
  EPRINTCLASS = {econ.EM},
  EPRINTTYPE = {arXiv},
  NUMBER = {arXiv:2504.13273},
  TITLE = {How {{Much Weak Overlap Can Doubly Robust T-Statistics Handle}}?},
  URLDATE = {2026-08-27},
}

@MISC{goldsmith-pinkhamEmpiricalMethodsTrends2026,
  AUTHOR = {{Goldsmith-Pinkham}, Paul},
  DATE = {2026},
  HOWPUBLISHED = {https://paulgp.com/econlit-pipeline/dashboard.html},
  TITLE = {Empirical {{Methods Trends}} - {{Economics Literature}}},
  URLDATE = {2026-09-04},
}

@REPORT{goldsmith-pinkhamTrackingCredibilityRevolution2026,
  AUTHOR = {{Goldsmith-Pinkham}, Paul},
  DATE = {2026},
  LANGID = {english},
  TITLE = {Tracking the {{Credibility Revolution}} across {{Fields}}},
  TYPE = {techreport},
}

@ARTICLE{goodman-baconDifferenceindifferencesVariationTreatment2021,
  AUTHOR = {{Goodman-Bacon}, Andrew},
  DATE = {2021-12},
  DOI = {10.1016/j.jeconom.2021.03.014},
  ISSN = {0304-4076},
  JOURNALTITLE = {Journal of Econometrics},
  NUMBER = {2},
  PAGES = {254--277},
  SERIES = {Themed {{Issue}}: {{Treatment Effect}} 1},
  TITLE = {Difference-in-Differences with Variation in Treatment Timing},
  URLDATE = {2024-09-05},
  VOLUME = {225},
}

@ARTICLE{gruberIncidenceMandatedMaternity1994,
  AUTHOR = {Gruber, Jonathan},
  PUBLISHER = {American Economic Association},
  DATE = {1994},
  EPRINT = {2118071},
  EPRINTTYPE = {jstor},
  ISSN = {0002-8282},
  JOURNALTITLE = {The American Economic Review},
  NUMBER = {3},
  PAGES = {622--641},
  TITLE = {The {{Incidence}} of {{Mandated Maternity Benefits}}},
  URLDATE = {2024-10-14},
  VOLUME = {84},
}

@ARTICLE{heilerValidInferenceTreatment2021,
  AUTHOR = {Heiler, Phillip and Kazak, Ekaterina},
  DATE = {2021-06},
  DOI = {10.1016/j.jeconom.2020.03.025},
  ISSN = {03044076},
  JOURNALTITLE = {Journal of Econometrics},
  LANGID = {english},
  NUMBER = {2},
  PAGES = {1083--1108},
  TITLE = {Valid Inference for Treatment Effect Parameters under Irregular Identification and Many Extreme Propensity Scores},
  URLDATE = {2026-08-27},
  VOLUME = {222},
}

@ARTICLE{hollandStatisticsCausalInference1986,
  AUTHOR = {Holland, Paul W.},
  PUBLISHER = {Taylor \& Francis},
  DATE = {1986-12},
  DOI = {10.1080/01621459.1986.10478354},
  ISSN = {0162-1459},
  JOURNALTITLE = {Journal of the American Statistical Association},
  NUMBER = {396},
  PAGES = {945--960},
  TITLE = {Statistics and {{Causal Inference}}},
  URLDATE = {2026-08-09},
  VOLUME = {81},
}

@ARTICLE{imaiStatisticalInferenceHeterogeneous2025,
  AUTHOR = {Imai, Kosuke and Li, Michael Lingzhi},
  PUBLISHER = {Taylor \& Francis},
  DATE = {2025-01},
  DOI = {10.1080/07350015.2024.2358909},
  ISSN = {0735-0015},
  JOURNALTITLE = {Journal of Business \& Economic Statistics},
  NUMBER = {1},
  PAGES = {256--268},
  TITLE = {Statistical {{Inference}} for {{Heterogeneous Treatment Effects Discovered}} by {{Generic Machine Learning}} in {{Randomized Experiments}}},
  URLDATE = {2026-08-12},
  VOLUME = {43},
}

@BOOK{imbensCausalInferenceStatistics2015,
  AUTHOR = {Imbens, Guido W. and Rubin, Donald B.},
  LOCATION = {Cambridge},
  PUBLISHER = {Cambridge University Press},
  DATE = {2015},
  ISBN = {978-0-521-88588-1},
  SHORTTITLE = {Causal {{Inference}} for {{Statistics}}, {{Social}}, and {{Biomedical Sciences}}},
  TITLE = {Causal {{Inference}} for {{Statistics}}, {{Social}}, and {{Biomedical Sciences}}: {{An Introduction}}},
  URLDATE = {2023-05-02},
}

@ARTICLE{kangDemystifyingDoubleRobustness2007,
  AUTHOR = {Kang, Joseph D. Y. and Schafer, Joseph L.},
  PUBLISHER = {Institute of Mathematical Statistics},
  DATE = {2007-11},
  DOI = {10.1214/07-STS227},
  ISSN = {0883-4237, 2168-8745},
  JOURNALTITLE = {Statistical Science},
  LANGID = {english},
  NUMBER = {4},
  PAGES = {523--539},
  SHORTTITLE = {Demystifying {{Double Robustness}}},
  TITLE = {Demystifying {{Double Robustness}}: {{A Comparison}} of {{Alternative Strategies}} for {{Estimating}} a {{Population Mean}} from {{Incomplete Data}}},
  URLDATE = {2026-08-16},
  VOLUME = {22},
}

@MISC{khanDoublyrobustHeteroscedasticityawareSample2024,
  AUTHOR = {Khan, Samir and Ugander, Johan},
  PUBLISHER = {arXiv},
  DATE = {2024-01},
  DOI = {10.48550/arXiv.2210.10171},
  EPRINTCLASS = {stat.ME},
  EPRINTTYPE = {arXiv},
  NUMBER = {arXiv:2210.10171},
  TITLE = {Doubly-Robust and Heteroscedasticity-Aware Sample Trimming for Causal Inference},
  URLDATE = {2026-08-27},
}

@ARTICLE{khanIrregularIdentificationSupport2010,
  AUTHOR = {Khan, Shakeeb and Tamer, Elie},
  DATE = {2010},
  DOI = {10.3982/ECTA7372},
  ISSN = {1468-0262},
  JOURNALTITLE = {Econometrica},
  LANGID = {english},
  NUMBER = {6},
  PAGES = {2021--2042},
  TITLE = {Irregular {{Identification}}, {{Support Conditions}}, and {{Inverse Weight Estimation}}},
  URLDATE = {2026-08-16},
  VOLUME = {78},
}

@ARTICLE{maRobustInferenceUsing2020,
  AUTHOR = {Ma, Xinwei and Wang, Jingshen},
  DATE = {2020-10},
  DOI = {10.1080/01621459.2019.1660173},
  ISSN = {0162-1459, 1537-274X},
  JOURNALTITLE = {Journal of the American Statistical Association},
  LANGID = {english},
  NUMBER = {532},
  PAGES = {1851--1860},
  TITLE = {Robust {{Inference Using Inverse Probability Weighting}}},
  URLDATE = {2026-08-27},
  VOLUME = {115},
}

@MISC{maDoublyRobustEstimators2026,
  AUTHOR = {Ma, Yukun and Sant'Anna, Pedro H C and Sasaki, Yuya and Ura, Takuya},
  DATE = {2026},
  LANGID = {english},
  TITLE = {Doubly {{Robust Estimators}} with {{Weak Overlap}}},
}

@ARTICLE{manskiMonotoneInstrumentalVariables2000a,
  AUTHOR = {Manski, Charles F. and Pepper, John V.},
  DATE = {2000},
  DOI = {10.1111/1468-0262.t01-1-00144a},
  ISSN = {1468-0262},
  JOURNALTITLE = {Econometrica},
  LANGID = {english},
  NUMBER = {4},
  PAGES = {997--1010},
  SHORTTITLE = {Monotone {{Instrumental Variables}}},
  TITLE = {Monotone {{Instrumental Variables}}: {{With}} an {{Application}} to the {{Returns}} to {{Schooling}}},
  URLDATE = {2022-03-19},
  VOLUME = {68},
}

@ARTICLE{matsaFemaleStyleCorporate2013b,
  AUTHOR = {Matsa, David A. and Miller, Amalia R.},
  DATE = {2013-07},
  DOI = {10.1257/app.5.3.136},
  ISSN = {1945-7782},
  JOURNALTITLE = {American Economic Journal: Applied Economics},
  LANGID = {english},
  NUMBER = {3},
  PAGES = {136--169},
  SHORTTITLE = {A {{Female Style}} in {{Corporate Leadership}}?},
  TITLE = {A {{Female Style}} in {{Corporate Leadership}}? {{Evidence}} from {{Quotas}}},
  URLDATE = {2025-04-04},
  VOLUME = {5},
}

@ARTICLE{oldenTripleDifferenceEstimator2022,
  AUTHOR = {Olden, Andreas and M{ø}en, Jarle},
  DATE = {2022-09},
  DOI = {10.1093/ectj/utac010},
  ISSN = {1368-4221},
  JOURNALTITLE = {The Econometrics Journal},
  NUMBER = {3},
  PAGES = {531--553},
  TITLE = {The Triple Difference Estimator},
  URLDATE = {2023-02-08},
  VOLUME = {25},
}

@MISC{ortiz-villavicencioBetterUnderstandingTriple2025,
  AUTHOR = {{Ortiz-Villavicencio}, Marcelo and Sant'Anna, Pedro H. C.},
  PUBLISHER = {arXiv},
  DATE = {2025-05},
  DOI = {10.48550/arXiv.2505.09942},
  EPRINTCLASS = {econ},
  EPRINTTYPE = {arXiv},
  NUMBER = {arXiv:2505.09942},
  TITLE = {Better {{Understanding Triple Differences Estimators}}},
  URLDATE = {2025-05-24},
}

@INPROCEEDINGS{pearlDirectIndirectEffects2001,
  AUTHOR = {Pearl, Judea},
  EDITOR = {Geffner, Hector and Dechter, Rina and Halpern, Joseph Y.},
  PUBLISHER = {ACM},
  BOOKTITLE = {{{UAI}}'01: {{Proceedings}} of the {{Seventeenth}} Conference on {{Uncertainty}} in Artificial Intellige},
  DATE = {2001},
  DOI = {10.1145/3501714.3501736},
  ISBN = {978-1-4503-9586-1},
  LANGID = {english},
  TITLE = {Direct and {{Indirect Effects}}},
  URLDATE = {2026-08-13},
}

@ARTICLE{robinsIdentifiabilityExchangeabilityDirect1992,
  AUTHOR = {Robins, James M. and Greenland, Sander},
  DATE = {1992},
  JOURNALTITLE = {Epidemiology},
  NUMBER = {2},
  PAGES = {143--155},
  TITLE = {Identifiability and Exchangeability for Direct and Indirect Effects},
  URLDATE = {2026-08-14},
  VOLUME = {3},
}

@ARTICLE{rothEfficientEstimationStaggered2023,
  AUTHOR = {Roth, Jonathan and Sant'Anna, Pedro H. C.},
  PUBLISHER = {The University of Chicago Press},
  DATE = {2023-11},
  DOI = {10.1086/726581},
  ISSN = {2832-9368},
  JOURNALTITLE = {Journal of Political Economy Microeconomics},
  NUMBER = {4},
  PAGES = {669--709},
  TITLE = {Efficient {{Estimation}} for {{Staggered Rollout Designs}}},
  URLDATE = {2024-12-03},
  VOLUME = {1},
}

@ARTICLE{rothWhatsTrendingDifferenceindifferences2023,
  AUTHOR = {Roth, Jonathan and Sant'Anna, Pedro H. C. and Bilinski, Alyssa and Poe, John},
  DATE = {2023-08},
  DOI = {10.1016/j.jeconom.2023.03.008},
  ISSN = {0304-4076},
  JOURNALTITLE = {Journal of Econometrics},
  NUMBER = {2},
  PAGES = {2218--2244},
  SHORTTITLE = {What's Trending in Difference-in-Differences?},
  TITLE = {What's Trending in Difference-in-Differences? {{A}} Synthesis of the Recent Econometrics Literature},
  URLDATE = {2025-01-06},
  VOLUME = {235},
}

@ARTICLE{santannaDoublyRobustDifferenceindifferences2020,
  AUTHOR = {Sant'Anna, Pedro H. C. and Zhao, Jun},
  DATE = {2020-11},
  DOI = {10.1016/j.jeconom.2020.06.003},
  ISSN = {0304-4076},
  JOURNALTITLE = {Journal of Econometrics},
  NUMBER = {1},
  PAGES = {101--122},
  TITLE = {Doubly Robust Difference-in-Differences Estimators},
  URLDATE = {2024-07-09},
  VOLUME = {219},
}

@ARTICLE{senRaceBundleSticks2016,
  AUTHOR = {Sen, Maya and Wasow, Omar},
  DATE = {2016-05},
  DOI = {10.1146/annurev-polisci-032015-010015},
  ISSN = {1094-2939, 1545-1577},
  JOURNALTITLE = {Annual Review of Political Science},
  LANGID = {english},
  NUMBER = {1},
  PAGES = {499--522},
  SHORTTITLE = {Race as a {{Bundle}} of {{Sticks}}},
  TITLE = {Race as a {{Bundle}} of {{Sticks}}: {{Designs}} That {{Estimate Effects}} of {{Seemingly Immutable Characteristics}}},
  URLDATE = {2026-08-09},
  VOLUME = {19},
}

@MISC{strezhnevDecomposingTripleDifferencesRegression2023,
  AUTHOR = {Strezhnev, Anton},
  PUBLISHER = {arXiv},
  DATE = {2023-07},
  EPRINT = {2307.02735},
  EPRINTCLASS = {stat},
  EPRINTTYPE = {arXiv},
  LANGID = {english},
  NUMBER = {arXiv:2307.02735},
  TITLE = {Decomposing {{Triple-Differences Regression}} under {{Staggered Adoption}}},
  URLDATE = {2023-10-08},
}

@ARTICLE{sunEstimatingDynamicTreatment2021,
  AUTHOR = {Sun, Liyang and Abraham, Sarah},
  DATE = {2021-12},
  DOI = {10.1016/j.jeconom.2020.09.006},
  ISSN = {0304-4076},
  JOURNALTITLE = {Journal of Econometrics},
  LANGID = {english},
  NUMBER = {2},
  PAGES = {175--199},
  SERIES = {Themed {{Issue}}: {{Treatment Effect}} 1},
  TITLE = {Estimating Dynamic Treatment Effects in Event Studies with Heterogeneous Treatment Effects},
  URLDATE = {2023-03-19},
  VOLUME = {225},
}

@MISC{usbureauofthecensusCurrentPopulationSurvey1992a,
  AUTHOR = {{US Bureau of the Census}},
  PUBLISHER = {ICPSR - Interuniversity Consortium for Political and Social Research},
  DATE = {1992},
  DOI = {10.3886/ICPSR07937.V1},
  LANGID = {english},
  SHORTTITLE = {Current {{Population Survey}}, {{May}} 1974},
  TITLE = {Current {{Population Survey}}, {{May}} 1974: {{Version}} 1},
  URLDATE = {2025-04-05},
}

@MISC{usbureauofthecensusCurrentPopulationSurvey1992b,
  AUTHOR = {{US Bureau of the Census}},
  PUBLISHER = {ICPSR - Interuniversity Consortium for Political and Social Research},
  DATE = {1992},
  DOI = {10.3886/ICPSR07938.V1},
  LANGID = {english},
  SHORTTITLE = {Current {{Population Survey}}, {{May}} 1975},
  TITLE = {Current {{Population Survey}}, {{May}} 1975: {{Version}} 1},
  URLDATE = {2025-04-05},
}

@MISC{usbureauofthecensusCurrentPopulationSurvey1992c,
  AUTHOR = {{US Bureau of the Census}},
  PUBLISHER = {ICPSR - Interuniversity Consortium for Political and Social Research},
  DATE = {1992},
  DOI = {10.3886/ICPSR07967.V1},
  LANGID = {english},
  SHORTTITLE = {Current {{Population Survey}}, {{May}} 1977},
  TITLE = {Current {{Population Survey}}, {{May}} 1977: {{Version}} 1},
  URLDATE = {2025-04-05},
}

@MISC{usbureauofthecensusCurrentPopulationSurvey1992,
  AUTHOR = {{US Bureau of the Census}},
  PUBLISHER = {ICPSR - Interuniversity Consortium for Political and Social Research},
  DATE = {1992},
  DOI = {10.3886/ICPSR07783.V1},
  LANGID = {english},
  SHORTTITLE = {Current {{Population Survey}}, {{May}} 1978},
  TITLE = {Current {{Population Survey}}, {{May}} 1978: {{Version}} 1},
  URLDATE = {2025-04-05},
}

@ARTICLE{vanderweeleUnificationMediationInteraction2014,
  AUTHOR = {VanderWeele, Tyler J.},
  DATE = {2014-09},
  DOI = {10.1097/EDE.0000000000000121},
  ISSN = {1044-3983},
  JOURNALTITLE = {Epidemiology (Cambridge, Mass.)},
  NUMBER = {5},
  PAGES = {749--761},
  SHORTTITLE = {A Unification of Mediation and Interaction},
  TITLE = {A Unification of Mediation and Interaction: A Four-Way Decomposition},
  URLDATE = {2026-08-14},
  VOLUME = {25},
}

@ARTICLE{xuFactorialDifferenceinDifferences2026,
  AUTHOR = {Xu, Yiqing and Zhao, Anqi and Ding, Peng},
  PUBLISHER = {Taylor \& Francis},
  DATE = {2026-04},
  DOI = {10.1080/01621459.2026.2628343},
  ISSN = {0162-1459},
  JOURNALTITLE = {Journal of the American Statistical Association},
  NUMBER = {554},
  PAGES = {963--975},
  TITLE = {Factorial {{Difference-in-Differences}}},
  URLDATE = {2026-09-06},
  VOLUME = {121},
}

@ARTICLE{yangPropensityScoreMatching2016,
  AUTHOR = {Yang, Shu and Imbens, Guido W. and Cui, Zhanglin and Faries, Douglas E. and Kadziola, Zbigniew},
  PUBLISHER = {John Wiley \& Sons, Ltd},
  DATE = {2016-12},
  DOI = {10.1111/biom.12505},
  ISSN = {1541-0420},
  JOURNALTITLE = {Biometrics},
  LANGID = {english},
  NUMBER = {4},
  PAGES = {1055--1065},
  TITLE = {Propensity Score Matching and Subclassification in Observational Studies with Multi-Level Treatments},
  URLDATE = {2026-08-16},
  VOLUME = {72},
}

@ARTICLE{yoshidaMultinomialExtensionPropensity2019,
  AUTHOR = {Yoshida, Kazuki and Solomon, Daniel H. and Haneuse, Sebastien and Kim, Seoyoung C. and Patorno, Elisabetta and Tedeschi, Sara K. and Lyu, Houchen and Franklin, Jessica M. and Stürmer, Til and {Hernández-Díaz}, Sonia and Glynn, Robert J.},
  DATE = {2019-03},
  DOI = {10.1093/aje/kwy263},
  ISSN = {1476-6256},
  JOURNALTITLE = {American Journal of Epidemiology},
  LANGID = {english},
  NUMBER = {3},
  PAGES = {609--616},
  SHORTTITLE = {Multinomial {{Extension}} of {{Propensity Score Trimming Methods}}},
  TITLE = {Multinomial {{Extension}} of {{Propensity Score Trimming Methods}}: {{A Simulation Study}}},
  VOLUME = {188},
}

\newpage
\appendix

\renewcommand\theproofn{\thesection.\arabic{proofn}}    
\renewcommand\theproposition{\thesection.\arabic{proposition}}    
\renewcommand*\thetable{\Alph{section}.\arabic{table}}
\renewcommand*\thefigure{\Alph{section}.\arabic{figure}}

\section{Proofs of results for panel data}\label{appendix_proofs}

\subsection{Identification}
\begin{proofn}{Proof of Proposition \ref{id_att_stag}: Identification of $DATT_{s-s'}(g,t)$, with not-yet-treated units used for comparison.}
    \label{proof:id_att_stag_ny}
\begin{align*}
   D &ATT_{s-s'}(g,t)  \\
   = & \mathbb{E}[Y_{t}(g) - Y_{t}(\infty) |G=g, S =s ] - \mathbb{E}[Y_{t}(g) - Y_{t}(\infty) | G=g, S=s' ] \\
   = & \mathbb{E}[Y_{t}(g) - Y_{t}(\infty) | G=g, S =s ] - \mathbb{E}[Y_{t}(g) - Y_{t}(\infty) | G=g, S=s' ] \\
   & + \mathbb{E} \Big[Y_{t}(\infty) - Y_{g-1}(\infty) | G=g, S=s\Big] - \mathbb{E} \Big[Y_{t}(\infty) - Y_{g-1}(\infty) | G=g, S=s' \Big] \\
   & - \Big(\mathbb{E} \Big[Y_{t}(\infty) - Y_{g-1}(\infty) | W_t = 0, G\neq g, S=s \Big] - \mathbb{E} \Big[Y_{t}(\infty) - Y_{g-1}(\infty) | W_t = 0, G \neq g, S=s' \Big]\Big) \\
   = & \mathbb{E}[Y_{t}(g) - Y_{g-1}(\infty) | G=g, S =s ] - \mathbb{E}[Y_{t}(g) - Y_{g-1}(\infty) | G=g, S=s' ] \\
   & - \Big(\mathbb{E} \Big[Y_{t}(\infty) - Y_{g-1}(\infty) | W_t = 0, G\neq g , S=s\Big] - \mathbb{E} \Big[Y_{t}(\infty) - Y_{g-1}(\infty) | W_t = 0,  G \neq g, S=s' \Big] \Big) \\
   = & \mathbb{E}[Y_{t} - Y_{g-1} | G=g, S=s] - \mathbb{E}[Y_{t} - Y_{g-1} | G=g, S=s']  \\
      & - \Big(\mathbb{E}[Y_{t} - Y_{g-1} | W_t=0, G \neq g, S=s] - \mathbb{E}[Y_{t} - Y_{g-1} | W_t=0, G \neq g, S=s']  \Big)
\end{align*}
where the first equality follows from the definition of the $DATT_{s-s'}(g,t)$, the second follows from Assumption \ref{cond_parallel_gaps_stag_ny} with degenerate $X$ for time periods $t \geq g$, the third from combining terms, and the last from Assumptions \ref{irreversible} and \ref{no_anticipation_stag}.

\end{proofn}

\begin{proofn}{Proof of Proposition \ref{id_att_stag}: Identification of $DATT_{s-s'}(g,t)$, with never-treated units used for comparison.}
    \label{proof:id_att_stag_nev}
\begin{align*}
   DATT_{s-s'}(g,t) &  \\
   = & \mathbb{E}[Y_{t}(g) - Y_{t}(\infty) | G=g, S =s] - \mathbb{E}[Y_{t}(g) - Y_{t}(\infty) |  G=g, S=s' ] \\
   = & \mathbb{E}[Y_{t}(g) - Y_{t}(\infty) | G=g, S =s ] - \mathbb{E}[Y_{t}(g) - Y_{t}(\infty) |  G=g, S=s' ] \\
   & + \mathbb{E} \Big[Y_{t}(\infty) - Y_{g-1}(\infty) | G=g, S=s \Big] - \mathbb{E} \Big[Y_{t}(\infty) - Y_{g-1}(\infty) | G=g, S=s' \Big] \\
   & - \Big(\mathbb{E} \Big[Y_{t}(\infty) - Y_{g-1}(\infty) | G = \infty, S=s  \Big] - \mathbb{E} \Big[Y_{t}(\infty) - Y_{g-1}(\infty) |  G =\infty, S=s' \Big]\Big) \\
   = & \mathbb{E}[Y_{t}(g) - Y_{g-1}(\infty) | G=g, S =s ] - \mathbb{E}[Y_{t}(g) - Y_{g-1}(\infty) | G=g, S=s' ] \\
   & - \Big(\mathbb{E} \Big[Y_{t}(\infty) - Y_{g-1}(\infty) | G=\infty, S=s \Big] - \mathbb{E} \Big[Y_{t}(\infty) - Y_{g-1}(\infty) | G=\infty, S=s'\Big] \Big)  \\
   = & \mathbb{E}[Y_{t} - Y_{g-1} | G=g, S=s] - \mathbb{E}[Y_{t} - Y_{g-1} | G=g, S=s']  \\
      & - \Big(\mathbb{E}[Y_{t} - Y_{g-1} | G = \infty, S=s] - \mathbb{E}[Y_{t} - Y_{g-1} | G=\infty , S=s']  \Big)
\end{align*}
where the first equality follows from the definition of the $DATT_{s-s'}(g,t)$, the second follows from Assumption \ref{cond_parallel_gaps_stag_nev} with degenerate $X$ for time periods $t \geq g$, the third from combining terms, and the last from the definitions of the potential and observed outcomes and Assumptions \ref{irreversible} and \ref{no_anticipation_stag}.

When conditioning on $X$, the results follow from those in \textcite{callawayDifferenceinDifferencesMultipleTime2021}. 
\end{proofn}

\begin{proofn}{Proof of Proposition \ref{id_satt_stag}: Identification of $CDATT_{s-s'}(g,t)$ with not-yet-treated group as comparison.}
    \label{proof:id_cdatt_stag_ny}
\begin{align*}
   C& DATT_{s-s'}(g,t)  \\
  = & \mathbb{E}[Y_{t}(g,s) - Y_{t}(\infty,s) | G=g, S =s ] - \mathbb{E}[Y_{t}(g,s') - Y_{t}(\infty,s') | G=g, S=s ] \\
   = & \mathbb{E}[Y_{t}(g) - Y_{t}(\infty) | G=g,S =s ] - \mathbb{E}[Y_{t}(g) - Y_{t}(\infty) | G=g,S=s'] \\
   = & \mathbb{E}[Y_{t}(g) - Y_{t}(\infty) | G=g,S =s] - \mathbb{E}[Y_{t}(g) - Y_{t}(\infty) | G=g,S=s'] \\
   & + \mathbb{E} \Big[Y_{t}(\infty) - Y_{g-1}(\infty) | G=g,S=s\Big] - \mathbb{E} \Big[Y_{t}(\infty) - Y_{g-1}(\infty) | G=g, S=s'\Big] \\
   & - \Big(\mathbb{E} \Big[Y_{t}(\infty) - Y_{g-1}(\infty) | W_t = 0, G\neq g, S=s \Big] - \mathbb{E} \Big[Y_{t}(\infty) - Y_{g-1}(\infty) | W_t = 0, G \neq g, S=s' \Big]\Big) \\
   = & \mathbb{E}[Y_{t}(g) - Y_{g-1}(\infty) | G=g, S =s ] - \mathbb{E}[Y_{t}(g) - Y_{g-1}(\infty) | G=g, S=s'] \\
   & - \Big(\mathbb{E} \Big[Y_{t}(\infty) - Y_{g-1}(\infty) | W_t = 0, G\neq g, S=s\Big] - \mathbb{E} \Big[Y_{t}(\infty) - Y_{g-1}(\infty) | W_t = 0, G \neq g, S=s' \Big] \Big) \\
   = & \mathbb{E}[Y_{t} - Y_{g-1} | G=g, S=s] - \mathbb{E}[Y_{t} - Y_{g-1} | G=g, S=s']  \\
      & - \Big(\mathbb{E}[Y_{t} - Y_{g-1} | W_t =0, G \neq g, S=s] - \mathbb{E}[Y_{t} - Y_{g-1} | W_t=0, G \neq g, S=s']  \Big)
\end{align*}
where the first equality follows from the definition of $CDATT_{s-s'}(g,t)$, the second from Assumption \ref{no_subgroup_selection_stag}, the third from Assumption \ref{cond_parallel_gaps_stag_ny}, the fourth from combining and simplifying terms as above, and the last from Assumptions \ref{irreversible} and \ref{no_anticipation_stag}.
\end{proofn}

\begin{proofn}{Proof of Proposition \ref{id_satt_stag}: Identification of $CDATT_{s-s'}(g,t)$ with never-treated group as comparison}
        \label{proof:id_cdatt_stag_nev}
\begin{align*}
   CDATT_{s-s'}(g,t) &  \\
  = & \mathbb{E}[Y_{t}(g,s) - Y_{t}(\infty,s) | G=g, S =s ] - \mathbb{E}[Y_{t}(g,s') - Y_{t}(\infty,s') | G=g, S=s ] \\
   = & \mathbb{E}[Y_{t}(g) - Y_{t}(\infty) | G=g, S =s ] - \mathbb{E}[Y_{t}(g) - Y_{t}(\infty) | G=g, S=s' ] \\
   = & \mathbb{E}[Y_{t}(g) - Y_{t}(\infty) | G=g, S =s ] - \mathbb{E}[Y_{t}(g) - Y_{t}(\infty) | G=g, S=s'] \\
   & + \mathbb{E} \Big[Y_{t}(\infty) - Y_{g-1}(\infty) | G=g, S=s\Big] - \mathbb{E} \Big[Y_{t}(\infty) - Y_{g-1}(\infty) | G=g, S=s' \Big] \\
   & - \Big(\mathbb{E} \Big[Y_{t}(\infty) - Y_{g-1}(\infty) | G=\infty, S=s \Big] - \mathbb{E} \Big[Y_{t}(\infty) - Y_{g-1}(\infty) | G= \infty, S=s' \Big]\Big) \\
   = & \mathbb{E}[Y_{t}(g) - Y_{g-1}(\infty) | G=g, S =s ] - \mathbb{E}[Y_{t}(g) - Y_{g-1}(\infty) | G=g, S=s' ] \\
   & - \Big(\mathbb{E} \Big[Y_{t}(\infty) - Y_{g-1}(\infty) | G = \infty, S=s \Big] - \mathbb{E} \Big[Y_{t}(\infty) - Y_{g-1}(\infty) | G=\infty, S=s' \Big] \Big) \\
   = & \mathbb{E}[Y_{t} - Y_{g-1} | G=g, S=s] - \mathbb{E}[Y_{t} - Y_{g-1} | G=g, S=s']  \\
      & - \Big(\mathbb{E}[Y_{t} - Y_{g-1} | G = \infty , S=s] - \mathbb{E}[Y_{t} - Y_{g-1} | G = \infty , S=s']  \Big)
\end{align*}
where the first equality follows from the definition of $CDATT_{s-s'}(g,t)$, the second from Assumption \ref{no_subgroup_selection_stag}, the third from Assumption \ref{cond_parallel_gaps_stag_ny}, the fourth from combining and simplifying terms as above, and the last from Assumptions \ref{irreversible} and \ref{no_anticipation_stag}.
\end{proofn}

\begin{proofn}
\label{proof_ipw}
To show $ CDATT^{IPW, c}_{s-s'}(g,t) = CDATT_{s-s'}(g,t)$. 

First, we can show that 
{ 
\begin{align*}
     \mathbb{E}\Big[ \Big( \frac{G_g S_s}{\mathbb{E}[G_g S_s]}\Big)(Y_{t} - Y_{g-1})\Big] & =  \frac{\mathbb{E}[G_g S_s] \mathbb{E}[Y_t - Y_{g-1} | G_g = 1, S_s = 1]}{\mathbb{E}[G_g S_s]} \\
     & =  \mathbb{E}[Y_t - Y_{g-1} | G_g = 1, S_s = 1]  \\
   \text{(using Assumptions \ref{irreversible} and \ref{no_anticipation_stag})} \quad   & =  \mathbb{E}[Y_t(g,s) - Y_{g-1}(\infty,s) | G_g = 1, S_s = 1] 
\end{align*}

By the same process, using the law of iterated expectations and the definition of the propensity score models,

\begin{align*}
    \mathbb{E}\Big[ \Big(\frac{G_gS_{s'}\pi_{g, s}(X)}{\pi_{g, s'}(X)\mathbb{E}[G_g S_{s'} \pi_{g, s}(X)/ \pi_{g, s'}(X) ]}\Big) & (Y_t - Y_{g-1}) \Big]   = \mathbb{E}\Big[\frac{\mathbb{E}[G_gS_s | X] \mathbb{E}[G_g S_{s'}(Y_t - Y_{g-1}) | X] }{\mathbb{E}[G_g S_{s'} | X]\mathbb{E}[G_g S_{s'}\mathbb{E}[G_g S_{s} | X]/\mathbb{E}[G_g S_{s'} | X]]} \Big] \\
    & = \mathbb{E}\Big[\frac{\mathbb{E}[G_g S_s | X] \mathbb{E}[G_g S_{s'} | X] \mathbb{E}[Y_t - Y_{g-1} | X, G_g =1, S_{s'} =1] }{\mathbb{E}[G_g S_{s'} | X]\mathbb{E}[G_g S_s]} \Big] \\
        \text{(using Assumptions \ref{irreversible} and \ref{no_anticipation_stag})} & = \mathbb{E}\Big[\frac{\mathbb{E}[G_gS_s|X]\mathbb{E}[(Y_t(g, s') - Y_{g-1}(\infty,s')) | X, G_g =1, S_{s'} =1] }{\mathbb{E}[G_g S_s]} \Big]\\
    \text{(using Assumption \ref{cond_subgroup_selection_stag})}& = \mathbb{E}\Big[\frac{\mathbb{E}[G_gS_s|X]\mathbb{E}[(Y_t(g,s') - Y_{g-1}(\infty,s')) | X, G_g =1, S_{s} =1] }{\mathbb{E}[G_g S_s]} \Big] \\
   & = \mathbb{E}[Y_t(g,s') - Y_{g-1}(\infty,s') | G_g =1, S_{s} =1 ] 
\end{align*}

Using a similar argument,
\begin{align*}
     \mathbb{E}\Big[\Big(\frac{ C_c S_s \pi_{g, s}(X)}{\pi_{c, s}(X)\mathbb{E}[C_c S_s \pi_{g,s}(X) / \pi_{c,s}(X)]}\Big) & (Y_{t} - Y_{g-1}) \Big] = \mathbb{E}\Big[\frac{\mathbb{E}[G_gS_s | X] \mathbb{E}[C_c S_s(Y_t - Y_{g-1}) | X] }{\mathbb{E}[C_c S_s | X]\mathbb{E}[C_c S_s \mathbb{E}[G_g S_s | X]/ \mathbb{E}[C_c S_s | X] ]} \Big] \\
     & = \mathbb{E}\Big[\frac{\mathbb{E}[G_g S_s | X] \mathbb{E}[C_c S_s | X] \mathbb{E}[Y_t - Y_{g-1} | X, C_c =1, S_s =1] }{\mathbb{E}[C_c S_s | X]\mathbb{E}[G_g S_s]} \Big] \\
& = \mathbb{E}\Big[\frac{\mathbb{E}[G_gS_s|X]\mathbb{E}[Y_t(\infty, s) - Y_{g-1}(\infty,s) | X, C_c =1, S_{s} =1] }{\mathbb{E}[G_g S_s]} \Big] \end{align*}

By the same process again,
{\footnotesize
 \begin{align*}
\mathbb{E}\Big[ \Big(\frac{C_c S_{s'}\pi_{g, s}(X)}{\pi_{c, s'}(X) \mathbb{E}[C_c S_{s'} \pi_{g,s}(X)/\pi_{c, s'}(X) ]} \Big) (Y_{t} - Y_{g-1})\Big] & = \mathbb{E}\Big[\frac{\mathbb{E}[G_gS_s|X]\mathbb{E}[Y_t(\infty, s') - Y_{g-1}(\infty,s') | X, C_c =1, S_{s'} =1] }{\mathbb{E}[G_g S_s]} \Big] 
\end{align*}
}
Applying Assumption \ref{cond_parallel_gaps_stag_ny} or \ref{cond_parallel_gaps_stag_nev}, conditional on $X$, finishes the proof:
{\scriptsize
\begin{align*}
\mathbb{E}&\Big[\frac{\mathbb{E}[G_gS_s|X]\mathbb{E}[Y_t(\infty, s) - Y_{g-1}(\infty,s) | X, C_c =1, S_{s} =1] }{\mathbb{E}[G_g S_s]} \Big] \\
&\quad \quad\quad - \mathbb{E}\Big[\frac{\mathbb{E}[G_gS_s|X]\mathbb{E}[Y_t(\infty, s') - Y_{g-1}(\infty,s') | X, C_c =1, S_{s'} =1] }{\mathbb{E}[G_g S_s]} \Big] \\
& = \mathbb{E}\Big[\frac{\mathbb{E}[G_gS_s|X](\mathbb{E}[Y_t(\infty, s) - Y_{g-1}(\infty,s) | X, C_c =1, S_{s} =1] - \mathbb{E}[Y_t(\infty, s') - Y_{g-1}(\infty,s') | X, C_c =1, S_{s'} =1])}{\mathbb{E}[G_g S_s]} \Big]  \\
\text{(using Assumption \ref{cond_parallel_gaps_stag_ny} or \ref{cond_parallel_gaps_stag_nev})} \quad & = \mathbb{E}\Big[\frac{\mathbb{E}[G_gS_s|X](\mathbb{E}[Y_t(\infty, s) - Y_{g-1}(\infty,s) | X, G_g =1, S_{s} =1] - \mathbb{E}[Y_t(\infty, s') - Y_{g-1}(\infty,s') | X, G_g =1, S_{s'} =1]) }{\mathbb{E}[G_g S_s]} \Big]  \\
\text{(using Assumption \ref{cond_subgroup_selection_stag})} \quad & = \mathbb{E}\Big[\frac{\mathbb{E}[G_gS_s|X](\mathbb{E}[Y_t(\infty, s) - Y_{g-1}(\infty,s) | X, G_g =1, S_{s} =1] - \mathbb{E}[Y_t(\infty, s') - Y_{g-1}(\infty,s') |  G_g =1, S_{s} =1]) }{\mathbb{E}[G_g S_s]} \Big]  \\
 & = \mathbb{E}[Y_t(\infty, s) - Y_{g-1}(\infty,s) | G_g =1, S_{s} =1] - \mathbb{E}[Y_t(\infty, s') - Y_{g-1}(\infty,s') | G_g =1, S_{s} =1] 
\end{align*}
}
Combining the terms returns the $CDATT_{s-s'}(g,t)$:
\begin{align*}
\mathbb{E}& [Y_t(g,s) - Y_{g-1}(\infty,s) | G_g =1, S_{s} =1 ]  - \mathbb{E}[Y_t(g,s') - Y_{g-1}(\infty,s') | G_g =1, S_{s} =1 ]\\
& - \Big(\mathbb{E}[Y_t(\infty, s) - Y_{g-1}(\infty,s) | G_g =1, S_{s} =1] - \mathbb{E}[Y_t(\infty, s') - Y_{g-1}(\infty,s') | G_g =1, S_{s} =1] \Big) \\
& = \mathbb{E}[Y_t(g,s) - Y_{t}(\infty,s) | G_g =1, S_{s} =1 ]  - \mathbb{E}[Y_t(g,s') - Y_{t}(\infty,s') | G_g =1, S_{s} =1 ]\\
& = CDATT_{s-s'}(g,t)
\end{align*}

}
\end{proofn}

\begin{proofn} To show $CDATT^{RA, c}_{s-s'}(g,t) =  CDATT_{s-s'}(g,t) $. 
    \label{proof_ra}
    
First note that, under Assumptions \ref{irreversible} and \ref{no_anticipation_stag},
\begin{align*}
\mathbb{E}\Big[ \frac{G_g S_s}{\mathbb{E}[G_g S_s]} (Y_{t}-Y_{g-1}) \Big]  = \mathbb{E}[Y_{t}(g,s)-Y_{g-1}(\infty,s) | G_g=1, S_s=1] 
\end{align*}
Using the same process and adding Assumption \ref{cond_subgroup_selection_stag},
\begin{align*}
   \mathbb{E}\Big[ \frac{G_g S_s}{\mathbb{E}[G_g S_s]} \mu_{g,t}^{s'}(X) \Big]  
   & = \mathbb{E}\Big[ \frac{G_g S_s}{\mathbb{E}[G_g S_s]}\mathbb{E}[Y_{t}(g,s')-Y_{g-1}(\infty,s') | G_g=1, S_{s'}=1, X] \Big] \\
   & =  \mathbb{E}[Y_{t}(g,s')-Y_{g-1}(\infty,s') | G_g=1, S_s=1] 
\end{align*}
Similarly, 
\begin{align*}
   \mathbb{E}\Big[ \frac{G_g S_s}{\mathbb{E}[G_g S_s]} \mu_{c, g, t}^{s}(X) \Big]  
   & = \mathbb{E}\Big[ \frac{G_g S_s}{\mathbb{E}[G_g S_s]}\mathbb{E}[Y_{t}(\infty,s)-Y_{g-1}(\infty,s) | C_c=1, S_s=1, X] \Big] \\
\end{align*}
and 
\begin{align*}
   \mathbb{E}\Big[ \frac{G_g S_s}{\mathbb{E}[G_g S_s]} \mu_{c, g, t}^{s'}(X) \Big]  
   & = \mathbb{E}\Big[ \frac{G_g S_s}{\mathbb{E}[G_g S_s]}\mathbb{E}[Y_{t}(\infty,s')-Y_{g-1}(\infty,s') | C_c=1, S_{s'}=1, X] \Big] \\
\end{align*}
Applying Assumption \ref{cond_parallel_gaps_stag_ny} or \ref{cond_parallel_gaps_stag_nev}, conditional on $X$, and Assumption \ref{cond_subgroup_selection_stag} finishes the proof.
\end{proofn}

\newpage

\begin{proofn} 
    \label{proof_dr} 
    To show $CDATT^{DR, c}_{s-s'}(g,t) = CDATT_{s-s'}(g,t)$, with
\begin{align*}
CDATT^{DR, c}_{s-s'}(g,t) = & \mathbb{E}\Big[ \frac{G_g S_s}{\mathbb{E}[G_g S_s]} \Big(Y_{t} - Y_{g-1} - \mu_{g,t}^{s'}(X) - \mu_{c,g,t}^{s}(X)+ \mu_{c,g,t}^{s'}(X) \Big)\Big] \\
&- \mathbb{E}\Big[w_2(G_g, S_s, X) (Y_t - Y_{g-1} - \mu_{g,t}^{s'}(X))\Big] \\
&- \mathbb{E}\Big[w_3^c(C_c, S_s, X)  (Y_t - Y_{g-1}- \mu_{c,g,t}^{s}(X))\Big] \\
&+\mathbb{E}\Big[w_4^c(C_c, S_{s'}, X)(Y_t-Y_{g-1}- \mu_{c,g,t}^{s'}(X))\Big]
\end{align*}

First, consider the case when the propensity scores are correctly specified. In this case, note that
\begin{align*}
\mathbb{E}&\Big[ \Big(\frac{G_g S_s}{\mathbb{E}[G_g S_s]}- w_2(G_g, S_s, X)\Big) \Big(\mu_{g,t}^{s'}(X)\Big)\Big] \\
    & = \mathbb{E}\Big[\Big(\frac{G_gS_s}{\mathbb{E}[G_g S_s]} - \frac{G_gS_{s'} \pi_{g,s}(X)}{\pi_{g,s'}(X)} \Big/ \mathbb{E} \Big[\frac{G_gS_{s'} \pi_{g,s}(X)}{\pi_{g,s'}(X)} \Big] \Big) \mu_{g,t}^{s'}(X) \Big] \\
    & = \mathbb{E}\Big[\Big(\frac{G_gS_s}{\mathbb{E}[G_g S_s]} - \frac{G_gS_{s'} \mathbb{E}[G_gS_{s} | X]}{\mathbb{E}[G_gS_{s'} | X]} \Big/ \mathbb{E} \Big[\frac{G_gS_{s'} \mathbb{E}[G_gS_{s} | X]}{\mathbb{E}[G_gS_{s'} | X]} \Big] \Big) \mu_{g,t}^{s'}(X) \Big] \\
    & = \frac{1}{\mathbb{E}[G_gS_{s}]} \mathbb{E}\Big[\Big(G_gS_s - \frac{G_gS_{s'} \mathbb{E}[G_gS_{s} | X]}{\mathbb{E}[G_gS_{s'} | X]} \Big) \mu_{g,t}^{s'}(X) \Big] \\
    & = \frac{1}{\mathbb{E}[G_gS_{s}]} \mathbb{E}\Big[\Big(\mathbb{E}[G_gS_s | X] - \mathbb{E}[G_gS_{s} | X] \Big) \mu_{g,t}^{s'}(X) \Big] \\
    & = 0 
\end{align*}
Similarly, 
\begin{align*}
    \mathbb{E}&\Big[\Big(\frac{G_g S_s}{\mathbb{E}[G_g S_s]}-w_3^c(C_c, S_s, X) \Big) \mu_{c,g,t}^{s}(X)\Big] \\
    & = \mathbb{E}\Big[\Big(\frac{G_gS_s}{\mathbb{E}[G_g S_s]}- \frac{C_c S_{s} \pi_{g,s}(X)}{\pi_{c,s}(X)} \Big/ \mathbb{E} \Big[\frac{C_c S_{s} \pi_{g,s}(X)}{\pi_{c,s}(X)}  \Big]  \Big) \mu_{c,g,t}^{s}(X) \Big] \\
    & = \frac{1}{\mathbb{E}[G_g S_{s}]} \mathbb{E}\Big[\Big(\mathbb{E}[G_gS_{s} | X] - \mathbb{E}[G_gS_{s} | X] \Big) \mu_{c,g,t}^{s}(X) \Big] \\
    & = 0 
\end{align*}
and
\begin{align*}
\mathbb{E}&\Big[\Big(\frac{G_g S_s}{\mathbb{E}[G_g S_s]}-w_4^c(C_c, S_{s'}, X) \Big) \mu_{c,g,t}^{s'}(X)\Big] = 0
\end{align*}

Then,
\begin{align*}
C& DATT^{DR, c}_{s-s'}(g,t) = CDATT^{IPW, c}_{s-s'}(g,t)  = CDATT_{s-s'}(g,t)
\end{align*}
where the first equality uses the results from above after rearranging terms and the second equality uses the result shown in Proof \ref{proof_ipw}.

On the other hand, consider the case when the outcome models are correct. In this case, 
\begin{align*}
\mathbb{E}&\Big[w_2(G_g, S_s, X) (Y_t - Y_{g-1} - \mu_{g,t}^{s'}(X))\Big] \\
& = \mathbb{E}\Big[w_2(G_g, S_s, X) (Y_t - Y_{g-1} - \mathbb{E}[Y_t - Y_{g-1} | G_g = 1, S_{s'} = 1, X])\Big] \\
& = \mathbb{E}\Big[\frac{G_gS_{s'} \pi_{g,s}(X)}{\pi_{g,s'}(X)} \Big/ \mathbb{E} \Big[\frac{G_gS_{s'} \pi_{g,s}(X)}{\pi_{g,s'}(X)}\Big] (Y_t - Y_{g-1} - \mathbb{E}[Y_t - Y_{g-1} | G_g = 1, S_{s'} = 1, X])\Big] \\
& = \mathbb{E}\Big[\frac{ \pi_{g,s}(X)}{\pi_{g,s'}(X)} \Big/ \mathbb{E} \Big[\frac{G_gS_{s'} \pi_{g,s}(X)}{\pi_{g,s'}(X)}\Big] (\mathbb{E}[G_gS_{s'}(Y_t - Y_{g-1})|X] - \mathbb{E}[G_gS_{s'}|X]\mathbb{E}[Y_t - Y_{g-1} | G_g = 1, S_{s'} = 1, X])\Big] \\
& = \mathbb{E}\Big[\frac{ \pi_{g,s}(X)}{\pi_{g,s'}(X)} \Big/ \mathbb{E} \Big[\frac{G_gS_{s'} \pi_{g,s}(X)}{\pi_{g,s'}(X)}\Big] (\mathbb{E}[G_gS_{s'}|X]\mathbb{E}[Y_t - Y_{g-1}|G_g=1, S_{s'} = 1, X] \\
& \quad \quad - \mathbb{E}[G_gS_{s'}|X]\mathbb{E}[Y_t - Y_{g-1} | G_g = 1, S_{s'} = 1, X])\Big] \\
& = 0
\end{align*}

Analogous arguments show 
\begin{align*}
\mathbb{E}\Big[w_3^c(C_c, S_s, X)  (Y_t - Y_{g-1}- \mu_{c,g,t}^{s}(X))\Big] & =0 \\
\mathbb{E}\Big[w_4^c(C_c, S_{s'}, X)(Y_t-Y_{g-1}- \mu_{c,g,t}^{s'}(X))\Big] & =0
\end{align*}

Then, 
\begin{align*}
C& DATT^{DR, c}_{s-s'}(g,t) = CDATT^{RA, c}_{s-s'}(g,t)  = CDATT_{s-s'}(g,t)
\end{align*}
where the first equality uses the results from above and the second equality uses the result shown in Proof \ref{proof_ra}.

\end{proofn}

\newpage

\subsection{Asymptotic properties}
\label{proof:asymptotic_properties}

\begin{assumption}Regularity conditions \\
\label{regularity}
Let $\norm{Z} = \sqrt{trace(Z'Z)}$ be the Euclidean norm. Let $j(X)$ be a generic notation for $e_{g,s}(X)$, $e_{g,s'}(X)$, $e_{c,s}(X)$, $e_{c,s'}(X)$, $m_{g,t}^{s'}(X)$, and $m_{c,g,t}^{s'}(X)$. Let $V = (Y_t, Y_{g-1}, G, t, X)$. Then,

\begin{itemize}
    \item $j(X) = j(X;\gamma)$ is a parametric model, where $\gamma \in \Theta \subset \mathbb{R}^k$ 
    \item $j(X;\gamma)$ is a.s. continuous at each $\gamma \in \Theta$
    \item There exists a unique pseudo-true parameter $\gamma^* \in int(\Theta)$
    \item $j(X; \gamma)$ is a.s. twice continuously differentiable in a neighborhood of $\gamma^*, \Theta^* \subset \Theta$
    \item The estimator $\widehat \gamma$ is strongly consistent for the $\gamma^*$ and satisfies the following linear expansion: 
    $$\sqrt{n}(\widehat\gamma - \gamma^*) = \frac{1}{\sqrt{n}} \sum_{i=1}^n l_j(V_i;\gamma^*) + o_p(1) $$
    with $l_j(\cdot; \gamma)$ being a $k \times 1$ vector such that $\mathbb{E}[l_j(V;\gamma^*)] = 0$, $\mathbb{E}[l_j(V; \gamma^*) l_j(V; \gamma^*)']$ exists and is positive definite and $\lim_{\delta\to 0} \mathbb{E}[\sup_{\gamma\in\Theta^*:\norm{\gamma-\gamma^*}\leq \delta}\norm{l_j(V;\gamma) - l_j(V;\gamma^*)}^2 = 0$.
    \item For some $\varepsilon > 0$ and for $a \in \{g,c\}$ and $b \in \{s,s'
\}$, $0 < e_{a,b}(X; \gamma) \leq 1 - \varepsilon$ a.s. for all $\theta_{a,b} \in int(\Theta^{ps}_{a,b})$ where $\Theta^{ps}_{a,b}$ denotes the parameter space of $\gamma_{a,b}$. 
\end{itemize}
\end{assumption}

\begin{proofn} Proof of Proposition \ref{asym_dist}. \\
\label{proof:asym_dist}
Recall that 

\begin{align*}
\widehat{CDATT}^{DR, c}_{s-s'}(g,t) = & \mathbb{E}_n\Big[ \widehat w_1(G_g,S_s) \Big(Y_{t} - Y_{g-1} - m_{g,t}^{s'}(X; \widehat\beta_{g,s'}) - m_{c,g,t}^{s}(X; \widehat\beta_{c,g,s})+ m_{c,g,t}^{s'}(X; \widehat\beta_{c,g,s'}) \Big)\Big] \\
&- \mathbb{E}_n\Big[\widehat w_2(G_g, S_s, X; \widehat\theta_{g}) (Y_t - Y_{g-1} - m_{g,t}^{s'}(X; \widehat\beta_{g,s'}))\Big] \\
&- \mathbb{E}_n\Big[\widehat w_3^c(C_c, S_s, X; \widehat\theta_{g})  (Y_t - Y_{g-1}- m_{c,g,t}^{s}(X; \widehat\beta_{c,g,s}))\Big] \\
&+\mathbb{E}_n\Big[\widehat w_4^c(C_c, S_{s'}, X; \widehat\theta_{g})(Y_t-Y_{g-1}- m_{c,g,t}^{s'}(X; \widehat\beta_{c,g,s'}))\Big]
\end{align*}
where $\widehat\theta_{g}, \widehat\beta_{g,s'}$, $\widehat\beta_{c,g,s}$, and $\widehat\beta_{c,g,s'}$ are estimators for the corresponding pseudo-true parameters $\theta_{g}^*, \beta_{g,s'}^*, \beta_{c,g,s}^*$, and $\beta_{c,g,s'}^*$, and
\begin{align*}
\widehat w_1(G_g, S_s) & = \frac{G_gS_s}{\mathbb{E}_n[G_gS_s]} \\
\widehat w_2(G_g, S_s, X;\widehat\theta_{g}) & = \frac{G_gS_{s'} e_{g,s}(X; \widehat\theta_{g})}{e_{g,s'}(X;\widehat\theta_{g})} \Big/ \mathbb{E}_n \Big[\frac{G_gS_{s'} e_{g,s}(X; \widehat\theta_{g})}{e_{g,s'}(X;\widehat\theta_{g})} \Big] \\
\widehat w_3^c(C_c, S_s, X;\widehat\theta_{g}) & = \frac{C_c S_{s} e_{g,s}(X; \widehat\theta_{g})}{e_{c,s}(X;\widehat\theta_{g})} \Big/ \mathbb{E}_n \Big[\frac{C_c S_{s} e_{g,s}(X; \widehat\theta_{g})}{e_{c,s}(X;\widehat\theta_{g})}  \Big]\\
\widehat w_4^c(C_c, S_{s'}, X;\widehat\theta_{g}) & = \frac{C_c S_{s'} e_{g,s}(X; \widehat\theta_{g})}{e_{c,s'}(X;\widehat\theta_{g})} \Big/ \mathbb{E}_n \Big[\frac{C_c S_{s'} e_{g,s}(X; \widehat\theta_{g})}{e_{c,s'}(X;\widehat\theta_{g})}  \Big]
\end{align*}
For notational simplicity, write these as $\widehat w_1$, $\widehat w_2(\widehat\theta_{g}), \widehat w_3^c(\widehat\theta_{g})$ and $\widehat w_4^c(\widehat\theta_{g})$, with analogues $w_2(\theta_{g}^*), w_3^c(\theta_{g}^*)$ and $w_4^c(\theta_{g}^*)$. 

Using the weak law of large numbers, as $n \to \infty$,
\begin{align*}
\mathbb{E}_n[G_gS_s] & \to^p \mathbb{E}[G_gS_s]  \\   
\mathbb{E}_n[G_gS_{s'}] & \to^p \mathbb{E}[G_gS_{s'}]  \\   
\mathbb{E}_n[C_c S_s] & \to^p \mathbb{E}[C_c S_s]  \\   
\mathbb{E}_n[C_c S_{s'}] & \to^p \mathbb{E}[C_c S_{s'}]  \\   
\mathbb{E}_n [\widehat w_2(\widehat\theta_{g})] & \to^p \mathbb{E} [w_2(\theta_{g}^*)]  \\
\mathbb{E}_n [\widehat w_3^c(\widehat\theta_{g})] & \to^p \mathbb{E} [ w_3^c(\theta_{g}^*)]  \\
\mathbb{E}_n [\widehat w_4^c(\widehat\theta_{g})] & \to^p \mathbb{E} [w_4^c(\theta_{g}^*)]  \\
\mathbb{E}_n [m_{g,t}^{s'}(X; \widehat\beta_{g,s'})] & \to^p \mathbb{E} [m_{g,t}^{s'}(X; \beta_{g,s'}^*)]  \\
\mathbb{E}_n [m_{c,g,t}^{s}(X; \widehat\beta_{c,g,s})] & \to^p \mathbb{E} [m_{c,g,t}^{s}(X; \beta_{c,g,s}^*)]  \\
\mathbb{E}_n [m_{c,g,t}^{s'}(X; \widehat\beta_{c,g,s'})] & \to^p \mathbb{E} [m_{c,g,t}^{s'}(X; \beta_{c,g,s'}^*)]  
\end{align*}

Using the continuous mapping theorem, this means 
\begin{align*}
\widehat{CDATT}^{DR, c}_{s-s'}(g,t) \to^p & \mathbb{E}\Big[ w_1 \Big(Y_{t} - Y_{g-1} - m_{g,t}^{s'}(X; \beta_{g,s'}^*) - m_{c,g,t}^{s}(X; \beta_{c,g,s}^*)+ m_{c,g,t}^{s'}(X; \beta_{c,g,s'}^*) \Big)\Big] \\
&- \mathbb{E}\Big[w_2(\theta_{g}^*) (Y_t - Y_{g-1} - m_{g,t}^{s'}(X; \beta_{g,s'}^*))\Big] \\
&- \mathbb{E}\Big[w_3^c(\theta_{g}^*)  (Y_t - Y_{g-1}- m_{c,g,t}^{s}(X; \beta_{c,g,s}^*))\Big] \\
&+\mathbb{E}\Big[w_4^c(\theta_{g}^*)(Y_t-Y_{g-1}- m_{c,g,t}^{s'}(X; \beta_{c,g,s'}^*))\Big]
\end{align*}

Following the approach in \textcite{santannaDoublyRobustDifferenceindifferences2020}, rewrite the $\widehat{CDATT}^{DR, c}_{s-s'}(g,t)$:
\begin{align*}
\widehat{CDATT}^{DR, c}_{s-s'}(g,t) =&  \mathbb{E}_n[ \widehat w_1 (Y_{t} - Y_{g-1})] - \mathbb{E}_n[\widehat w_1 m_{g,t}^{s'}(X; \widehat\beta_{g,s'})] \\
& - \mathbb{E}_n[\widehat w_1 m_{c,g,t}^{s}(X; \widehat\beta_{c,g,s})]  + \mathbb{E}_n[\widehat w_1 m_{c,g,t}^{s'}(X; \widehat\beta_{c,g,s'})]  \\
&- \mathbb{E}_n[\widehat w_2(\widehat\theta_{g}) (Y_t - Y_{g-1})]  + \mathbb{E}_n[\widehat w_2(\widehat\theta_{g}) m_{g,t}^{s'}(X; \widehat\beta_{g,s'})] \\
&- \mathbb{E}_n[\widehat w_3^c(\widehat\theta_{g})  (Y_t - Y_{g-1})]  + \mathbb{E}_n[\widehat w_3^c(\widehat\theta_{g}) m_{c,g,t}^{s}(X; \widehat\beta_{c,g,s})] \\
&+\mathbb{E}_n[\widehat w_4^c(\widehat\theta_{g})(Y_t-Y_{g-1})] -\mathbb{E}_n[\widehat w_4^c(\widehat\theta_{g}) m_{c,g,t}^{s'}(\widehat\beta_{c,g,s'})]
\\
\equiv \widehat{D_1^{g,s}}  & - \widehat{D_{RA1}^{g,s'}} - \widehat{D_{RA1}^{c,s}} + \widehat{ D_{RA1}^{c,s'}}  - \widehat{D_1^{g,s'}} +\widehat {D_{RA2}^{g,s'}} - \widehat{D_1^{c,s}} +\widehat {D_{RA3}^{c,s}} + \widehat{D_1^{c,s'}} -\widehat {D_{RA4}^{c,s'}}
\end{align*}
Thus, 
\begin{align*}
  \sqrt{n}(\widehat{CDATT}_{s-s'}^{DR,c}(g,t)- {CDATT}_{s-s'}^{DR,c}(g,t)) = & \sqrt{n}(\widehat{D_1^{g,s}} - D_1^{g,s}) - \sqrt{n}(\widehat{D_{RA1}^{g,s'}} - D_{RA1}^{g,s'}) \\
  & - \sqrt{n}(\widehat{D_{RA1}^{c,s}} - D_{RA1}^{c,s})  +  \sqrt{n}(\widehat{D_{RA1}^{c,s'}} - D_{RA1}^{c,s'}) \\
  & - \sqrt{n}(\widehat{D_1^{g,s'}} - D_1^{g,s'}) + \sqrt{n}(\widehat{D_{RA2}^{g,s'}} - D_{RA2}^{g,s'})\\
  & - \sqrt{n}(\widehat{D_1^{c,s}} - D_1^{c,s}) + \sqrt{n}(\widehat{D_{RA3}^{c,s}} - D_{RA3}^{c,s})\\
  & + \sqrt{n}(\widehat{D_1^{c,s'}} - D_1^{c,s'}) - \sqrt{n}(\widehat{D_{RA4}^{c,s'}} - D_{RA4}^{c,s'})\\
\end{align*}

Now, we will study each term in turn. First, notice that, using the central limit theorem, 
\begin{align*}
    \mathbb{E}_n[G_gS_s] - \mathbb{E}[G_gS_s] & = O_p(n^{-1/2}) \\
    \Longrightarrow     (\mathbb{E}_n[G_gS_s] - \mathbb{E}[G_gS_s])^2 & = O_p(n^{-1}) = o_p(n^{-1/2})\\
    \mathbb{E}_n[\frac{G_gS_s}{\mathbb{E}[G_gS_s]^2}] - \mathbb{E}[\frac{G_gS_s}{\mathbb{E}[G_gS_s]^2}] & = O_p(n^{-1/2}) = o_p(1)
\end{align*}

Using a second-order Taylor expansion of $\widehat{D}_1^{g,s}$ around $\mathbb{E}[G_gS_s]$,
\begin{align*}
    \widehat{D_1^{g,s}}  = & \mathbb{E}_n \Big[ \frac{G_gS_s}{\mathbb{E}_n[G_gS_s]} (Y_t - Y_{g-1})\Big] \\
    = & \mathbb{E}_n \Big[ \frac{G_gS_s}{\mathbb{E}[G_gS_s]} (Y_t - Y_{g-1})\Big] - (\mathbb{E}_n[G_gS_s] - \mathbb{E}[G_gS_s])\mathbb{E}_n \Big[\frac{G_gS_s}{\mathbb{E}[G_gS_s]^2}(Y_t - Y_{g-1}) \Big] \\
    & + O_p((\mathbb{E}_n[G_gS_s] - \mathbb{E}[G_gS_s])^2) \\
    = & \mathbb{E}_n \Big[ \frac{G_gS_s}{\mathbb{E}[G_gS_s]} (Y_t - Y_{g-1})\Big] - (\mathbb{E}_n[G_gS_s] - \mathbb{E}[G_gS_s])\mathbb{E}_n \Big[\frac{G_gS_s}{\mathbb{E}[G_gS_s]^2}(Y_t - Y_{g-1}) \Big] + o_p(n^{-1/2})
\end{align*}
Then, we have 
\begin{align*}
    & \sqrt{n}(\widehat{D_1^{g,s}} - D_1^{g,s})  \\
     = & \sqrt{n}\Big(\mathbb{E}_n \Big[ \frac{G_gS_s}{\mathbb{E}[G_gS_s]} (Y_t - Y_{g-1})\Big] \\
    & - (\mathbb{E}_n[G_gS_s] - \mathbb{E}[G_gS_s])\mathbb{E}_n \Big[\frac{G_gS_s}{\mathbb{E}[(G_gS_s)]^2}(Y_t - Y_{g-1}) \Big] - \mathbb{E}\Big[ \frac{G_gS_s}{\mathbb{E}[G_gS_s]}(Y_t - Y_{g-1})\Big]\Big) + o_p(1) \\
     =& \sqrt{n}\Big( \mathbb{E}_n \Big[ \frac{G_gS_s}{\mathbb{E}[G_gS_s]} (Y_t - Y_{g-1}) - \frac{G_gS_s}{\mathbb{E}[G_gS_s]}\mathbb{E}\Big[ \frac{G_gS_s}{\mathbb{E}[G_gS_s]}(Y_t - Y_{g-1})\Big]\Big] \Big) + o_p(1)  \\
     = & \sqrt{n}\mathbb{E}_n \Big[ w_1(G_g, S_s)\Big(Y_t - Y_{g-1} - \mathbb{E}[w_1(G_g, S_s)(Y_t - Y_{g-1})]\Big) \Big]+ o_p(1) 
\end{align*}

The remaining $\widehat{D_1}$ terms follow similarly, but require an additional step due to estimation of the propensity score models. For example, repeating the same process by expanding $\sqrt{n}(\widehat{D_1^{g,s'}}- D_1^{g,s'})$ around $\mathbb{E}[G_gS_{s'}]$, we can find that 
\begin{align*}
\sqrt{n}(\widehat{D_1^{g,s'}} - D_1^{g,s'}) = \sqrt{n}\mathbb{E}_n \Big[  \Tilde{w_2}(\widehat\theta_{g})  \Big(Y_t - Y_{g-1} - \mathbb{E}[w_2(\theta_{g}^*)(Y_t - Y_{g-1})] \Big) \Big]+ o_p(1) 
\end{align*}
where 
\begin{align*}
    \Tilde{w_2}(\widehat\theta_{g}) = \frac{G_g S_{s'} e_{g,s}(X; \widehat\theta_{g}) }{e_{g,s'}(X; \widehat\theta_{g}) } \Big/ \mathbb{E}\Big[\frac{e_{g,s}(X; \theta_{g}^*) G_g S_{s'}}{e_{g,s'}(X; \theta_{g}^*) }  \Big]
\end{align*}

Then, we can Taylor-expand around the pseudo-true $\theta_{g}^*$ to write 
\begin{align*}
\sqrt{n}(\widehat{D_1^{g,s'}} - D_1^{g,s'}) & = \sqrt{n}\mathbb{E}_n \Big[ Y_t - Y_{g-1} - \mathbb{E}[w_2(\theta_{g}^*)(Y_t - Y_{g-1})] \Big] \\
& + \sqrt{n}(\widehat \theta_{g} - \theta_{g}^*)' \mathbb{E}_n \Big[\dot w_{2}(\theta_{g}^*) \Big(Y_t - Y_{g-1}  - \mathbb{E}[w_2(\theta_{g}^*)(Y_t - Y_{g-1}) ] \Big) \Big] \\
& + o_p(1) 
\end{align*}
where $\dot w_{2}(\theta_{g}^*)$ is the derivative of $\Tilde{w_2}(\widehat\theta_{g})$ with respect to $\theta_g$ evaluated at $\theta_g^*$. 

By the regularity assumptions on the propensity score models, we have 
\begin{align*}
    \sqrt{n}(\widehat \theta_{g} - \theta_{g}^*) & = \sqrt{n}\mathbb{E}_n [l_{e}(G_g, S_s, X; \theta_{g}^*)] + o_p(1) \\
\end{align*}

For notational simplicity, these will be written $l_{e}( \theta_{g}^*)$ 
Next, combining and rearranging terms,
{\footnotesize
\begin{align*}
\sqrt{n}&(\widehat{D_1^{g,s'}} - D_1^{g,s'}) =\\
& \sqrt{n}\mathbb{E}_n \Big[w_2(\theta_{g}^*) \Big(Y_t - Y_{g-1} - \mathbb{E}[w_2(\theta_{g}^*)(Y_t - Y_{g-1})] \Big) \Big] \\
& + \sqrt{n}\mathbb{E}_n \Big[l_{e}(\theta_{g}^*)'  \mathbb{E}\Big[\dot w_{2}(\theta_{g}^*) \Big(Y_t - Y_{g-1} - \mathbb{E}[w_2(\theta_{g}^*)(Y_t - Y_{g-1}) ] \Big)\Big] \Big] + o_p(1)  \\
    \end{align*}
}

Next, we can use the same arguments to show that 
\begin{align*}
\sqrt{n}&(\widehat{D_1^{c,s}} - D_1^{c,s}) =\\
 & \sqrt{n} \mathbb{E}_n \Big[ w_3^c(\theta_{g}^*) \Big(Y_t - Y_{g-1} - \mathbb{E}[w_3^c(\theta_{g}^*)(Y_t - Y_{g-1})] \Big)  \\
& + l_{e}(\theta_{g}^*)' \mathbb{E} \Big[ \dot w_3^c(\theta_{g}^*) \Big(Y_t - Y_{g-1} - \mathbb{E}[w_3^c(\theta_{g}^*)(Y_t - Y_{g-1})] \Big) \Big]  
 \Big]+ o_p(1)  \\
\sqrt{n}&(\widehat{D_1^{c,s'}} - D_1^{c,s'}) =\\
 & \sqrt{n} \mathbb{E}_n \Big[ w_4^c(\theta_{g}^*) \Big(Y_t - Y_{g-1} - \mathbb{E}[w_4^c(\theta_{g}^*)(Y_t - Y_{g-1})] \Big)  \\
& + l_{e}(\theta_{g}^*)' \mathbb{E} \Big[ \dot w_4^c(\theta_{g}^*) \Big(Y_t - Y_{g-1} - \mathbb{E}[w_4^c(\theta_{g}^*)(Y_t - Y_{g-1})] \Big) \Big]  
\Big]+ o_p(1) 
\end{align*}
with terms analogous to the above. 

Now, we can address the RA components. Using the same argument by Taylor expanding around $\mathbb{E}[G_gS_s]$ as above, we can show 
\begin{align*}
\sqrt{n}(\widehat{D_{RA1}^{g,s}} - D_{RA1}^{g,s}) = \sqrt{n}\mathbb{E}_n \Big[w_1(G_g, S_s)\Big( m_{g,t}^{s'}(X; \widehat\beta_{g,s'}) - \mathbb{E}[w_1(G_g, S_s) m_{g,t}^{s'}(X; \beta_{g,s'}^*)]\Big) \Big] + o_p(1) 
\end{align*}
Using a second order Taylor expansion around $\beta^*$ and the regularity conditions on $m$, 
\begin{align*}
\sqrt{n}(\widehat{D_{RA1}^{g,s}} - D_{RA1}^{g,s}) = & \sqrt{n}\mathbb{E}_n \Big[w_1\Big( m_{g,t}^{s'}(X; \beta_{g,s'}^*) - \mathbb{E}[w_1 m_{g,t}^{s'}(X; \beta_{g,s'}^*)]\Big) \Big] \\
& + \sqrt{n}(\widehat \beta_{g,s'} - \beta_{g,s'}^*)' \mathbb{E}_n [w_1\dot m_{g,t}^{s'}(X; \beta_{g,s'}^*) ] + o_p(1) \\
 = & \sqrt{n}\mathbb{E}_n \Big[w_1 \Big( m_{g,t}^{s'}(X; \beta_{g,s'}^*) - \mathbb{E}[w_1 m_{g,t}^{s'}(X; \beta_{g,s'}^*)]\Big) \Big] \\
& + \sqrt{n}\mathbb{E}_n [l_{m_{g,t}^{s'}}(\beta_{g,s'}^*)' \mathbb{E}[w_1 \dot m_{g,t}^{s'}(X; \beta_{g,s'}^*) ]] + o_p(1) \\
 = & \sqrt{n} \mathbb{E}_n \Big[w_1 \Big( m_{g,t}^{s'}(X; \beta_{g,s'}^*) - \mathbb{E}[w_1 m_{g,t}^{s'}(X; \beta_{g,s'}^*)]\Big)  \\
& + l_{m_{g,t}^{s'}}(\beta_{g,s'}^*)'  \mathbb{E}[w_1 \dot m_{g,t}^{s'}(X; \beta_{g,s'}^*) ] \Big] + o_p(1) 
\end{align*}
By the same argument, we can obtain:
\begin{align*}
\sqrt{n}(\widehat{D_{RA1}^{c,s}} - D_{RA1}^{c,s}) = & \\
 & \sqrt{n} \mathbb{E}_n \Big[w_1 \Big( m_{c,g,t}^{s}(X; \beta_{c,g,s}^*) - \mathbb{E}[w_1 m_{c,g,t}^{s}(X; \beta_{c,g,s}^*)]\Big)  \\
& + l_{m_{c,g,t}^{s}}(\beta_{c,g,s}^*)'  \mathbb{E}[w_1 \dot m_{c,g,t}^{s}(X; \beta_{c,g,s}^*) ] \Big] + o_p(1) \\
\sqrt{n}(\widehat{D_{RA1}^{c,s'}} - D_{RA1}^{c,s'}) = & \\
  & \sqrt{n} \mathbb{E}_n \Big[w_1 \Big( m_{c,g,t}^{s'}(X; \beta_{c,g,s'}^*) - \mathbb{E}[w_1 m_{c,g,t}^{s'}(X; \beta_{c,g,s'}^*)]\Big)  \\
& + l_{m_{c,g,t}^{s'}}(\beta_{c,g,s'}^*)'  \mathbb{E}[w_1 \dot m_{c,g,t}^{s'}(X; \beta_{c,g,s'}^*) ] \Big] + o_p(1) 
\end{align*}

Next, for $D_{RA2}^{g,s'}$, using the same process, 
\begin{align*}
    \sqrt{n}&(\widehat{D_{RA2}^{g,s'}} - D_{RA2}^{g,s'}) \\
     = & \sqrt{n}\mathbb{E}_n \Big[w_2(\theta_{g}^*)\Big( m_{g,t}^{s'}(X;\beta_{g,s'}^*) - \mathbb{E}[w_2(\theta_{g}^*) m_{g,t}^{s'}(X; \beta_{g,s'}^*)]\Big)   \\
    & + l_{e}(\theta_{g}^*)' \mathbb{E}\Big[\dot w_2(\theta_{g}^*) \Big (m_{g,t}^{s'}(X;\beta_{g,s'}^*) - \mathbb{E}[w_2(\theta_{g}^*)m_{g,t}^{s'}(X;\beta_{g,s'}^*)]  \Big) \Big] \\
    & + l_{m_{g,t}^{s'}}(\beta_{g,s'}^*)'  \mathbb{E}[w_2(\theta_{g}^*) \dot m_{g,t}^{s'}(X; \beta_{g,s'}^*) ] \Big] + o_p(1)
\end{align*}

$D_{RA3}^{c,s}$ and $D_{RA4}^{c,s'}$ follow similarly:
\begin{align*}
    \sqrt{n}&(\widehat{D_{RA3}^{c,s}} - D_{RA3}^{c,s}) \\
     = & \sqrt{n}\mathbb{E}_n \Big[w_3^c(\theta_{g}^*)\Big( m_{c,g,t}^{s}(X;\beta_{c,g,s}^*) - \mathbb{E}[w_3^c(\theta_{g}^*) m_{c,g,t}^{s}(X; \beta_{c,g,s}^*)]\Big)   \\
    & + l_{e}(\theta_{g}^*)' \mathbb{E}\Big[\dot w_3^c(\theta_{g}^*) \Big (m_{c,g,t}^{s}(X;\beta_{c,g,s}^*) - \mathbb{E}[w_3^c(\theta_{g}^*)m_{c,g,t}^{s}(X;\beta_{c,g,s}^*)]  \Big) \Big] \\
    & + l_{m_{c,g,t}^{s}}(\beta_{c,g,s}^*)'  \mathbb{E}[w_3^c(\theta_{g}^*) \dot m_{c,g,t}^{s}(X; \beta_{c,g,s}^*) ] \Big] + o_p(1) \\
    \sqrt{n}&(\widehat{D_{RA4}^{c,s'}} - D_{RA4}^{c,s'}) \\
     = & \sqrt{n}\mathbb{E}_n \Big[w_4^c(\theta_{g}^*)\Big( m_{c,g,t}^{s'}(X;\beta_{c,g,s'}^*) - \mathbb{E}[w_4^c(\theta_{g}^*) m_{c,g,t}^{s'}(X; \beta_{c,g,s'}^*)]\Big)   \\
    & + l_{e}(\theta_{g}^*)' \mathbb{E}\Big[\dot w_4^c(\theta_{g}^*) \Big (m_{c,g,t}^{s'}(X;\beta_{c,g,s'}^*) - \mathbb{E}[w_4^c(\theta_{g}^*)m_{c,g,t}^{s'}(X;\beta_{c,g,s'}^*)]  \Big) \Big] \\
    & + l_{m_{c,g,t}^{s'}}(\beta_{c,g,s'}^*)'  \mathbb{E}[w_4^c(\theta_{g,s}^*, \theta_{c,s'}^*) \dot m_{c,g,t}^{s'}(X; \beta_{c,g,s'}^*) ] \Big] + o_p(1)
\end{align*}

Define the estimation term arising from the estimation of the propensity scores as
\begin{align*}
\eta_{g,t}^{c, est(e)} & (Y,G,S,X; \theta^*, \beta^*, w_2) = \\
 & l_{e}(\theta_{g}^*)' \mathbb{E}\Big[ \dot w_2(\theta_{g}^*) \Big(\Delta Y - m_{g,t}^{s'}(X; \beta_{g,s'}^*) -  \mathbb{E}[w_2(\theta_{g}^*)(\Delta Y - m_{g,t}^{s'}(X; \beta_{g,s'}^*))] \Big) \Big]
\end{align*}

and analogously for 
$\eta_{g,t}^{c, est_e}(Y,G,S,X; \theta^*, \beta^*, w_3)$, and so on, with $\Delta Y = Y_t - Y_{g-1}$.

Also define the estimation terms arising from the estimation of the outcome models as 
\begin{align*}
\eta_{g,t}^{c, est(m_{g,t}^{s'})}(Y,G,S,X; \theta^*, \beta^*) & = l_{m_{g,t}^{s'}}(\beta_{g,s'}^*)'  \mathbb{E}[(w_1 - w_2(\theta_{g}^*)) \dot m_{g,t}^{s'}(X; \beta_{g,s'}^*) ] \\
\eta_{c,g,t}^{c, est(m_{c,g,t}^{s})}(Y,G,S,X; \theta^*, \beta^*) & = l_{m_{c,g,t}^{s}}(\beta_{c,g,s}^*)'  \mathbb{E}[(w_1 - w_3^c(\theta_{g}^*)) \dot m_{c,g,t}^{s}(X; \beta_{c,g,s}^*) ] \\
\eta_{c,g,t}^{c, est(m_{c,g,t}^{s'})}(Y,G,S,X; \theta^*, \beta^*) & = l_{m_{c,g,t}^{s'}}(\beta_{c,g,s'}^*)'  \mathbb{E}[(w_1 - w_4^c(\theta_{g}^*)) \dot m_{c,g,t}^{s'}(X; \beta_{c,g,s'}^*) ] \\
\end{align*}

Putting everything together and combining terms, 
\begin{align*}
\sqrt{n} & (\widehat{CDATT}_{s-s'}^{DR,c}(g,t)  - {CDATT}_{s-s'}^{DR,c}(g,t)) = \\
 & \sqrt{n}\mathbb{E}_n \Big[ w_1(G_g, S_s) \Big(\Delta Y - m_{g,t}^{s'}(X;\beta^*_{g,s'}) - m_{c,g,t}^s(X;\beta^*_{c,g,s}) + m_{c,g,t}^{s'}(X;\beta^*_{c,g,s'}) \\
& \quad -\mathbb{E}\Big[w_1(G_g, S_s)\Big(\Delta Y - m_{g,t}^{s'}(X;\beta^*_{g,s'}) - m_{c,g,t}^s(X;\beta^*_{c,g,s}) + m_{c,g,t}^{s'}(X;\beta^*_{c,g,s'})\Big)\Big] \Big) \\
& -   \eta_{g,t}^{c, est(m_{g,t}^{s'})}(Y,G,S,X; \theta^*, \beta^*) - \eta_{c,g,t}^{c, est(m_{c,g,t}^{s})}(Y,G,S,X; \theta^*, \beta^*) \\
&  +  \eta_{g,t}^{c, est(m_{c,g,t}^{s'})}(Y,G,S,X; \theta^*, \beta^*)  \\
- &  w_2(\theta_{g}^*) \Big(\Delta Y - m_{g,t}^{s'}(X;\beta_{g,s'}^*) - \mathbb{E}[w_2(\theta_{g}^*)(\Delta Y - m_{g,t}^{s'}(X; \beta_{g,s'}^*))] \Big) \\
& - \eta_{g,t}^{c, est(e)}(Y,G,S,X; \theta^*, \beta^*, w_2)
\\
- &  w_3^c(\theta_{g}^*) \Big(\Delta Y - m_{c,g,t}^{s}(X;\beta_{c,g,s}^*) - \mathbb{E}[w_3^c(\theta_{g}^*)(\Delta Y - m_{c,g,t}^{s}(X; \beta_{c,g,s}^*))] \Big) \\
& - \eta_{g,t}^{c, est(e)}(Y,G,S,X; \theta^*, \beta^*, w_3)  
\\
+ &  w_4^c(\theta_{g}^*) \Big(\Delta Y - m_{c,g,t}^{s'}(X;\beta_{c,g,s'}^*) - \mathbb{E}[w_4^c(\theta_{g}^*)(\Delta Y - m_{c,g,t}^{s'}(X; \beta_{c,g,s'}^*))] \Big) \\
& + \eta_{g,t}^{c, est(e)}(Y,G,S,X; \theta^*, \beta^*, w_4) 
\Big]   + o_p(1) \\
\equiv & \sqrt{n} \mathbb{E}_n \Big[ \eta_{g,t}^{DR,c}(Y,G,S,X; \theta^*, \beta^*, CDATT)\Big] + o_p(1)
\end{align*}

This establishes the asymptotic linear representation of $\widehat{CDATT}_{s-s'}^{DR,c}(g,t)$ for each $g$ and $t$. The result for $\widehat{CDATT}_{t \geq g}^{DR,c}$ follows by applying the Lindeberg-Lévy central limit theorem. 
\end{proofn}
\newpage 
\section{Results for repeated cross-section data}
\label{appendix_rc}

Results for the case of repeated cross-section data appear in this appendix. In this case, we observe, for each observation at time $t$, $(Y, G, W, S, X, t)$. First, I describe modified assumptions needed for this case. Then, I discuss the identification and estimation of the doubly-robust estimator for the $CDATT$ in this case.

\subsection{Assumptions}
Following \textcite{abadieSemiparametricDifferenceinDifferencesEstimators2005, santannaDoublyRobustDifferenceindifferences2020, callawayDifferenceinDifferencesMultipleTime2021}, I assume that treatment status, subgroup status, and covariates do not vary with time.\footnote{ A case where covariates are time-varying is addressed by \textcite{caetanoDifferenceDifferencesTimeVarying2024}.}

\begin{assumption}[Sampling: repeated-cross section.]
\label{rc_sampling}
For each $t \in \{1, \dots, \mathcal{T}\}$, the draws from $(Y_t, G, W, S, X)$ are independently and identically distributed and invariant to $t$. 
\end{assumption}
This assumption is adapted from Assumption B.1 in \textcite{callawayDifferenceinDifferencesMultipleTime2021} and is discussed there.

\subsection{Identification of \texorpdfstring{$CDATT_{s-s'}$}{CDATT} with repeated cross-sections }
\setcounter{proofn}{0}
\setcounter{proposition}{0}

Define $T_a$ to be a binary variable that equals 1 if $t = a$ and 0 otherwise. Let $c \in \{nev, ny\}$ represent the comparison group of interest, so that $C_{nev}$ is a binary variable that equals 1 if a unit belongs to the never-treated group and $C_{ny}$ is a binary variable which equals 1 if a unit belongs to the not-yet-treated group. 

Define the following weights 
\begin{align*}
w_1^{rc}(G_g, S_s, t) & = \frac{T_t G_gS_s}{\mathbb{E}[T_t G_gS_s]} \\
w_2^{rc}(G_g, S_{s'}, X,t) & = \frac{T_t G_gS_{s'} \pi_{g,s}(X)}{ \pi_{g,s'}(X)} \Big/ \mathbb{E} \Big[\frac{T_t G_gS_{s'} \pi_{g,s}(X)}{\pi_{g,s'}(X)} \Big] \\
w_3^{c, rc}(C_c, S_s, X, t) & = \frac{T_t C_c S_{s} \pi_{g,s}(X)}{\pi_{c,s}(X)} \Big/ \mathbb{E} \Big[\frac{T_t C_c S_{s} \pi_{g,s}(X)}{\pi_{c,s}(X)}  \Big]\\
w_4^{c,rc}(C_c, S_{s'}, X, t) & = \frac{T_t C_c S_{s'} \pi_{g,s}(X)}{\pi_{c,s'}(X)} \Big/ \mathbb{E} \Big[\frac{T_t C_c S_{s'} \pi_{g,s}(X)}{\pi_{c,s'}(X)}  \Big] 
\end{align*}
To simplify notation, these will be referred to as $w_1^{rc}(t)$, $w_2^{rc}(t)$, $w_3^{c, rc}(t)$, and $w_4^{c,rc}(t)$, respectively. 

and outcome functions
\begin{align*}
\mu^{rc, \sigma}_{g,t}(X) & = \mathbb{E}[Y | G=g, S=\sigma, X, T=t] \\
\mu^{rc, \sigma}_{c, g, t}(X) & =\mathbb{E}[Y | C_c=1, S=\sigma, X, T=t]  \\
\end{align*}
for $\sigma \in \{s, s'\}$ and for all $t$ and $g$. \\

Then, define
\begin{align*}
CDATT^{DR-RC, c}_{s-s'}(g,t) =&  \mathbb{E} \Big[ \frac{G_g S_s}{\mathbb{E}[G_g S_s]}\Big(\mu^{rc, s}_{g,t}(X) - \mu^{rc, s}_{g,g-1}(X) - \Big(\mu^{rc, s'}_{g,t}(X) - \mu^{rc, s'}_{g,g-1}(X) \Big)\Big)\Big]  \\
 & -  \mathbb{E}\Big[  \frac{G_g S_s}{\mathbb{E}[G_g S_s]} \Big(\mu^{rc, s}_{c,g,t}(X) - \mu^{rc, s}_{c,g,g-1}(X) - \Big(\mu^{rc, s'}_{c,g,t}(X) - \mu^{rc, s'}_{c,g,g-1}(X) \Big)\Big)\Big]  \\ 
 & + \mathbb{E}\Big[w_1^{rc}(t)(Y - \mu^{rc, s}_{g,t}(X)) -  w_1^{rc}(g-1)(Y - \mu^{rc, s}_{g,g-1}(X))\Big] \\
 & - \mathbb{E}\Big[w_2^{rc}(t)(Y - \mu^{rc, s'}_{g,t}(X)) -  w_2^{rc}(g-1)(Y - \mu^{rc, s'}_{g,g-1}(X))\Big] \\
 & - \mathbb{E}\Big[w_3^{c,rc}(t)(Y - \mu^{rc, s}_{c,g,t}(X)) -  w_3^{c,rc}(g-1)(Y - \mu^{rc, s}_{c,g,g-1}(X))\Big] \\
 & + \mathbb{E}\Big[w_4^{c,rc}(t)(Y - \mu^{rc, s'}_{c,g,t}(X)) -  w_4^{c,rc}(g-1)(Y - \mu^{rc, s'}_{c,g,g-1}(X))\Big] 
\end{align*}
This estimator is the natural counterpart to the panel data estimator and follows in a style similar to the difference-in-difference estimators proposed by \textcite{santannaDoublyRobustDifferenceindifferences2020,callawayDifferenceinDifferencesMultipleTime2021}.

\begin{proposition}[Identification of $CDATT_{s-s'}(g,t)$ with observable treatment effect heterogeneity, repeated cross-section] \label{id_satt_stag_cond_rc}

If Assumptions \ref{irreversible}, \ref{no_anticipation_stag}, \ref{cond_parallel_gaps_stag_ny} (applied to the $s'$ potential outcomes), \ref{cond_subgroup_selection_stag}, and \ref{overlap_cdatt} hold for subgroup $s$ and $s'$, then, the $CDATT_{s-s'}(g,t)$ is identified for $t \geq 2$, and 
\begin{align*}
CDATT_{s-s'}(g,t) = CDATT^{DR-RC, ny}_{s-s'}(g,t)
\end{align*}

If Assumptions \ref{irreversible}, \ref{no_anticipation_stag}, \ref{cond_parallel_gaps_stag_nev} (applied to the $s'$ potential outcomes), \ref{cond_subgroup_selection_stag}, and \ref{overlap_cdatt} hold for subgroup $s$ and $s'$, then, the $CDATT_{s-s'}(g,t)$ is identified for $t \geq 2$, and 
\begin{align*}
CDATT_{s-s'}(g,t) = CDATT^{DR-RC, nev}_{s-s'}(g,t)
\end{align*}
\end{proposition}
The proof follows.\\
\begin{proofn} 
To show $CDATT^{DR-RC, c}_{s-s'}(g,t) = CDATT_{s-s'}(g,t)$, first note that

\begin{align*}
    \mathbb{E}&\Big[w_1^{rc}(t)(Y-\mu_{g,t}^{rc,s}(X))\Big] \\
    & = \mathbb{E}\Big[\frac{T_t G_gS_s}{\mathbb{E}[T_t G_gS_s]}(Y-\mathbb{E}[Y | G=g, S=s, X, T=t])\Big]\\
    & = \mathbb{E}\Big[(\mathbb{E}[Y | G=g, S=s, T=t]-\mathbb{E}[Y | G=g, S=s,T=t])\Big] = 0    
\end{align*}
by using the law of iterated expectations. The result follows analogously to show 
\begin{align*}
     \mathbb{E}\Big[w_1^{rc}(g-1)(Y-\mu_{g,g-1}^{rc,s}(X))\Big] & = 0
\end{align*}
Similarly, 
\begin{align*}
    \mathbb{E}&\Big[w_2^{rc}(t)(Y-\mu_{g,t}^{rc,s'}(X))\Big] \\
    & = \mathbb{E}\Big[\frac{T_t G_gS_{s'} \pi_{g,s}(X)}{ \pi_{g,s'}(X)} \Big/ \mathbb{E} \Big[\frac{T_t \mathbb{E}[G_gS_{s'}\pi_{g,s}(X) | X, T=t] }{\pi_{g,s'}(X)} \Big](Y-\mathbb{E}[Y | G=g, S=s', X,T=t])\Big] \\
    & = \mathbb{E}\Big[\frac{T_t G_gS_{s'} \pi_{g,s}(X)}{ \pi_{g,s'}(X)} \Big/ \mathbb{E} \Big[T_t G_g S_{s}\Big] (Y-\mathbb{E}[Y | G=g, S=s', X, T=t])\Big] \\
    & = \mathbb{E}\Big[(\mathbb{E}[Y | G=g, S=s', X, T=t]-\mathbb{E}[Y | G=g, S=s', X, T=t])\Big] \\
    & = 0
\end{align*}
where the second and third equalities use Assumption \ref{rc_sampling} (the time invariance with respect to $T$) and the fourth uses the law of iterated expectations. 

The analogous results show that 
\begin{align*}
 \mathbb{E}\Big[w_2^{rc}(g-1)(Y-\mu_{g,g-1}^{rc,s'}(X))\Big] & = 0\\
  \mathbb{E}\Big[w_3^{c,rc}(t)(Y-\mu_{c,g,t}^{rc,s}(X))\Big] & = 0\\
  \mathbb{E}\Big[w_3^{c,rc}(g-1)(Y-\mu_{c,g,g-1}^{rc,s}(X))\Big] & = 0 \\
  \mathbb{E}\Big[w_4^{c,rc}(t)(Y-\mu_{c,g,t}^{rc,s'}(X))\Big] & = 0\\
  \mathbb{E}\Big[w_4^{c,rc}(g-1)(Y-\mu_{c,g,g-1}^{rc,s'}(X))\Big] & = 0\\
\end{align*}

Next, we can see that 
\begin{align*}
    \mathbb{E} & \Big[ \frac{G_g S_s}{\mathbb{E}[G_g S_s]}\Big(\mu^{rc, s}_{g,t}(X) - \mu^{rc, s}_{g,g-1}(X)) \Big) \Big]  \\
    & = \mathbb{E} \Big[ \frac{G_g S_s}{\mathbb{E}[G_g S_s]}\Big(\mathbb{E}[Y | G=g, S=s, X, T=t]  - \mathbb{E}[Y | G=g, S=s, X, T=g-1] \Big) \Big] \\
    & = \mathbb{E} \Big[ \frac{G_g S_s}{\mathbb{E}[G_g S_s]}\Big(\mathbb{E}[Y_t(g,s) | G=g, S=s, X, T=t]  - \mathbb{E}[Y_{g-1}(\infty,s) | G=g, S=s, X, T=g-1] \Big) \Big] \\
    & = \mathbb{E} \Big[ Y_t(g,s) - Y_{g-1}(\infty,s)  | G=g, S=s \Big]
\end{align*}
where the first equality uses the definition of the outcome functions, the second uses the definition of the potential outcomes, Assumption \ref{irreversible}, and Assumption \ref{no_anticipation_stag}, and the last uses the law of iterated expectations. \\

Following the same process, we can also find that 
\begin{align*}
    \mathbb{E} & \Big[ \frac{G_g S_s}{\mathbb{E}[G_g S_s]}\Big(\mu^{rc, s'}_{g,t}(X) - \mu^{rc, s'}_{g,g-1}(X)) \Big) \Big]  \\
    & = \mathbb{E} \Big[ \frac{G_g S_s}{\mathbb{E}[G_g S_s]}\Big(\mathbb{E}[Y | G=g, S=s', X, T=t]  - \mathbb{E}[Y | G=g, S=s', X, T=g-1] \Big) \Big] \\
    & = \mathbb{E} \Big[ \frac{G_g S_s}{\mathbb{E}[G_g S_s]}\Big(\mathbb{E}[Y_t(g,s') | G=g, S=s', X, T=t]  - \mathbb{E}[Y_{g-1}(\infty,s') | G=g, S=s', X,  T=g-1] \Big) \Big] \\
    & = \mathbb{E} \Big[ \frac{G_g S_s}{\mathbb{E}[G_g S_s]}\Big(\mathbb{E}[Y_t(g,s') | G=g, S=s, X, T=t]  - \mathbb{E}[Y_{g-1}(\infty,s') | G=g, S=s, X, T=g-1] \Big) \Big] \\
    & = \mathbb{E} \Big[ Y_t(g,s') - Y_{g-1}(\infty,s')  | G=g, S=s \Big]
\end{align*}
where the first equality uses the definition of the outcome functions, the second uses the definition of the potential outcomes, Assumption \ref{irreversible}, and Assumption \ref{no_anticipation_stag}, the third uses Assumption \ref{cond_subgroup_selection_stag}, and the last uses the law of iterated expectations. \\

The same process and an application of \ref{cond_parallel_gaps_stag_ny} or \ref{cond_parallel_gaps_stag_nev} gives 
\begin{align*}
    \mathbb{E} & \Big[ \frac{G_g S_s}{\mathbb{E}[G_g S_s]}\Big(\mu^{rc, s}_{c,g,t}(X) - \mu^{rc, s}_{c,g,g-1}(X)) \Big) \Big]  -\mathbb{E} \Big[ \frac{G_g S_s}{\mathbb{E}[G_g S_s]}\Big(\mu^{rc, s'}_{c,g,t}(X) - \mu^{rc, s'}_{c,g,g-1}(X)) \Big) \Big]\\
     & = \mathbb{E} \Big[ Y_t(\infty,s) - Y_{g-1}(\infty,s)  | G_g=1, S=s \Big] - \mathbb{E} \Big[ Y_t(\infty,s') - Y_{g-1}(\infty,s')  | G_g=1, S=s \Big]
\end{align*}

Then, we can see that
\begin{align*}
CD& ATT^{DR-RC, c}_{s-s'}(g,t) = \mathbb{E} \Big[ \frac{G_g S_s}{\mathbb{E}[G_g S_s]}\Big(\mu^{rc, s}_{g,t}(X) - \mu^{rc, s}_{g,g-1}(X) - \Big(\mu^{rc, s'}_{g,t}(X) - \mu^{rc, s'}_{g,g-1}(X) \Big)\Big)\Big]  \\
 & -  \mathbb{E}\Big[  \frac{G_g S_s}{\mathbb{E}[G_g S_s]} \Big(\mu^{rc, s}_{c,g,t}(X) - \mu^{rc, s}_{c,g,g-1}(X) - \Big(\mu^{rc, s'}_{c,g,t}(X) - \mu^{rc, s'}_{c,g,g-1}(X) \Big)\Big)\Big]  \\ 
 & + \mathbb{E}\Big[w_1^{rc}(t)(Y - \mu^{rc, s}_{g,t}(X)) -  w_1^{rc}(g-1)(Y - \mu^{rc, s}_{g,g-1}(X))\Big] \\
 & - \mathbb{E}\Big[w_2^{rc}(t)(Y - \mu^{rc, s'}_{g,t}(X)) -  w_2^{rc}(g-1)(Y - \mu^{rc, s'}_{g,g-1}(X))\Big] \\
 & - \mathbb{E}\Big[w_3^{c,rc}(t)(Y - \mu^{rc, s}_{c,g,t}(X)) -  w_3^{c,rc}(g-1)(Y - \mu^{rc, s}_{c,g,g-1}(X))\Big] \\
 & + \mathbb{E}\Big[w_4^{c,rc}(t)(Y - \mu^{rc, s'}_{c,g,t}(X)) -  w_4^{c,rc}(g-1)(Y - \mu^{rc, s'}_{c,g,g-1}(X))\Big]  \\
  = &  \mathbb{E} \Big[ \frac{G_g S_s}{\mathbb{E}[G_g S_s]}\Big(\mu^{rc, s}_{g,t}(X) - \mu^{rc, s}_{g,g-1}(X) - \Big(\mu^{rc, s'}_{g,t}(X) - \mu^{rc, s'}_{g,g-1}(X) \Big)\Big)\Big]  \\
 & -  \mathbb{E}\Big[  \frac{G_g S_s}{\mathbb{E}[G_g S_s]} \Big(\mu^{rc, s}_{c,g,t}(X) - \mu^{rc, s}_{c,g,g-1}(X) - \Big(\mu^{rc, s'}_{c,g,t}(X) - \mu^{rc, s'}_{c,g,g-1}(X) \Big)\Big)\Big] \\
= & \mathbb{E} \Big[ Y_t(g,s) - Y_{g-1}(\infty,s)  | G=g, S=s \Big] -  \mathbb{E} \Big[ Y_t(g,s') - Y_{g-1}(\infty,s')  | G=g, S=s \Big]  \\
 & -  \Big(\mathbb{E} \Big[ Y_t(\infty,s) - Y_{g-1}(\infty,s)  | G=g, S=s \Big] - \mathbb{E} \Big[ Y_t(\infty,s') - Y_{g-1}(\infty,s')  | G=g, S=s \Big]\Big)  \\
= & CDATT_{s-s'}(g,t)
\end{align*}

\end{proofn}

\subsection{Estimation and inference for repeated cross-sections}
As in the panel case, consider parametric models for the propensity scores $e_{g,s}, e_{g,s'}, e_{c,s}$ and $e_{c,s'}$ that take the form $e_{a,b}(X; \theta_{g})$ for $a \in \{g,c\}$ and $b \in \{s,s'\}$. Consider also parametric models for the outcome functions of the form $m_{a,\tau}^{rc,b}(X; \beta_{a, \tau}^{b})$ for $a \in\{g,(c,g)\}$, $b \in \{s, s'\}$, and $\tau \in \{1, \dots, \mathcal{T} \}$. These parametric models are estimated by $e_{a,b}(X; \widehat\theta_{g})$ and $m_{a,\tau}^{rc,b}(X; \widehat\beta_{a,\tau}^b)$, respectively. 

Let $\mathbb{E}_n [X] = \frac{1}{n} \sum_{i=1}^n X_i$. Then, the following estimates $CDATT^{DR-RC,c}_{s-s'}(g,t)$, for $c \in \{ny, nev\}$ : 
\begin{align*}
\widehat{CDATT}&^{DR-RC, c}_{s-s'}(g,t) \\
= & \mathbb{E}_n \Big[ \frac{G_g S_s}{\mathbb{E}_n[G_g S_s]}\Big(m^{rc, s}_{g,t}(X; \widehat \beta_{g,t}^s) - m^{rc, s}_{g,g-1}(X; \widehat \beta_{g,g-1}^{s}) - \Big(m^{rc, s'}_{g,t}(X; \widehat \beta_{g,t}^{s'}) - m^{rc, s'}_{g,g-1}(X; \widehat \beta_{g,g-1}^{s'}) \Big)\Big)\Big]  \\
 & -  \mathbb{E}_n\Big[  \frac{G_g S_s}{\mathbb{E}_n[G_g S_s]} \Big(m^{rc, s}_{c,g,t}(X; \widehat \beta_{c,g,t}^s) - m^{rc, s}_{c,g,g-1}(X; \widehat \beta_{c,g, g-1}^s) - \Big(m^{rc, s'}_{c,g,t}(X; \widehat \beta_{c,g,t}^{s'}) - m^{rc, s'}_{c,g,g-1}(X; \widehat \beta_{c, g,g-1}^{s'}) \Big)\Big)\Big]  \\ 
 & + \mathbb{E}_n\Big[\widehat{w_1}^{rc}(t)(Y - m^{rc, s}_{g,t}(X; \widehat \beta_{g,t}^s)) -  \widehat{w_1}^{rc}(g-1)(Y - m^{rc, s}_{g,g-1}(X; \widehat \beta_{g, g-1}^{s}))\Big] \\
 & - \mathbb{E}_n\Big[\widehat{w_2}^{rc}(t;\widehat\theta_{g})(Y - m^{rc, s'}_{g,t}(X; \widehat \beta_{g,t}^{s'})) -  \widehat{w_2}^{rc}(g-1;\widehat\theta_{g})(Y - m^{rc, s'}_{g,g-1}(X; \widehat \beta_{g,g-1}^{s'}))\Big] \\
 & - \mathbb{E}_n\Big[\widehat{w_3}^{c,rc}(t; \widehat\theta_{g} )(Y - m^{rc, s}_{c,g,t}(X; \widehat \beta_{c,g,t}^s)) -  \widehat{w_3}^{c,rc}(g-1; \widehat\theta_{g})(Y - m^{rc, s}_{c,g,g-1}(X; \widehat\beta_{c,g,g-1}^s))\Big] \\
 & + \mathbb{E}_n\Big[\widehat{w_4}^{c,rc}(t; \widehat\theta_{g})(Y - m^{rc, s'}_{c,g,t}(X; \widehat \beta_{c,g,t}^{s'})) -  \widehat{w_4}^{c,rc}(g-1; \widehat\theta_{g})(Y - m^{rc, s'}_{c,g,g-1}(X; \widehat \beta_{c,g,g-1}^{s'}))\Big]  
\end{align*}
where 
\begin{align*}
\widehat w_1^{rc}(t) & = \frac{T_t G_gS_s}{\mathbb{E}_n[T_t G_gS_s]} \\
\widehat w_2^{rc}(t;\widehat\theta_{g}) & = \frac{T_t G_gS_{s'} e_{g,s}(X; \widehat\theta_{g})}{e_{g,s'}(X;\widehat\theta_{g})} \Big/ \mathbb{E}_n \Big[\frac{T_t G_gS_{s'} e_{g,s}(X; \widehat\theta_{g})}{e_{g,s'}(X;\widehat\theta_{g})} \Big] \\
\widehat w_3^{c, rc}(t;\widehat\theta_{g}) & = \frac{T_t C_c S_{s} e_{g,s}(X; \widehat\theta_{g})}{e_{c,s}(X;\widehat\theta_{g})} \Big/ \mathbb{E}_n \Big[\frac{T_t C_c S_{s} e_{g,s}(X; \widehat\theta_{g})}{e_{c,s}(X;\widehat\theta_{g})}  \Big]\\
\widehat w_4^{c,rc}(t;\widehat\theta_{g}) & = \frac{T_t C_c S_{s'} e_{g,s}(X; \widehat\theta_{g})}{e_{c,s'}(X;\widehat\theta_{g})} \Big/ \mathbb{E}_n \Big[\frac{T_t C_c S_{s'} e_{g,s}(X; \widehat\theta_{g})}{e_{c,s'}(X;\widehat\theta_{g})}  \Big]
\end{align*}

The asymptotic properties of this estimator follow from a process very similar to the panel data case.

\newpage

\section{Identification of \texorpdfstring{$ATT$}{ATT} in a triple difference design}\label{appendix_unaffected_subgroup}
\setcounter{figure}{0}    
\setcounter{table}{0}   
\setcounter{assumption}{0}    
\renewcommand*\theassumption{\Alph{section}.\arabic{assumption}}

\FloatBarrier
As described above, the triple difference $DATT_{s-s'}$ is equivalent to the difference in the $ATT$ for subgroup $s$ and the $ATT$ for subgroup $s'$. However, individually, neither the $ATT$ for subgroup $s$ nor the $ATT$ for subgroup $s'$ can be identified under the parallel gaps assumption. 

To recover the $ATT$ using a triple difference design, we must impose at least one assumption on the treated counterfactual outcomes. One common approach would be to assume that one subgroup was unaffected by the treatment \parencite{ortiz-villavicencioBetterUnderstandingTriple2025,oldenTripleDifferenceEstimator2022}. For example, to recover the impact of a policy on married individuals, we might assume that unmarried individuals would be unaffected by the policy. 

\begin{assumption}[Unaffected subgroup]
    \label{control_subgroup}
    For some subgroup $s'$, 
    $$ \mathbb{E}[Y_t(g) - Y_t(\infty) | G=g, S=s'] = 0 $$
\end{assumption}

Under Assumption \ref{control_subgroup} for subgroup $s'$, $DATT_{s-s'} = \mathbb{E}[Y_t(g) - Y_t(\infty) | G=g, S=s]$. This assumption is made implicitly by \textcite{oldenTripleDifferenceEstimator2022,ortiz-villavicencioBetterUnderstandingTriple2025}, as they assume that $Y_t(\infty)$ is observed for treated units with $S=s'$, that is, they assume that units in $s'$ are unaffected by the treatment. 

An important violation of the unaffected subgroup assumption arises in many practical applications when there is the possibility of spillovers between subgroups. For example, a researcher using a design like that in \textcite{gruberIncidenceMandatedMaternity1994} might assume that unmarried men are unaffected by a policy requiring insurance coverage of the costs of childbirth. However, this will not be true if employers substitute between the groups of workers. 

Although the unaffected subgroup assumption is not directly testable, some authors have attempted to assess evidence of spillovers between subgroups eligible and ineligible for a program. For example, \textcite{desiereHowEffectiveAre2022} examine the impacts of a hiring subsidy on job outcomes by comparing a subgroup of older workers (age 45-48) who were eligible for the subsidy and a subgroup of younger workers who were not (age 40-43). The unaffected subgroup assumption will be violated if employers substitute between the older and younger workers. The authors test for evidence of this by comparing impacts on these subgroups with a subgroup of even younger workers. If employers substitute between workers around the eligibility cutoff, then the just-ineligible workers would experience different effects than the even-younger workers. 

Alternatively, if Assumption \ref{control_subgroup} is not satisfied, the $ATT$ for subgroup $s$ can be bounded under an assumption limiting the magnitude of the $ATT$ for subgroup $s'$. For example, this might be relevant in cases where it is believed that the comparison subgroup is not unaffected but is only affected a small amount.  

\newpage
\section{Additional results from simulation study}
\setcounter{figure}{0}    
\setcounter{table}{0}    
\FloatBarrier

Tables \ref{tab:sim_study_param_ps} and \ref{tab:sim_study_mod_y} present the calibrated parameters used in the main simulation study described in section \ref{section:simulation_design}. 

\begin{table}[!h]
    \caption{Coefficients for propensity score models in simulation study}
    \label{tab:sim_study_param_ps}
    \centering
    \resizebox{0.4\textwidth}{!}{
    \begin{tabular}{lcc}
    \hline \\
    \multicolumn{3}{l}{$\mathbf{P(G=1, S=1)}$} \\ \hline
    % latex table generated in R 4.4.3 by xtable 1.8-4 package
% Fri Sep  4 02:35:01 2026
 & Estimate & Std. Error \\ 
  \hline
Intercept & -6.676 & 0.790 \\ 
  Education & -0.134 & 0.014 \\ 
  Non-white & -0.202 & 0.089 \\ 
  Age & 0.533 & 0.059 \\ 
  Age$^2$ & -0.006 & 0.001 \\ 
  Union & 0.527 & 0.066 \\ 
  White collar & 1.297 & 0.064 \\ 
   \hline
 \\
    \multicolumn{3}{l}{$\mathbf{P(G=1, S=0)}$} \\ \hline
    % latex table generated in R 4.4.3 by xtable 1.8-4 package
% Fri Sep  4 02:35:01 2026
 & Estimate & Std. Error \\ 
  \hline
Intercept & -0.640 & 0.934 \\ 
  Education & 0.025 & 0.014 \\ 
  Non-white & -0.067 & 0.105 \\ 
  Age & -0.011 & 0.070 \\ 
  Age$^2$ & 0.001 & 0.001 \\ 
  Union & 0.432 & 0.076 \\ 
  White collar & 0.316 & 0.075 \\ 
   \hline
 \\
    \multicolumn{3}{l}{$\mathbf{P(G=\infty, S=1)}$} \\ \hline
    % latex table generated in R 4.4.3 by xtable 1.8-4 package
% Fri Sep  4 02:35:01 2026
 & Estimate & Std. Error \\ 
  \hline
Intercept & -6.552 & 0.767 \\ 
  Education & -0.207 & 0.014 \\ 
  Non-white & -0.203 & 0.089 \\ 
  Age & 0.626 & 0.058 \\ 
  Age$^2$ & -0.008 & 0.001 \\ 
  Union & 0.154 & 0.066 \\ 
  White collar & 1.015 & 0.065 \\ 
   \hline \hline
 \\
    % latex table generated in R 4.4.3 by xtable 1.8-4 package
% Fri Sep  4 02:35:01 2026
 $P(G=1,S=1)$ & 0.275 \\ 
  $P(G=1,S=0)$ & 0.204 \\ 
  $P(G=\infty,S=1)$ & 0.401 \\ 
  $P(G=\infty,S=0)$ & 0.120 \\ 
   \hline \\ 
    \end{tabular}
}
{\\ \raggedright \textbf{Notes:} This table presents the parameters used in the generation of propensity scores in the simulation study presented above. The parameters are estimated from a logit regression of each outcome (eg, an indicator for a unit having $G=1$ and $S=1$, relative to being in the comparison subgroup) on the given covariates in the original, unaltered dataset used to calibrate the simulation. \par }
\end{table}

\begin{table}[!h]
    \caption{Coefficients for outcome models in simulation study}
    \label{tab:sim_study_mod_y}
    \centering
    \resizebox{0.5\textwidth}{!}{
    \begin{tabular}{lcc}\hline
    % latex table generated in R 4.4.3 by xtable 1.8-4 package
% Fri Sep  4 02:35:01 2026
 & Estimate & Std. Error \\ 
  \hline
Intercept & -1.185 & 0.065 \\ 
  Education & 0.051 & 0.001 \\ 
  Non-white & -0.092 & 0.008 \\ 
  Age & 0.117 & 0.005 \\ 
  Age$^2$ & -0.002 & 0.000 \\ 
  Union & 0.259 & 0.006 \\ 
  White collar & 0.071 & 0.006 \\ 
   \hline \hline
 \\
    % latex table generated in R 4.4.3 by xtable 1.8-4 package
% Fri Sep  4 02:35:01 2026
 SD of residual & 0.408 &  \\ 
  SD of $\hat Y$ & 0.876 &  \\ 
   \hline %
    \end{tabular}
}
{\\ \raggedright \textbf{Notes:} This table presents the parameters used in the generation of propensity scores in the simulation study presented above. The parameters are estimated from regressing the log of hourly wages on the given covariates in the original, unaltered dataset used to calibrate the simulation. SD of residual represents the standard deviations of residuals from this regression and SD of $\hat Y$ represents the standard deviation of predicted values from this dataset. \par }
\end{table}

\FloatBarrier
\subsection{Overlap and trimming}\label{appendix_overlap}

In order to understand the importance of the overlap assumption, I present a series of results which highlight issues that arise when overlap is limited and demonstrate the performance of the proposed estimators when using trimming as a method for improving overlap. 

\begin{figure}[h]
    \caption{Distribution of estimated propensity scores, main simulation specification}
    \centering
    \includegraphics[width=0.5\linewidth]{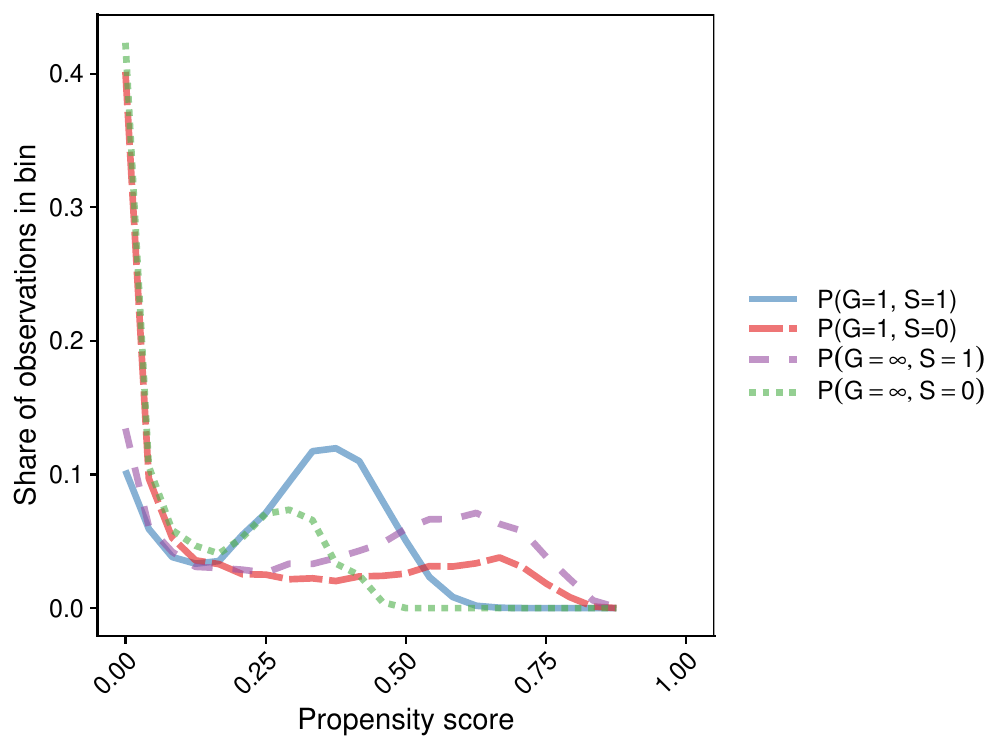}
    \label{fig:propensity_score_density_AGE40_fullx_alpha1_beta1}
    {\\ \raggedright \textbf{Notes:} This figure presents the estimated propensity scores using the data-generating processes described in Section \ref{section:simulation_design}. Each line plots the share of observations within a propensity score bin, with 22 bins.  \par }

\end{figure}

In the original design by \textcite{gruberIncidenceMandatedMaternity1994}, the subgroups of interest are married men age 20-40, married women age 20-40, and single women age 20-40, while the comparison subgroup is single men age 20-40 and all individuals over age 40. This design creates a clear case of limited overlap in a potentially important covariate: age. Individuals over age 40 can only belong to the comparison subgroup. Because of this, in my main simulation sample, I only consider individuals between the ages of 20 and 40. Even so, marriage is highly correlated with age. The estimated propensity scores for each subgroup are plotted in Figure \ref{fig:propensity_score_density_AGE40_fullx_alpha1_beta1}. The nature of the overlap problem is seen by the concentration of units at very small propensity scores.

Limited overlap can create substantial issues, in particular, when using IPW and doubly-robust estimators. In these estimators, extremely small propensity scores can cause large weight to be placed on a small number of observations, resulting in imprecise estimates, large standard errors, and slow convergence to asymptotic properties \parencite{khanIrregularIdentificationSupport2010, maRobustInferenceUsing2020, heilerValidInferenceTreatment2021}.

To improve performance of IPW and doubly-robust estimators in this case, several authors have proposed trimming propensity scores, that is, dropping observations with extreme propensity scores (eg, \textcite{crumpDealingLimitedOverlap2009a}).\footnote{Other approaches include stabilizing the weights by replacing the (potentially extreme) individually-estimated propensity scores with averages for strata of similar propensity scores \parencite{kangDemystifyingDoubleRobustness2007}. } Although trimming changes the population over which the estimates are generated (for example, in this application, by potentially dropping individuals over age 40, who may not overlap with the subgroups of interest) and may reduce the double-robustness property \parencite{maDoublyRobustEstimators2026}, trimming has the potential to substantially improve the performance of these estimators. Popular rule-of-thumb trimming approaches in two-treatment designs include trimming all propensity scores more extreme than 0.1 and 0.9 \parencite{crumpDealingLimitedOverlap2009a}. 

Relevant to the $CDATT$, for which propensity scores are estimated for at least four possible combinations of treatment status and subgroup status, are approaches dealing with multivalued treatments. In the case where propensity scores are estimated for multivalued treatments, \textcite{yangPropensityScoreMatching2016} develop optimal trimming rules for the estimation of ATTs in multi-treatment designs. However, this optimality is based on minimizing the variance of the parameter of interest and so does not translate directly to estimating the $CDATT$. \textcite{yoshidaMultinomialExtensionPropensity2019} propose multinomial extensions to popular rules of thumb, for example, arguing that, in the multinomial case, the 0.1 rule of thumb proposed by \textcite{crumpDealingLimitedOverlap2009a} translates to a smaller threshold. Because of this, in the main simulation design, when estimating the CDATT, I trim observations with extreme propensity scores (less than 0.025). In the results below, I test sensitivity to trimming propensity scores less than 0.01, 0.025, 0.05, and 0.1 as a range of plausible trimming values covering established rules of thumb. 

There are three insights from these results. 

\begin{figure}[h]
    \caption{Estimator performance under various trimming scenarios when overlap is moderately limited}
    \centering
    \includegraphics[width=0.9\linewidth]{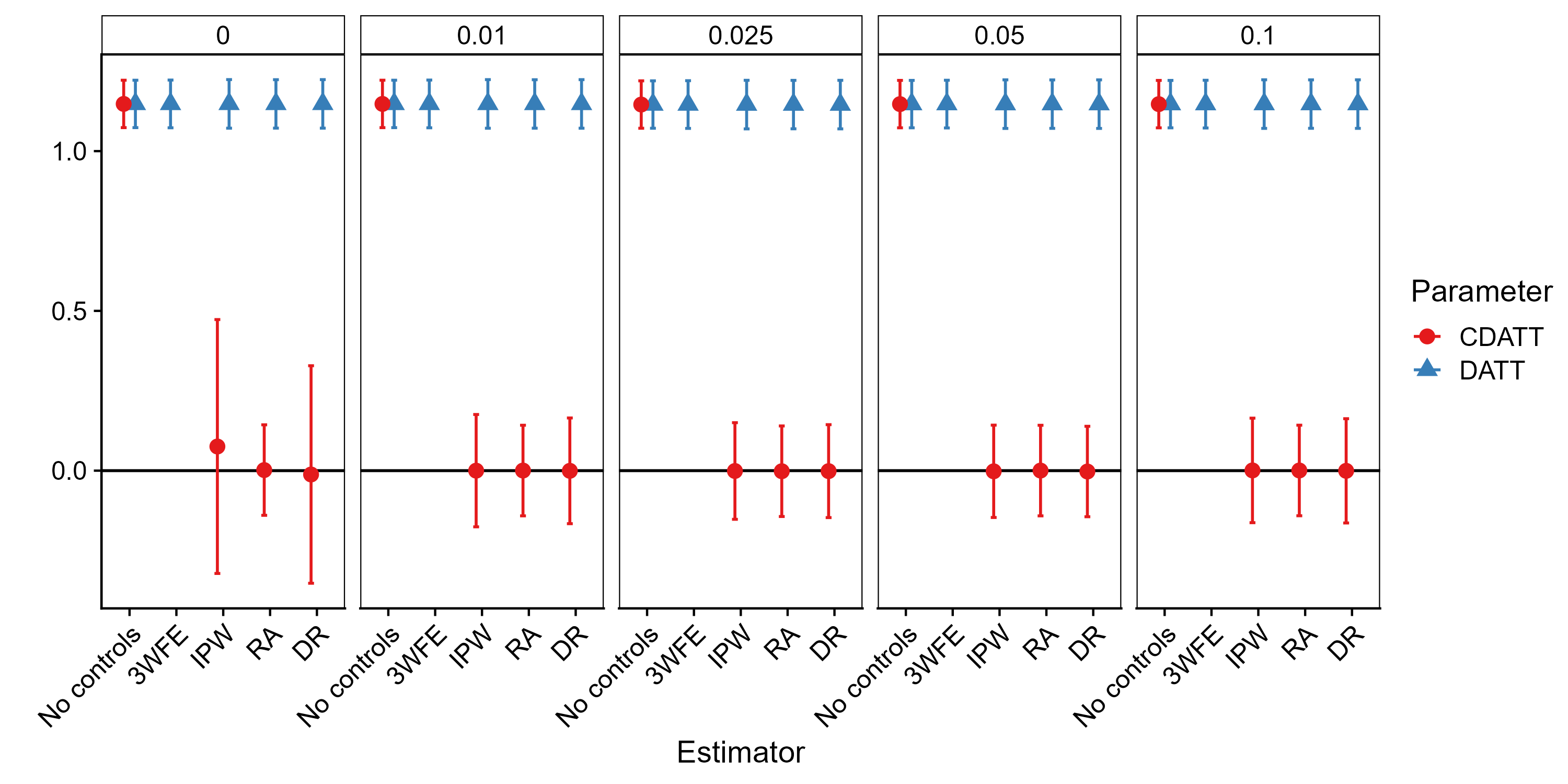}
    \label{fig:calibrated_by_trim}
    {\\ \raggedright \textbf{Notes:} This figure presents the results of 2000 Monte Carlo simulations using the data-generating processes described above, for various trimming thresholds. The points represent the average value of the estimate across 2000 simulations. 95\% confidence intervals are shown. \par }

\end{figure}

First, without trimming, limited overlap causes substantial imprecision when estimating the $CDATT$ using either the IPW or DR estimators. The results appear in Table \ref{tab:sim_study_att_by_trim} and Figure \ref{fig:calibrated_by_trim}. When using either the IPW or DR estimators of the CDATT, the standard errors become quite large, as large as 0.20, relative to a standard error of 0.072 when using regression adjustment. This results in large confidence intervals, whose length can be up to 70\% of the size of the bias when not using controls. Although the total possible sample size is larger when allowing individuals with extreme propensity scores to be included in the sample, this does not offer much or any improvement in precision due to the limited overlap. 

Second, without trimming, coverage of the IPW and DR estimators suffers somewhat. This finding is consistent with prior work on other IPW and DR estimators, which has shown that limited overlap can cause severe underestimation of standard errors by conventional methods \parencite{heilerValidInferenceTreatment2021, maRobustInferenceUsing2020}. Readers are referred to \textcite{heilerValidInferenceTreatment2021, dornHowMuchWeak2025, khanDoublyrobustHeteroscedasticityawareSample2024, maRobustInferenceUsing2020} for recent suggestions on alternative approaches that perform better when there are overlap concerns. 

Third, trimming offers substantial efficiency gains in these estimates. Across all propensity score thresholds tested, trimming results in reduces the standard error of the IPW and doubly-robust estimators of the CDATT to less than 0.1. These standard errors are much closer to those for the RA estimator, and they result in correct coverage. The results also highlight the tradeoff between trimming observations with extreme propensity scores and reducing the sample size. The standard errors are smallest when using a moderate level of trimming (trimming observations with propensity scores less than 0.025 or 0.05) and are slightly larger when trimming more observations.

\newpage 
\FloatBarrier
\begin{table}[h]
    \caption{Simulation study results based on 2000 Monte Carlo trials, moderately limited overlap and trimming}
    \label{tab:sim_study_att_by_trim}
    \centering
    \resizebox{0.6\textwidth}{!}{
    \begin{tabular}{lccccccc}
    \multicolumn{5}{l}{\textbf{No trimming}} \\ \hline
    % latex table generated in R 4.4.3 by xtable 1.8-4 package
% Fri Sep  4 02:35:08 2026
Estimator & Avg. bias & Med. bias & RMSE & Asym. SE & Cover & CI Len \\ 
  \hline
No controls & 1.148 & 1.148 & 1.148 & 0.038 & 0.000 & 0.148 \\ 
 % $DATT^{3WFE}$ & 1.148 & 1.148 & 1.148 & 0.038 & 0.000 & 0.149 \\ 
 % $DATT^{RA}$ & 1.148 & 1.148 & 1.148 & 0.039 & 0.000 & 0.152 \\ 
 % $DATT^{IPW}$ & 1.148 & 1.148 & 1.148 & 0.039 & 0.000 & 0.152 \\ 
 % $DATT^{DR}$ & 1.148 & 1.148 & 1.148 & 0.039 & 0.000 & 0.152 \\ 
  $CDATT^{RA}$ & 0.002 & 0.001 & 0.074 & 0.072 & 0.938 & 0.283 \\ 
  $CDATT^{IPW}$ & 0.076 & 0.140 & 0.350 & 0.203 & 0.726 & 0.795 \\ 
  $CDATT^{DR}$ & -0.012 & -0.017 & 0.269 & 0.174 & 0.878 & 0.681 \\ 
   \hline
 \\
    \multicolumn{5}{l}{\textbf{Trimmed propensity scores $< 0.01$}} \\ \hline
    % latex table generated in R 4.4.3 by xtable 1.8-4 package
% Fri Sep  4 02:35:08 2026
Estimator & Avg. bias & Med. bias & RMSE & Asym. SE & Cover & CI Len \\ 
  \hline
No controls & 1.148 & 1.147 & 1.148 & 0.038 & 0.000 & 0.148 \\ 
% $DATT^{3WFE}$ & 1.148 & 1.147 & 1.148 & 0.038 & 0.000 & 0.149 \\ 
%  $DATT^{RA}$ & 1.148 & 1.147 & 1.148 & 0.039 & 0.000 & 0.152 \\ 
%  $DATT^{IPW}$ & 1.148 & 1.147 & 1.148 & 0.039 & 0.000 & 0.152 \\ 
  %$DATT^{DR}$ & 1.148 & 1.148 & 1.148 & 0.039 & 0.000 & 0.152 \\ 
  $CDATT^{RA}$ & 0.000 & 0.000 & 0.074 & 0.072 & 0.948 & 0.284 \\ 
  $CDATT^{IPW}$ & -0.000 & 0.001 & 0.087 & 0.090 & 0.952 & 0.352 \\ 
  $CDATT^{DR}$ & -0.001 & 0.002 & 0.084 & 0.084 & 0.945 & 0.331 \\ 
   \hline
 \\
    \multicolumn{5}{l}{\textbf{Trimmed propensity scores $< 0.025$}} \\ \hline
    % latex table generated in R 4.4.3 by xtable 1.8-4 package
% Fri Sep  4 02:35:08 2026
Estimator & Avg. bias & Med. bias & RMSE & Asym. SE & Cover & CI Len \\ 
  \hline
No controls & 1.146 & 1.146 & 1.146 & 0.038 & 0.000 & 0.148 \\ 
 % $DATT^{3WFE}$ & 1.146 & 1.146 & 1.146 & 0.038 & 0.000 & 0.149 \\ 
 % $DATT^{RA}$ & 1.146 & 1.146 & 1.146 & 0.039 & 0.000 & 0.152 \\ 
 % $DATT^{IPW}$ & 1.146 & 1.146 & 1.146 & 0.039 & 0.000 & 0.152 \\ 
 % $DATT^{DR}$ & 1.146 & 1.146 & 1.146 & 0.039 & 0.000 & 0.152 \\ 
  $CDATT^{RA}$ & -0.002 & -0.000 & 0.071 & 0.072 & 0.952 & 0.283 \\ 
  $CDATT^{IPW}$ & -0.001 & 0.000 & 0.078 & 0.077 & 0.951 & 0.302 \\ 
  $CDATT^{DR}$ & -0.001 & 0.000 & 0.074 & 0.074 & 0.948 & 0.291 \\ 
   \hline
 \\
    \multicolumn{5}{l}{\textbf{Trimmed propensity scores $< 0.05$}} \\ \hline
    % latex table generated in R 4.4.3 by xtable 1.8-4 package
% Fri Sep  4 02:35:08 2026
Estimator & Avg. bias & Med. bias & RMSE & Asym. SE & Cover & CI Len \\ 
  \hline
No controls & 1.147 & 1.148 & 1.148 & 0.038 & 0.000 & 0.148 \\ 
 % $DATT^{3WFE}$ & 1.147 & 1.148 & 1.148 & 0.038 & 0.000 & 0.149 \\ 
 % $DATT^{RA}$ & 1.147 & 1.148 & 1.148 & 0.039 & 0.000 & 0.152 \\ 
 % $DATT^{IPW}$ & 1.147 & 1.147 & 1.148 & 0.039 & 0.000 & 0.152 \\ 
 % $DATT^{DR}$ & 1.147 & 1.148 & 1.148 & 0.039 & 0.000 & 0.152 \\ 
  $CDATT^{RA}$ & 0.000 & -0.002 & 0.073 & 0.072 & 0.947 & 0.283 \\ 
  $CDATT^{IPW}$ & -0.002 & -0.001 & 0.077 & 0.074 & 0.935 & 0.289 \\ 
  $CDATT^{DR}$ & -0.003 & -0.001 & 0.076 & 0.072 & 0.942 & 0.283 \\ 
   \hline
 \\
    \multicolumn{5}{l}{\textbf{Trimmed propensity scores $< 0.1$}} \\ \hline
    % latex table generated in R 4.4.3 by xtable 1.8-4 package
% Fri Sep  4 02:35:08 2026
Estimator & Avg. bias & Med. bias & RMSE & Asym. SE & Cover & CI Len \\ 
  \hline
No controls & 1.147 & 1.146 & 1.148 & 0.038 & 0.000 & 0.148 \\ 
 % $DATT^{3WFE}$ & 1.147 & 1.146 & 1.148 & 0.038 & 0.000 & 0.149 \\ 
 % $DATT^{RA}$ & 1.147 & 1.148 & 1.148 & 0.039 & 0.000 & 0.152 \\ 
 % $DATT^{IPW}$ & 1.147 & 1.147 & 1.148 & 0.039 & 0.000 & 0.152 \\ 
 % $DATT^{DR}$ & 1.147 & 1.147 & 1.148 & 0.039 & 0.000 & 0.152 \\ 
  $CDATT^{RA}$ & 0.001 & 0.002 & 0.070 & 0.072 & 0.958 & 0.283 \\ 
  $CDATT^{IPW}$ & 0.001 & 0.001 & 0.084 & 0.083 & 0.950 & 0.327 \\ 
  $CDATT^{DR}$ & -0.001 & -0.001 & 0.083 & 0.083 & 0.949 & 0.326 \\ 
   \hline
 \\
    \end{tabular}
}
{\\ \raggedright \textbf{Notes:} This table presents the results of 2000 Monte Carlo simulations using the data-generating processes described above. Avg. bias refers to the average value of the estimate across 2000 simulations. Median bias refers to the median estimate. RMSE refers to the root mean squared error. SE refers to the average standard error across the trials. Cover refers to the 95\% confidence interval coverage rate. CI Len refers to the average length of the 95\% confidence interval. \par }
\end{table}
\newpage

\begin{figure}[h]
    \caption{Estimator performance under various trimming scenarios when overlap is severely limited}
    \centering
    \includegraphics[width=0.9\linewidth]{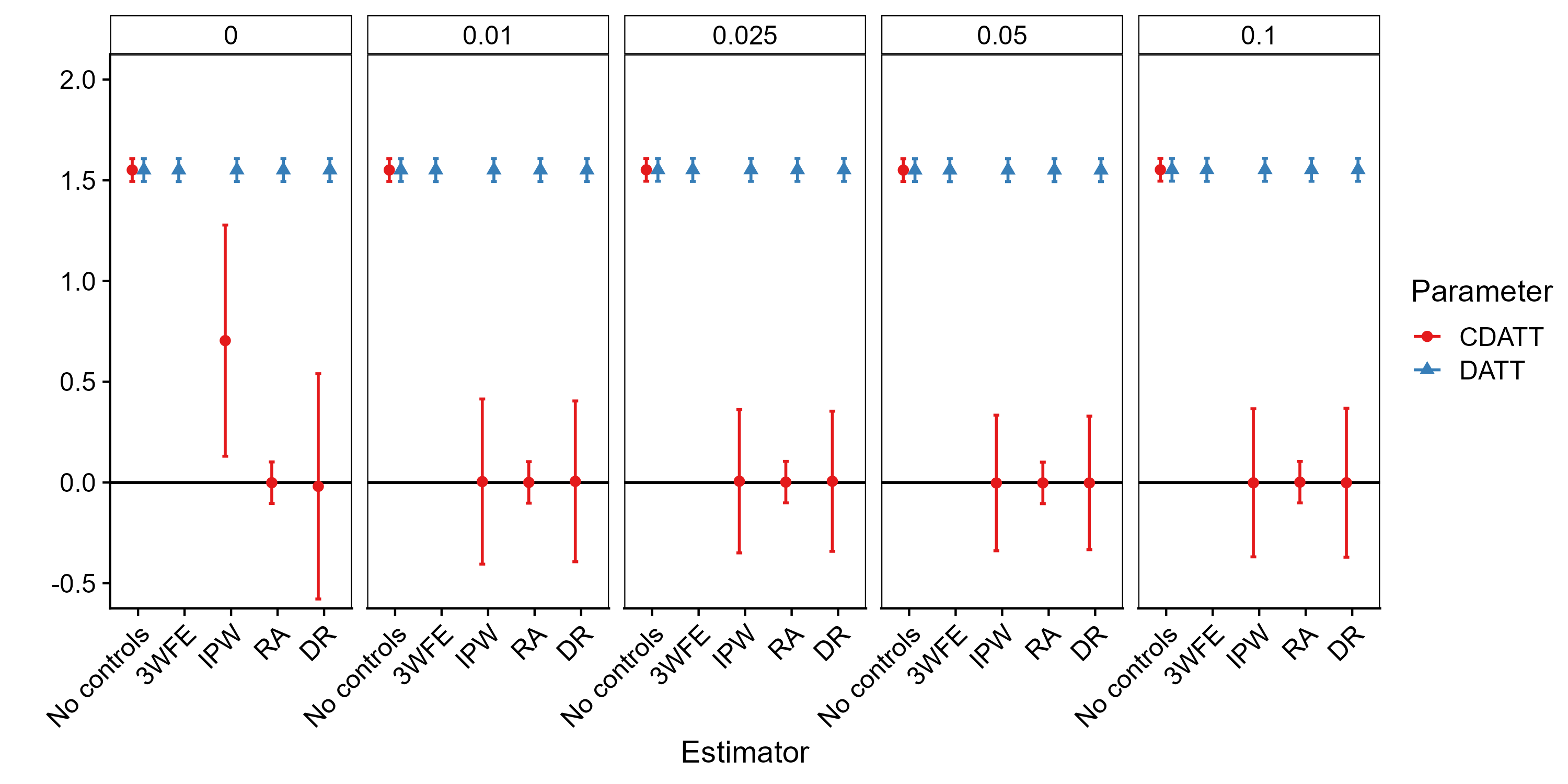}
    \label{fig:calibrated_att_by_overlap_zoomed}
    {\\ \raggedright \textbf{Notes:} This figure presents the results of 2000 Monte Carlo simulations using the data-generating processes described above, for various trimming thresholds. The points represent the average value of the estimate across 2000 simulations. 95\% confidence intervals are shown. \par }

\end{figure}

Next, to better understand the role of limited overlap, I modify the simulation design to highlight a case of even more extremely limited overlap. I use the same data generating process described in Section \ref{section:simulation_design}, but allow individuals up to age 60 to enter the sample. The results from this exercise are shown in Figure \ref{fig:calibrated_att_by_overlap_zoomed} and Table \ref{tab:sim_study_att_by_overlap}. As above, trimming is reasonably successful in improving estimator performance, both reducing estimated standard errors and improving coverage. 

Thus, although conventional trimming approaches have the potential weakness of altering the population over which average treatment effects, DATTs, or CDATTs are identified, they can offer useful rule-of-thumb approaches to improve estimator performance.

\begin{table}[h]
    \caption{Simulation study results based on 2000 Monte Carlo trials, severely limited overlap and trimming}
    \label{tab:sim_study_att_by_overlap}
    \centering
    \resizebox{0.6\textwidth}{!}{
    \begin{tabular}{lccccccc}
    \multicolumn{5}{l}{\textbf{No trimming}} \\ \hline
    % latex table generated in R 4.4.3 by xtable 1.8-4 package
% Fri Sep  4 02:35:14 2026
Estimator & Avg. bias & Med. bias & RMSE & Asym. SE & Cover & CI Len \\ 
  \hline
No controls & 1.551 & 1.551 & 1.551 & 0.029 & 0.000 & 0.113 \\ 
%  $DATT^{3WFE}$ & 1.551 & 1.551 & 1.551 & 0.029 & 0.000 & 0.115 \\ 
%  $DATT^{RA}$ & 1.551 & 1.551 & 1.552 & 0.029 & 0.000 & 0.114 \\ 
%  $DATT^{IPW}$ & 1.551 & 1.551 & 1.552 & 0.029 & 0.000 & 0.114 \\ 
%  $DATT^{DR}$ & 1.551 & 1.551 & 1.552 & 0.029 & 0.000 & 0.114 \\ 
  $CDATT^{RA}$ & -0.001 & -0.001 & 0.051 & 0.053 & 0.956 & 0.206 \\ 
  $CDATT^{IPW}$ & 0.704 & 0.738 & 0.882 & 0.293 & 0.320 & 1.147 \\ 
  $CDATT^{DR}$ & -0.019 & 0.002 & 0.509 & 0.285 & 0.771 & 1.118 \\ 
   \hline
 \\
    \multicolumn{5}{l}{\textbf{Trimmed propensity scores $< 0.01$}} \\ \hline
    % latex table generated in R 4.4.3 by xtable 1.8-4 package
% Fri Sep  4 02:35:14 2026
Estimator & Avg. bias & Med. bias & RMSE & Asym. SE & Cover & CI Len \\ 
  \hline
No controls & 1.551 & 1.551 & 1.552 & 0.029 & 0.000 & 0.113 \\ 
 % $DATT^{3WFE}$ & 1.551 & 1.551 & 1.552 & 0.029 & 0.000 & 0.115 \\ 
 % $DATT^{RA}$ & 1.551 & 1.551 & 1.551 & 0.029 & 0.000 & 0.114 \\ 
 % $DATT^{IPW}$ & 1.551 & 1.551 & 1.551 & 0.029 & 0.000 & 0.114 \\ 
  %$DATT^{DR}$ & 1.551 & 1.551 & 1.551 & 0.029 & 0.000 & 0.114 \\ 
  $CDATT^{RA}$ & 0.001 & 0.001 & 0.054 & 0.053 & 0.946 & 0.206 \\ 
  $CDATT^{IPW}$ & 0.005 & 0.002 & 0.227 & 0.209 & 0.925 & 0.820 \\ 
  $CDATT^{DR}$ & 0.006 & -0.001 & 0.218 & 0.204 & 0.926 & 0.798 \\ 
   \hline
 \\
    \multicolumn{5}{l}{\textbf{Trimmed propensity scores $< 0.025$}} \\ \hline
    % latex table generated in R 4.4.3 by xtable 1.8-4 package
% Fri Sep  4 02:35:14 2026
Estimator & Avg. bias & Med. bias & RMSE & Asym. SE & Cover & CI Len \\ 
  \hline
No controls & 1.552 & 1.552 & 1.552 & 0.029 & 0.000 & 0.113 \\ 
  %$DATT^{3WFE}$ & 1.552 & 1.552 & 1.552 & 0.029 & 0.000 & 0.115 \\ 
  %$DATT^{RA}$ & 1.552 & 1.552 & 1.552 & 0.029 & 0.000 & 0.114 \\ 
  %$DATT^{IPW}$ & 1.552 & 1.552 & 1.552 & 0.029 & 0.000 & 0.114 \\ 
  %$DATT^{DR}$ & 1.552 & 1.552 & 1.552 & 0.029 & 0.000 & 0.114 \\ 
  $CDATT^{RA}$ & 0.002 & 0.000 & 0.053 & 0.053 & 0.943 & 0.206 \\ 
  $CDATT^{IPW}$ & 0.006 & 0.004 & 0.187 & 0.181 & 0.943 & 0.711 \\ 
  $CDATT^{DR}$ & 0.006 & 0.004 & 0.180 & 0.177 & 0.945 & 0.696 \\ 
   \hline
 \\
    \multicolumn{5}{l}{\textbf{Trimmed propensity scores $< 0.05$}} \\ \hline
    % latex table generated in R 4.4.3 by xtable 1.8-4 package
% Fri Sep  4 02:35:14 2026
Estimator & Avg. bias & Med. bias & RMSE & Asym. SE & Cover & CI Len \\ 
  \hline
No controls & 1.551 & 1.551 & 1.551 & 0.029 & 0.000 & 0.113 \\ 
 % $DATT^{3WFE}$ & 1.551 & 1.551 & 1.551 & 0.029 & 0.000 & 0.115 \\ 
 % $DATT^{RA}$ & 1.551 & 1.551 & 1.551 & 0.029 & 0.000 & 0.114 \\ 
 % $DATT^{IPW}$ & 1.551 & 1.551 & 1.551 & 0.029 & 0.000 & 0.114 \\ 
 % $DATT^{DR}$ & 1.551 & 1.551 & 1.551 & 0.029 & 0.000 & 0.114 \\ 
  $CDATT^{RA}$ & -0.002 & -0.003 & 0.052 & 0.053 & 0.951 & 0.206 \\ 
  $CDATT^{IPW}$ & -0.002 & -0.007 & 0.186 & 0.172 & 0.926 & 0.673 \\ 
  $CDATT^{DR}$ & -0.002 & -0.010 & 0.182 & 0.169 & 0.931 & 0.663 \\ 
   \hline
 \\
    \multicolumn{5}{l}{\textbf{Trimmed propensity scores $< 0.1$}} \\ \hline
    % latex table generated in R 4.4.3 by xtable 1.8-4 package
% Fri Sep  4 02:35:14 2026
Estimator & Avg. bias & Med. bias & RMSE & Asym. SE & Cover & CI Len \\ 
  \hline
No controls & 1.552 & 1.553 & 1.553 & 0.029 & 0.000 & 0.113 \\ 
  %$DATT^{3WFE}$ & 1.552 & 1.553 & 1.553 & 0.029 & 0.000 & 0.115 \\ 
  %$DATT^{RA}$ & 1.552 & 1.552 & 1.553 & 0.029 & 0.000 & 0.114 \\ 
  %$DATT^{IPW}$ & 1.552 & 1.552 & 1.553 & 0.029 & 0.000 & 0.114 \\ 
  %$DATT^{DR}$ & 1.552 & 1.552 & 1.553 & 0.029 & 0.000 & 0.114 \\ 
  $CDATT^{RA}$ & 0.002 & 0.003 & 0.053 & 0.053 & 0.948 & 0.206 \\ 
  $CDATT^{IPW}$ & -0.001 & 0.003 & 0.190 & 0.188 & 0.947 & 0.735 \\ 
  $CDATT^{DR}$ & -0.001 & 0.004 & 0.189 & 0.189 & 0.947 & 0.739 \\ 
   \hline
 \\
    \end{tabular}
}
{\\ \raggedright \textbf{Notes:} This table presents the results of 2000 Monte Carlo simulations using the data-generating processes described above. Avg. bias refers to the average value of the estimate across 2000 simulations. Median bias refers to the median estimate. RMSE refers to the root mean squared error. SE refers to the average standard error across the trials. Cover refers to the 95\% confidence interval coverage rate. CI Len refers to the average length of the 95\% confidence interval. \par }
\end{table}

\FloatBarrier

\subsection{Simulation with balanced subgroups}\label{appendix_simulation_balanced}
\FloatBarrier

\begin{figure}[h]
    \caption{Estimator performance under various treatment effect heterogeneity scenarios, balanced subgroups}
    \centering
    \includegraphics[width=0.9\linewidth]{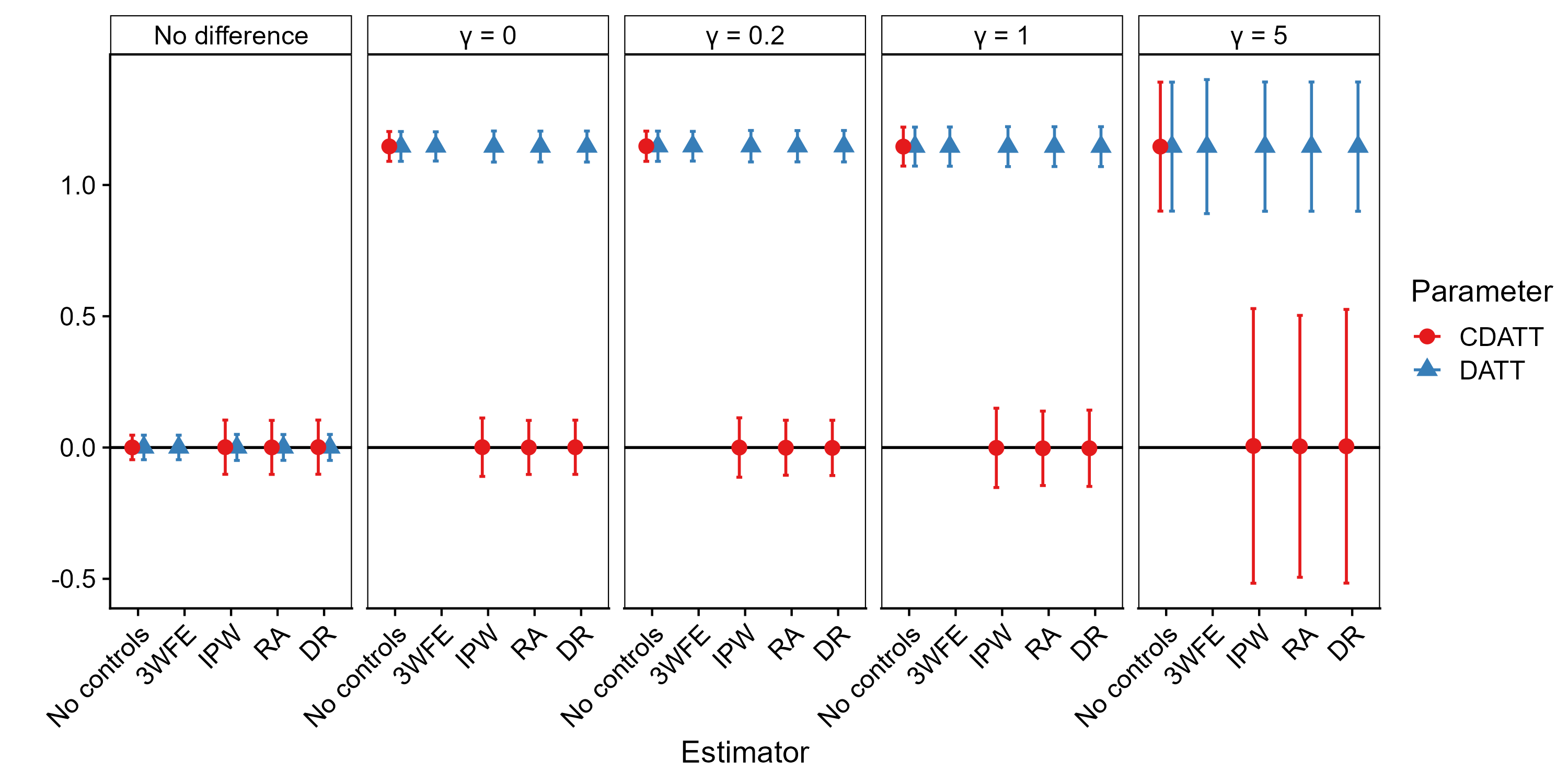}
    \label{fig:calibrated_by_het_with_intercept}
    {\\ \raggedright \textbf{Notes:} This figure presents the results of 2000 Monte Carlo simulations using the data-generating processes described in Section \ref{section:simulation_design}, with subgroup probabilities re-normalized to ensure balance between them. The points represent the average value of the estimate across 2000 simulations. 95\% confidence intervals are shown. ``No difference'' refers to a scenario with no average treatment effect and no difference between subgroups. \par }

\end{figure}

In this section, results are presented for a simulation specification that mirrors that in \ref{section:simulation_design} with subgroup probabilities re-normalized to ensure that each subgroup receives 25\% of the sample. Figure \ref{fig:calibrated_by_het_with_intercept} and Table \ref{tab:sim_study_att_by_het_balanced} show that this change in balance of the subgroups does not substantially change the results. 

\newpage
\begin{table}[H]
    \caption{Simulation study results based on 2000 Monte Carlo trials, balanced subgroups}
    \label{tab:sim_study_att_by_het_balanced}
    \centering
    \resizebox{0.55\textwidth}{!}{
    \begin{tabular}{lccccccc}
    \multicolumn{5}{l}{\textbf{No average treatment effect on the treated}} \\ \hline
    % latex table generated in R 4.4.3 by xtable 1.8-4 package
% Fri Sep  4 02:35:10 2026
Estimator & Avg. bias & Med. bias & RMSE & SE & Cover & CI Len \\ 
  \hline
No controls & 0.000 & 0.001 & 0.024 & 0.024 & 0.950 & 0.093 \\ 
  $DATT^{3WFE}$ & 0.000 & 0.001 & 0.024 & 0.024 & 0.949 & 0.093 \\ 
  $DATT^{RA}$ & 0.000 & 0.001 & 0.025 & 0.025 & 0.947 & 0.098 \\ 
  $DATT^{IPW}$ & 0.000 & 0.001 & 0.025 & 0.025 & 0.944 & 0.099 \\ 
  $DATT^{DR}$ & 0.000 & 0.001 & 0.025 & 0.025 & 0.943 & 0.099 \\ 
  $CDATT^{RA}$ & 0.001 & 0.002 & 0.052 & 0.052 & 0.953 & 0.206 \\ 
  $CDATT^{IPW}$ & 0.001 & 0.001 & 0.052 & 0.053 & 0.951 & 0.207 \\ 
  $CDATT^{DR}$ & 0.001 & 0.002 & 0.052 & 0.053 & 0.955 & 0.206 \\ 
   \hline
 \\
    \multicolumn{5}{l}{\textbf{$\mathbf{\gamma = 0}$ }} \\ \hline
    % latex table generated in R 4.4.3 by xtable 1.8-4 package
% Fri Sep  4 02:35:10 2026
Estimator & Avg. bias & Med. bias & RMSE & SE & Cover & CI Len \\ 
  \hline
No controls & 1.147 & 1.147 & 1.147 & 0.029 & 0.000 & 0.113 \\ 
  $DATT^{3WFE}$ & 1.147 & 1.147 & 1.147 & 0.028 & 0.000 & 0.111 \\ 
  $DATT^{RA}$ & 1.146 & 1.146 & 1.147 & 0.030 & 0.000 & 0.118 \\ 
  $DATT^{IPW}$ & 1.146 & 1.146 & 1.147 & 0.030 & 0.000 & 0.118 \\ 
  $DATT^{DR}$ & 1.146 & 1.146 & 1.147 & 0.030 & 0.000 & 0.118 \\ 
  $CDATT^{RA}$ & 0.000 & 0.000 & 0.053 & 0.052 & 0.949 & 0.206 \\ 
  $CDATT^{IPW}$ & 0.001 & 0.000 & 0.058 & 0.057 & 0.948 & 0.223 \\ 
  $CDATT^{DR}$ & 0.001 & 0.001 & 0.053 & 0.053 & 0.947 & 0.207 \\ 
   \hline
 \\
    \multicolumn{5}{l}{\textbf{$\mathbf{\gamma = 0.2}$}} \\ \hline
    % latex table generated in R 4.4.3 by xtable 1.8-4 package
% Fri Sep  4 02:35:10 2026
Estimator & Avg. bias & Med. bias & RMSE & SE & Cover & CI Len \\ 
  \hline
No controls & 1.146 & 1.146 & 1.147 & 0.038 & 0.000 & 0.148 \\ 
  $DATT^{3WFE}$ & 1.146 & 1.146 & 1.147 & 0.038 & 0.000 & 0.149 \\ 
  $DATT^{RA}$ & 1.146 & 1.147 & 1.147 & 0.039 & 0.000 & 0.152 \\ 
  $DATT^{IPW}$ & 1.146 & 1.146 & 1.147 & 0.039 & 0.000 & 0.152 \\ 
  $DATT^{DR}$ & 1.146 & 1.147 & 1.147 & 0.039 & 0.000 & 0.152 \\ 
  $CDATT^{RA}$ & -0.003 & -0.003 & 0.074 & 0.072 & 0.947 & 0.283 \\ 
  $CDATT^{IPW}$ & -0.001 & -0.001 & 0.079 & 0.077 & 0.942 & 0.302 \\ 
  $CDATT^{DR}$ & -0.003 & -0.002 & 0.077 & 0.074 & 0.945 & 0.291 \\ 
   \hline
 \\
    \multicolumn{5}{l}{\textbf{$\mathbf{\gamma = 1}$}} \\ \hline
    % latex table generated in R 4.4.3 by xtable 1.8-4 package
% Fri Sep  4 02:35:10 2026
Estimator & Avg. bias & Med. bias & RMSE & SE & Cover & CI Len \\ 
  \hline
No controls & 1.148 & 1.147 & 1.148 & 0.029 & 0.000 & 0.115 \\ 
  $DATT^{3WFE}$ & 1.148 & 1.147 & 1.148 & 0.029 & 0.000 & 0.113 \\ 
  $DATT^{RA}$ & 1.148 & 1.147 & 1.148 & 0.030 & 0.000 & 0.119 \\ 
  $DATT^{IPW}$ & 1.148 & 1.147 & 1.148 & 0.031 & 0.000 & 0.120 \\ 
  $DATT^{DR}$ & 1.148 & 1.147 & 1.148 & 0.031 & 0.000 & 0.120 \\ 
  $CDATT^{RA}$ & -0.001 & -0.000 & 0.053 & 0.053 & 0.948 & 0.209 \\ 
  $CDATT^{IPW}$ & 0.000 & 0.000 & 0.059 & 0.058 & 0.949 & 0.226 \\ 
  $CDATT^{DR}$ & -0.001 & -0.000 & 0.055 & 0.054 & 0.944 & 0.211 \\ 
   \hline
 \\
    \multicolumn{5}{l}{\textbf{$\mathbf{\gamma = 5}$}} \\ \hline
    % latex table generated in R 4.4.3 by xtable 1.8-4 package
% Fri Sep  4 02:35:10 2026
Estimator & Avg. bias & Med. bias & RMSE & SE & Cover & CI Len \\ 
  \hline
No controls & 1.146 & 1.150 & 1.153 & 0.125 & 0.000 & 0.491 \\ 
  $DATT^{3WFE}$ & 1.146 & 1.150 & 1.153 & 0.130 & 0.000 & 0.510 \\ 
  $DATT^{RA}$ & 1.146 & 1.150 & 1.153 & 0.126 & 0.000 & 0.492 \\ 
  $DATT^{IPW}$ & 1.146 & 1.150 & 1.153 & 0.126 & 0.000 & 0.492 \\ 
  $DATT^{DR}$ & 1.146 & 1.150 & 1.153 & 0.126 & 0.000 & 0.492 \\ 
  $CDATT^{RA}$ & 0.005 & 0.011 & 0.255 & 0.254 & 0.955 & 0.997 \\ 
  $CDATT^{IPW}$ & 0.006 & 0.017 & 0.273 & 0.267 & 0.946 & 1.046 \\ 
  $CDATT^{DR}$ & 0.005 & 0.016 & 0.272 & 0.266 & 0.946 & 1.042 \\ 
   \hline
 \\
    \end{tabular}
}
{\\ \raggedright \textbf{Notes:} This table presents the results of 2000 Monte Carlo simulations using the data-generating processes described in Section \ref{section:simulation_design}, with subgroup probabilities re-normalized to ensure balance between them. Avg. bias refers to the average value of the estimate across 2000 simulations. Median bias refers to the median estimate. RMSE refers to the root mean squared error. SE refers to the average standard error across the trials. Cover refers to the 95\% confidence interval coverage rate. CI Len refers to the average length of the 95\% confidence interval. \par }
\end{table}

\FloatBarrier

\clearpage
\section{Supplementary tables and figures}
\setcounter{figure}{0}    
\setcounter{table}{0}    
\FloatBarrier

\begin{figure}[!h]
    \caption{Results from re-analyzing impacts of childbirth coverage, static estimation}
    \label{fig:gruber_full_static}
    \centering
\begin{subfigure}{.45\textwidth}
    \caption{Impacts on log of hourly wages}
    \centering
    \includegraphics[width=\textwidth]{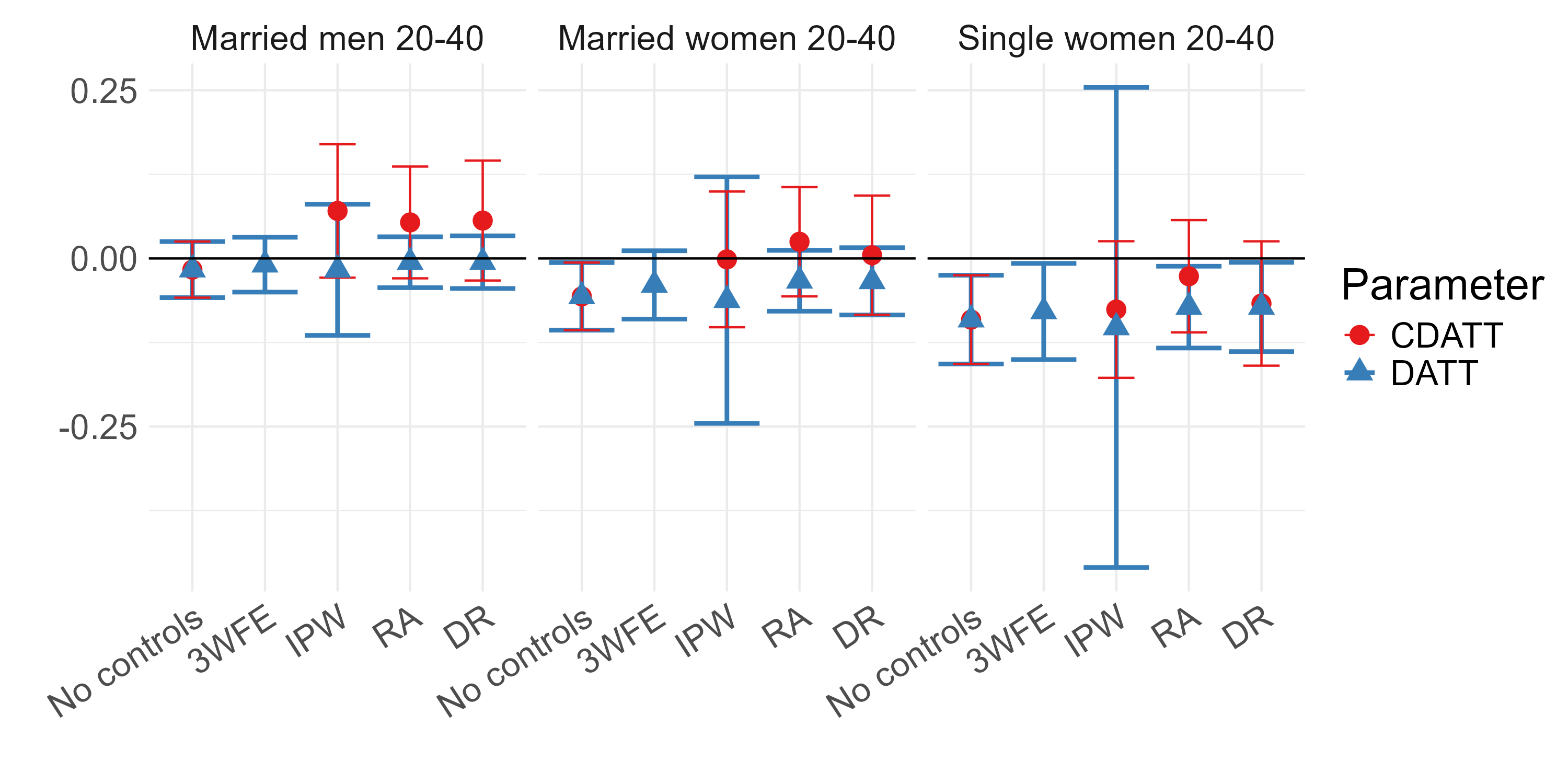}
    \label{fig:gruber_wages}
\end{subfigure}
\begin{subfigure}{.45\textwidth}
    \caption{Impacts on log of hours worked}
    \centering
    \includegraphics[width=\textwidth]{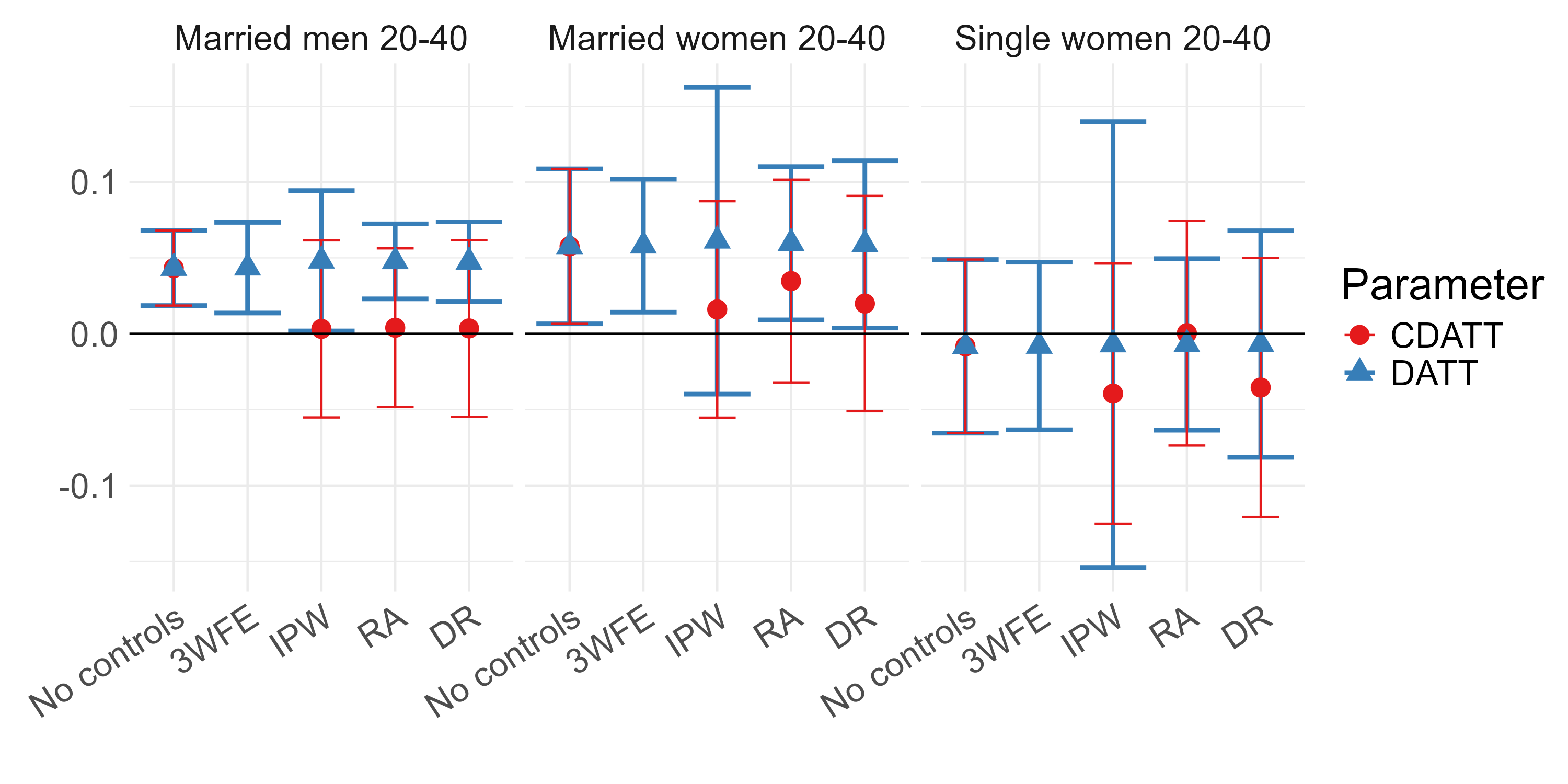}
    \label{fig:gruber_hours}
\end{subfigure}
\begin{subfigure}{.45\textwidth}
    \caption{Impacts on employment}
    \centering
    \includegraphics[width=\textwidth]{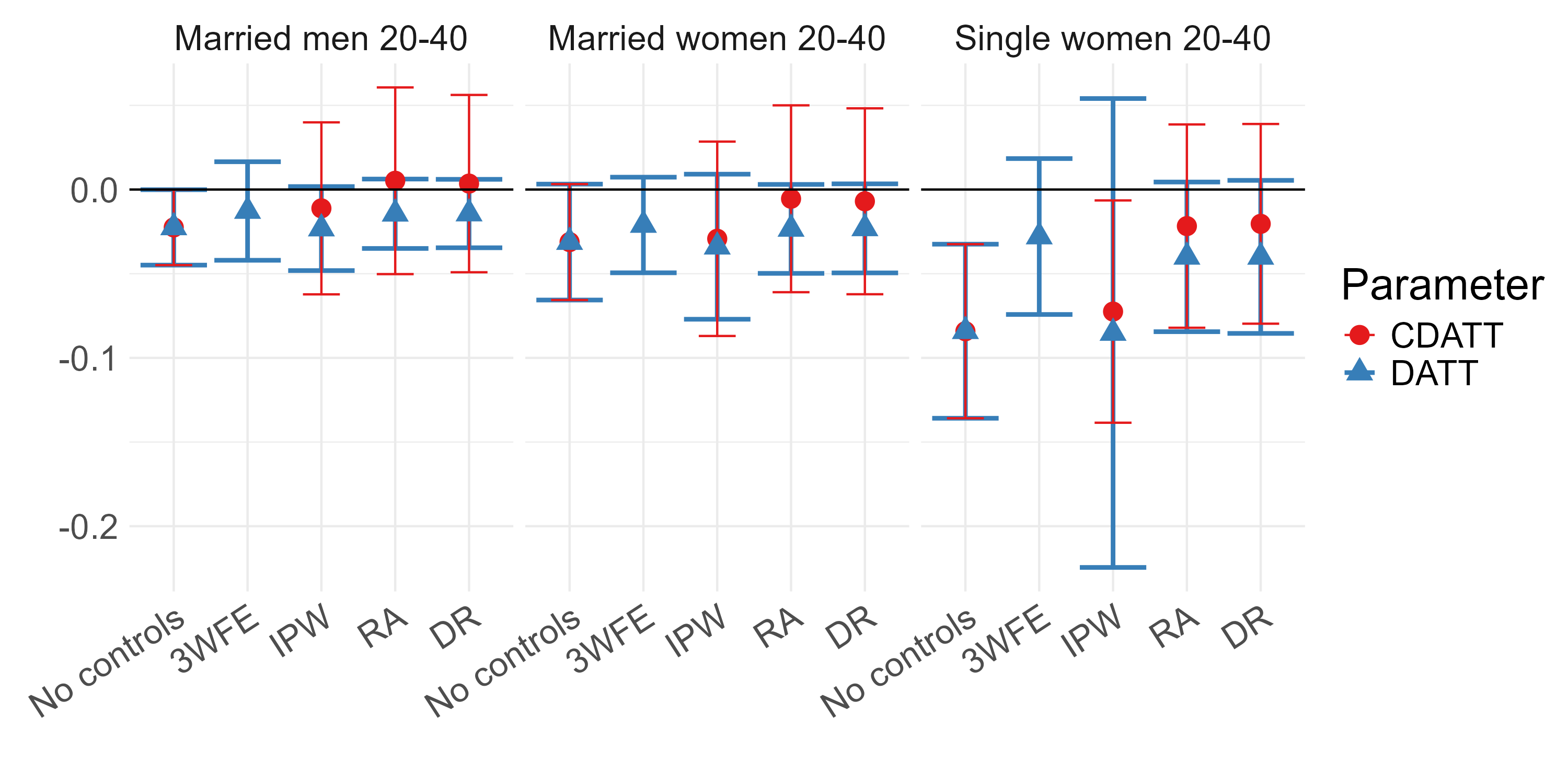}
    \label{fig:gruber_emp}
\end{subfigure}
{\\ \raggedright \textbf{Notes:} This figure plots estimated effects from the re-analysis of \textcite{gruberIncidenceMandatedMaternity1994} using pooled years as in the original paper. 3WFE represents the 3WFE estimator most closely replicating results from the original paper. IPW, RA, and DR represents the inverse propensity score weighting, regression adjustment, and doubly-robust estimators of the DATT and CDATT described in this paper. Effects are plotted for each of the three subgroups of interest (married men 20-40, married women 20-40, and single women 20-40) relative to the comparison subgroup, and for each of three outcomes (the log of hourly wages, the log of hours worked, and employment). 95\% confidence intervals are shown. \par }

\end{figure}

\begin{figure}[!h]
    \caption{Event-study estimates of impacts of childbirth coverage on employment, 3WFE specification}
    \label{fig:gruber_es_3WFE}
    \centering
\begin{subfigure}{.45\textwidth}
    \caption{Impacts on the log of hourly wages}
    \centering
    \includegraphics[width=\textwidth]{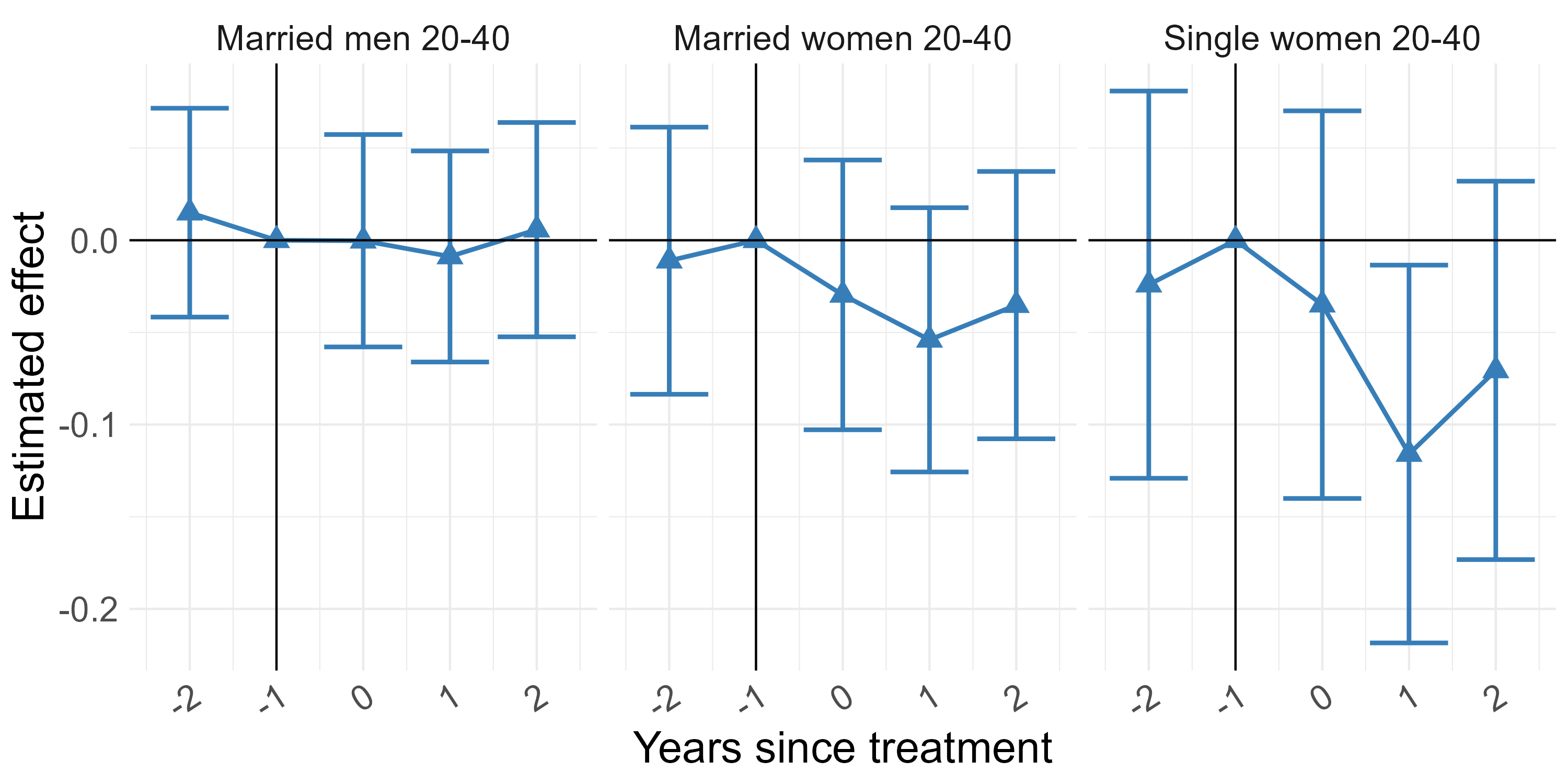}
    \label{fig:gruber_wages_es_3WFE}
\end{subfigure}
\begin{subfigure}{.45\textwidth}
    \caption{Impacts on log of hours worked}
    \centering
    \includegraphics[width=\textwidth]{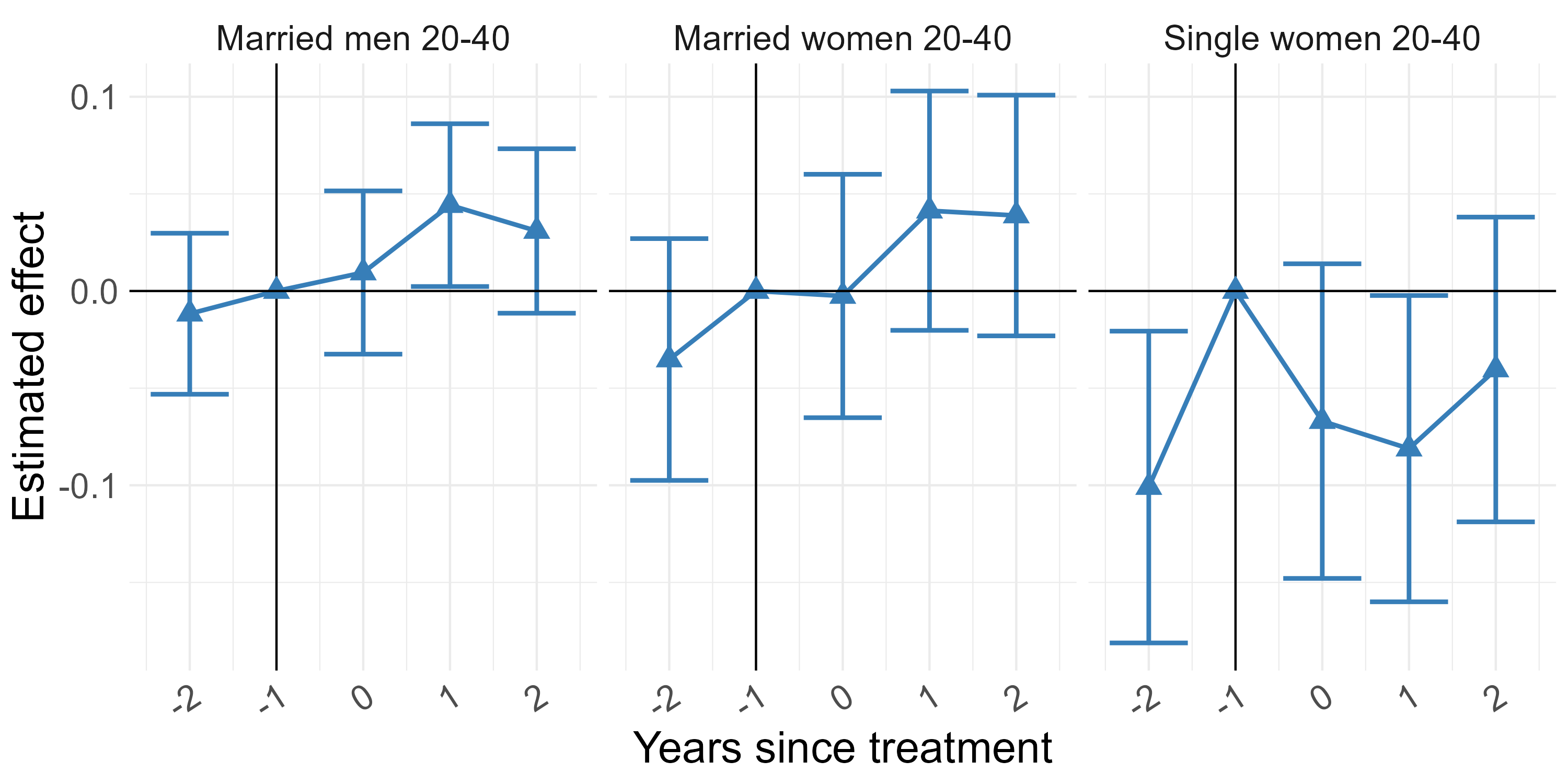}
    \label{fig:gruber_hours_es_3WFE}
\end{subfigure}
\begin{subfigure}{.45\textwidth}
    \caption{Impacts on employment}
    \centering
    \includegraphics[width=\textwidth]{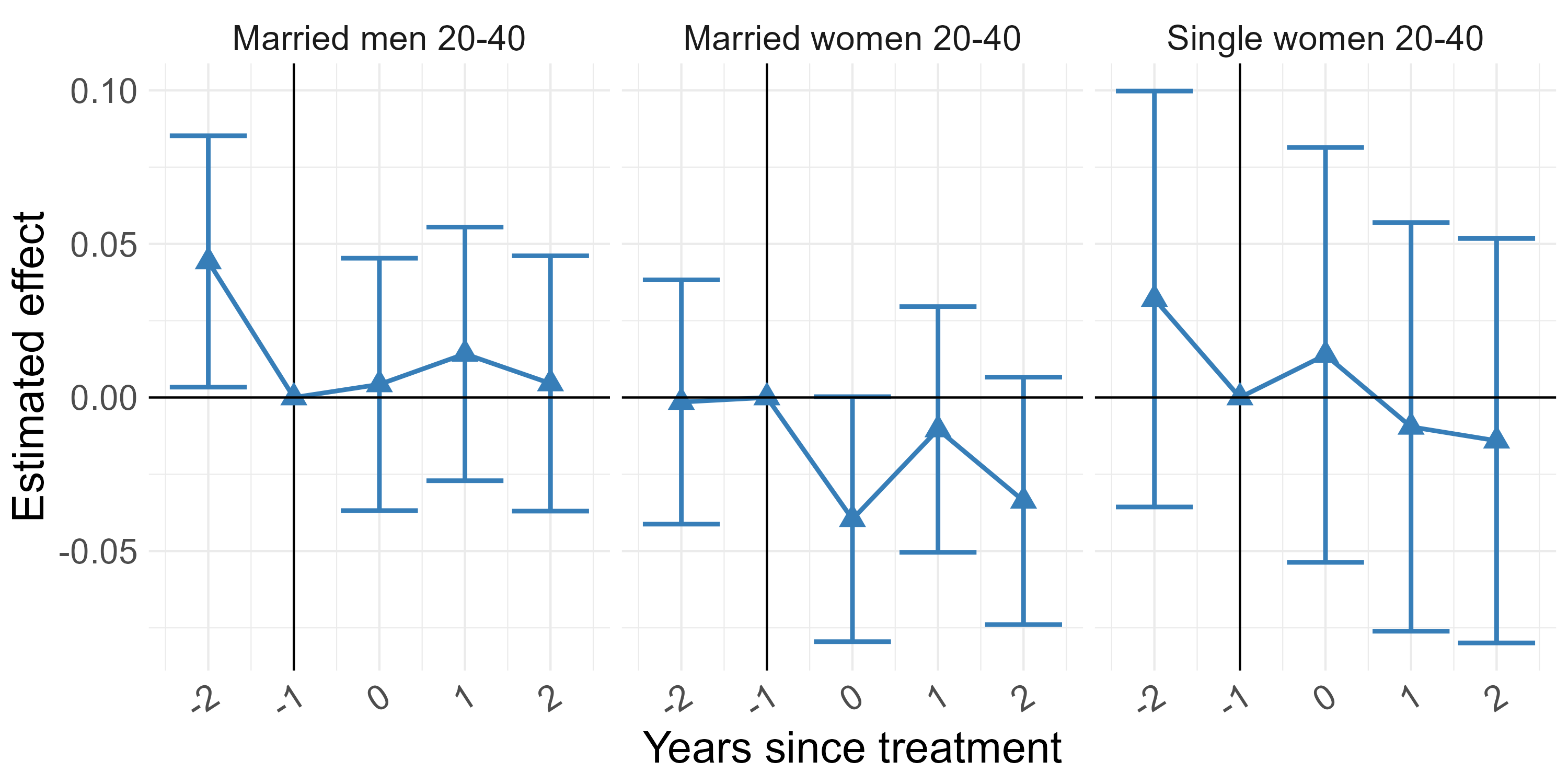}
    \label{fig:gruber_emp_es_3WFE}
\end{subfigure}
{ \\ \raggedright \textbf{Notes:} This figure plots estimated effects of childbirth coverage mandates on employment using an event study approach. Points represent event study estimates drawn from the 3WFE estimators of the DATT and CDATT. Effects are plotted for each of the three subgroups of interest (married men 20-40, married women 20-40, and single women 20-40) and for each outcome of interest. 95\% confidence intervals are shown. \par }
\end{figure}

\end{document}